\documentclass[longauth]{aa}

\usepackage{graphicx,colortbl}
\usepackage{natbib}
\usepackage{scalerel}

\usepackage[version=4]{mhchem}
\usepackage[table]{xcolor}

\usepackage{txfonts,textcomp}
\usepackage{hyperref}
\hypersetup{
    colorlinks=true,
    linkcolor=blue,
    filecolor=magenta,      
    urlcolor=blue,
    citecolor=blue
}
\makeatletter
\renewcommand*\aa@pageof{, page \thepage{} of \pageref*{LastPage}}
\makeatother

\usepackage{euclid}

\newcommand*{\Ih}{\textrm{I}\textsubscript{h}\xspace}
\newcommand*{\Ic}{\textrm{I}\textsubscript{c}\xspace}
\newcommand*{\Isd}{\textrm{I}\textsubscript{sd}\xspace}
\newcommand*{\Iah}{\textrm{I}\textsubscript{ah}\xspace}
\newcommand*{\Ial}{\textrm{I}\textsubscript{al}\xspace}
\newcommand*{\Iar}{\textrm{I}\textsubscript{ar}\xspace}

\newcommand*{\Cassini}{\textit{Cassini}\xspace}

\begin{document}

   \title{\Euclid preparation. XXIX. Water ice in spacecraft part I:\\ The physics of ice formation and contamination}

\author{%\normalsize
Euclid Collaboration: M.~Schirmer\orcid{0000-0003-2568-9994}$^{1}$\thanks{\email{schirmer@mpia.de}}, K.~Th\"{u}rmer\orcid{0000-0002-3078-7372}$^{2}$, B.~Bras$^{3}$, M.~Cropper\orcid{0000-0003-4571-9468}$^{4}$, J.~Martin-Fleitas$^{5}$, Y.~Goueffon$^{6}$, R.~Kohley$^{7}$, A.~Mora\orcid{0000-0002-1922-8529}$^{8}$, M.~Portaluppi$^{3}$, G.~D.~Racca$^{3}$
% ...
}

%% please do not edit the affiliation list -- contact ECEB Bureau for changes
\institute{$^{1}$ Max-Planck-Institut f\"ur Astronomie, K\"onigstuhl 17, 69117 Heidelberg, Germany\\
$^{2}$ Sandia National Laboratories, Livermore, California 94550, USA\\
$^{3}$ European Space Agency/ESTEC, Keplerlaan 1, 2201 AZ Noordwijk, The Netherlands\\
$^{4}$ Mullard Space Science Laboratory, University College London, Holmbury St Mary, Dorking, Surrey RH5 6NT, UK\\
$^{5}$ Aurora Technology B.V., Zwarteweg 39, 2201 AA Noordwijk, The Netherlands\\
$^{6}$ Airbus Defence \& Space SAS, Toulouse, France\\
$^{7}$ ESAC/ESA, Camino Bajo del Castillo, s/n., Urb. Villafranca del Castillo, 28692 Villanueva de la Ca\~nada, Madrid, Spain\\
$^{8}$ Aurora Technology for European Space Agency (ESA), Camino bajo del Castillo, s/n, Urbanizacion Villafranca del Castillo, Villanueva de la Ca\~nada, 28692 Madrid, Spain\\
$^{9}$ Astronomisches Rechen-Institut, Zentrum f\"ur Astronomie der Universit\"at Heidelberg, M\"onchhofstr. 12-14, 69120 Heidelberg, Germany\\
$^{10}$ SYRTE, Observatoire de Paris, Universit\'{e} PSL, CNRS, Sorbonne Universit\'{e}, LNE, 61 avenue de l'Observatoire 75014 Paris, France\\
$^{11}$ INAF-Osservatorio Astrofisico di Torino, Via Osservatorio 20, 10025 Pino Torinese (TO), Italy\\
$^{12}$ Institut de Ci\`{e}ncies del Cosmos (ICCUB), Universitat de Barcelona (IEEC-UB), Mart\'{i} i Franqu\`{e}s 1, 08028 Barcelona, Spain\\
$^{13}$ Institut d'Estudis Espacials de Catalunya (IEEC), Carrer Gran Capit\'a 2-4, 08034 Barcelona, Spain\\
$^{14}$ Max Planck Institute for Extraterrestrial Physics, Giessenbachstr. 1, 85748 Garching, Germany\\
$^{15}$ University Observatory, Faculty of Physics, Ludwig-Maximilians-Universit{\"a}t, Scheinerstr. 1, 81679 Munich, Germany\\
$^{16}$ Departament de F\'isica Qu\`antica i Astrof\'isica (FQA), Universitat de Barcelona (UB), Mart\'{\i} i Franqu\`es 1, E-08028 Barcelona, Spain\\
$^{17}$ Universit\'e Paris-Saclay, CNRS, Institut d'astrophysique spatiale, 91405, Orsay, France\\
$^{18}$ Institute of Cosmology and Gravitation, University of Portsmouth, Portsmouth PO1 3FX, UK\\
$^{19}$ Institut f\"ur Theoretische Physik, University of Heidelberg, Philosophenweg 16, 69120 Heidelberg, Germany\\
$^{20}$ Dipartimento di Fisica e Astronomia "Augusto Righi" - Alma Mater Studiorum Universit\'a di Bologna, via Piero Gobetti 93/2, 40129 Bologna, Italy\\
$^{21}$ INAF-Osservatorio di Astrofisica e Scienza dello Spazio di Bologna, Via Piero Gobetti 93/3, 40129 Bologna, Italy\\
$^{22}$ INFN-Sezione di Bologna, Viale Berti Pichat 6/2, 40127 Bologna, Italy\\
$^{23}$ Dipartimento di Fisica, Universit\'a di Genova, Via Dodecaneso 33, 16146, Genova, Italy\\
$^{24}$ INFN-Sezione di Genova, Via Dodecaneso 33, 16146, Genova, Italy\\
$^{25}$ Department of Physics "E. Pancini", University Federico II, Via Cinthia 6, 80126, Napoli, Italy\\
$^{26}$ INAF-Osservatorio Astronomico di Capodimonte, Via Moiariello 16, 80131 Napoli, Italy\\
$^{27}$ Instituto de Astrof\'isica e Ci\^encias do Espa\c{c}o, Universidade do Porto, CAUP, Rua das Estrelas, PT4150-762 Porto, Portugal\\
$^{28}$ Dipartimento di Fisica, Universit\'a degli Studi di Torino, Via P. Giuria 1, 10125 Torino, Italy\\
$^{29}$ INFN-Sezione di Torino, Via P. Giuria 1, 10125 Torino, Italy\\
$^{30}$ INAF-IASF Milano, Via Alfonso Corti 12, 20133 Milano, Italy\\
$^{31}$ Institut de F\'{i}sica d'Altes Energies (IFAE), The Barcelona Institute of Science and Technology, Campus UAB, 08193 Bellaterra (Barcelona), Spain\\
$^{32}$ Port d'Informaci\'{o} Cient\'{i}fica, Campus UAB, C. Albareda s/n, 08193 Bellaterra (Barcelona), Spain\\
$^{33}$ INAF-Osservatorio Astronomico di Roma, Via Frascati 33, 00078 Monteporzio Catone, Italy\\
$^{34}$ INFN section of Naples, Via Cinthia 6, 80126, Napoli, Italy\\
$^{35}$ Dipartimento di Fisica e Astronomia "Augusto Righi" - Alma Mater Studiorum Universit\'a di Bologna, Viale Berti Pichat 6/2, 40127 Bologna, Italy\\
$^{36}$ Centre National d'Etudes Spatiales -- Centre spatial de Toulouse, 18 avenue Edouard Belin, 31401 Toulouse Cedex 9, France\\
$^{37}$ Institut national de physique nucl\'eaire et de physique des particules, 3 rue Michel-Ange, 75794 Paris C\'edex 16, France\\
$^{38}$ Institute for Astronomy, University of Edinburgh, Royal Observatory, Blackford Hill, Edinburgh EH9 3HJ, UK\\
$^{39}$ Jodrell Bank Centre for Astrophysics, Department of Physics and Astronomy, University of Manchester, Oxford Road, Manchester M13 9PL, UK\\
$^{40}$ European Space Agency/ESRIN, Largo Galileo Galilei 1, 00044 Frascati, Roma, Italy\\
$^{41}$ University of Lyon, Univ Claude Bernard Lyon 1, CNRS/IN2P3, IP2I Lyon, UMR 5822, 69622 Villeurbanne, France\\
$^{42}$ Institute of Physics, Laboratory of Astrophysics, Ecole Polytechnique F\'ed\'erale de Lausanne (EPFL), Observatoire de Sauverny, 1290 Versoix, Switzerland\\
$^{43}$ Departamento de F\'isica, Faculdade de Ci\^encias, Universidade de Lisboa, Edif\'icio C8, Campo Grande, PT1749-016 Lisboa, Portugal\\
$^{44}$ Instituto de Astrof\'isica e Ci\^encias do Espa\c{c}o, Faculdade de Ci\^encias, Universidade de Lisboa, Campo Grande, 1749-016 Lisboa, Portugal\\
$^{45}$ Department of Astronomy, University of Geneva, ch. d'Ecogia 16, 1290 Versoix, Switzerland\\
$^{46}$ INAF-Istituto di Astrofisica e Planetologia Spaziali, via del Fosso del Cavaliere, 100, 00100 Roma, Italy\\
$^{47}$ INFN-Padova, Via Marzolo 8, 35131 Padova, Italy\\
$^{48}$ Universit\'e Paris-Saclay, Universit\'e Paris Cit\'e, CEA, CNRS, Astrophysique, Instrumentation et Mod\'elisation Paris-Saclay, 91191 Gif-sur-Yvette, France\\
$^{49}$ INAF-Osservatorio Astronomico di Trieste, Via G. B. Tiepolo 11, 34143 Trieste, Italy\\
$^{50}$ Aix-Marseille Universit\'e, CNRS/IN2P3, CPPM, Marseille, France\\
$^{51}$ Institute of Theoretical Astrophysics, University of Oslo, P.O. Box 1029 Blindern, 0315 Oslo, Norway\\
$^{52}$ Leiden Observatory, Leiden University, Niels Bohrweg 2, 2333 CA Leiden, The Netherlands\\
$^{53}$ Jet Propulsion Laboratory, California Institute of Technology, 4800 Oak Grove Drive, Pasadena, CA, 91109, USA\\
$^{54}$ von Hoerner \& Sulger GmbH, Schlo{\ss}Platz 8, 68723 Schwetzingen, Germany\\
$^{55}$ Technical University of Denmark, Elektrovej 327, 2800 Kgs. Lyngby, Denmark\\
$^{56}$ Cosmic Dawn Center (DAWN), Denmark\\
$^{57}$ Universit\'e de Gen\`eve, D\'epartement de Physique Th\'eorique and Centre for Astroparticle Physics, 24 quai Ernest-Ansermet, CH-1211 Gen\`eve 4, Switzerland\\
$^{58}$ Department of Physics, P.O. Box 64, 00014 University of Helsinki, Finland\\
$^{59}$ Helsinki Institute of Physics, Gustaf H{\"a}llstr{\"o}min katu 2, University of Helsinki, Helsinki, Finland\\
$^{60}$ NOVA optical infrared instrumentation group at ASTRON, Oude Hoogeveensedijk 4, 7991PD, Dwingeloo, The Netherlands\\
$^{61}$ Argelander-Institut f\"ur Astronomie, Universit\"at Bonn, Auf dem H\"ugel 71, 53121 Bonn, Germany\\
$^{62}$ Department of Physics, Institute for Computational Cosmology, Durham University, South Road, DH1 3LE, UK\\
$^{63}$ Universit\'e Paris Cit\'e, CNRS, Astroparticule et Cosmologie, 75013 Paris, France\\
$^{64}$ Institut d'Astrophysique de Paris, 98bis Boulevard Arago, 75014, Paris, France\\
$^{65}$ Institut d'Astrophysique de Paris, UMR 7095, CNRS, and Sorbonne Universit\'e, 98 bis boulevard Arago, 75014 Paris, France\\
$^{66}$ CEA Saclay, DFR/IRFU, Service d'Astrophysique, Bat. 709, 91191 Gif-sur-Yvette, France\\
$^{67}$ Kapteyn Astronomical Institute, University of Groningen, PO Box 800, 9700 AV Groningen, The Netherlands\\
$^{68}$ Universit\'e Paris-Saclay, Universit\'e Paris Cit\'e, CEA, CNRS, AIM, 91191, Gif-sur-Yvette, France\\
$^{69}$ Space Science Data Center, Italian Space Agency, via del Politecnico snc, 00133 Roma, Italy\\
$^{70}$ Institute of Space Science, Str. Atomistilor, nr. 409 M\u{a}gurele, Ilfov, 077125, Romania\\
$^{71}$ Dipartimento di Fisica e Astronomia "G. Galilei", Universit\'a di Padova, Via Marzolo 8, 35131 Padova, Italy\\
$^{72}$ Dipartimento di Fisica e Astronomia, Universit\'a di Bologna, Via Gobetti 93/2, 40129 Bologna, Italy\\
$^{73}$ Universit\"ats-Sternwarte M\"unchen, Fakult\"at f\"ur Physik, Ludwig-Maximilians-Universit\"at M\"unchen, Scheinerstrasse 1, 81679 M\"unchen, Germany\\
$^{74}$ Departamento de F\'isica, FCFM, Universidad de Chile, Blanco Encalada 2008, Santiago, Chile\\
$^{75}$ Institut de Ciencies de l'Espai (IEEC-CSIC), Campus UAB, Carrer de Can Magrans, s/n Cerdanyola del Vall\'es, 08193 Barcelona, Spain\\
$^{76}$ Centre for Electronic Imaging, Open University, Walton Hall, Milton Keynes, MK7~6AA, UK\\
$^{77}$ Centro de Investigaciones Energ\'eticas, Medioambientales y Tecnol\'ogicas (CIEMAT), Avenida Complutense 40, 28040 Madrid, Spain\\
$^{78}$ Instituto de Astrof\'isica e Ci\^encias do Espa\c{c}o, Faculdade de Ci\^encias, Universidade de Lisboa, Tapada da Ajuda, 1349-018 Lisboa, Portugal\\
$^{79}$ Universidad Polit\'ecnica de Cartagena, Departamento de Electr\'onica y Tecnolog\'ia de Computadoras,  Plaza del Hospital 1, 30202 Cartagena, Spain\\
$^{80}$ Institut de Recherche en Astrophysique et Plan\'etologie (IRAP), Universit\'e de Toulouse, CNRS, UPS, CNES, 14 Av. Edouard Belin, 31400 Toulouse, France\\
$^{81}$ INFN-Bologna, Via Irnerio 46, 40126 Bologna, Italy\\
$^{82}$ Infrared Processing and Analysis Center, California Institute of Technology, Pasadena, CA 91125, USA\\
$^{83}$ IFPU, Institute for Fundamental Physics of the Universe, via Beirut 2, 34151 Trieste, Italy\\
$^{84}$ INAF-Osservatorio Astronomico di Brera, Via Brera 28, 20122 Milano, Italy\\
$^{85}$ Instituto de Astrof\'isica de Canarias, Calle V\'ia L\'actea s/n, 38204, San Crist\'obal de La Laguna, Tenerife, Spain\\
$^{86}$ Department of Physics and Helsinki Institute of Physics, Gustaf H\"allstr\"omin katu 2, 00014 University of Helsinki, Finland\\
$^{87}$ Dipartimento di Fisica "Aldo Pontremoli", Universit\'a degli Studi di Milano, Via Celoria 16, 20133 Milano, Italy\\
$^{88}$ INFN-Sezione di Milano, Via Celoria 16, 20133 Milano, Italy\\
$^{89}$ Junia, EPA department, 41 Bd Vauban, 59800 Lille, France\\
$^{90}$ Instituto de F\'isica Te\'orica UAM-CSIC, Campus de Cantoblanco, 28049 Madrid, Spain\\
$^{91}$ CERCA/ISO, Department of Physics, Case Western Reserve University, 10900 Euclid Avenue, Cleveland, OH 44106, USA\\
$^{92}$ Laboratoire de Physique de l'\'Ecole Normale Sup\'erieure, ENS, Universit\'e PSL, CNRS, Sorbonne Universit\'e, 75005 Paris, France\\
$^{93}$ Observatoire de Paris, Universit\'e PSL, Sorbonne Universit\'e, LERMA, 750 Paris, France\\
$^{94}$ Astrophysics Group, Blackett Laboratory, Imperial College London, London SW7 2AZ, UK\\
$^{95}$ SISSA, International School for Advanced Studies, Via Bonomea 265, 34136 Trieste TS, Italy\\
$^{96}$ INFN, Sezione di Trieste, Via Valerio 2, 34127 Trieste TS, Italy\\
$^{97}$ Dipartimento di Fisica e Scienze della Terra, Universit\'a degli Studi di Ferrara, Via Giuseppe Saragat 1, 44122 Ferrara, Italy\\
$^{98}$ Istituto Nazionale di Fisica Nucleare, Sezione di Ferrara, Via Giuseppe Saragat 1, 44122 Ferrara, Italy\\
$^{99}$ NASA Ames Research Center, Moffett Field, CA 94035, USA\\
$^{100}$ Kavli Institute for Particle Astrophysics \& Cosmology (KIPAC), Stanford University, Stanford, CA 94305, USA\\
$^{101}$ INAF, Istituto di Radioastronomia, Via Piero Gobetti 101, 40129 Bologna, Italy\\
$^{102}$ Universit\'e C\^{o}te d'Azur, Observatoire de la C\^{o}te d'Azur, CNRS, Laboratoire Lagrange, Bd de l'Observatoire, CS 34229, 06304 Nice cedex 4, France\\
$^{103}$ Institute for Theoretical Particle Physics and Cosmology (TTK), RWTH Aachen University, 52056 Aachen, Germany\\
$^{104}$ Institute for Astronomy, University of Hawaii, 2680 Woodlawn Drive, Honolulu, HI 96822, USA\\
$^{105}$ Department of Physics \& Astronomy, University of California Irvine, Irvine CA 92697, USA\\
$^{106}$ University of Lyon, UCB Lyon 1, CNRS/IN2P3, IUF, IP2I Lyon, 4 rue Enrico Fermi, 69622 Villeurbanne, France\\
$^{107}$ Canada-France-Hawaii Telescope, 65-1238 Mamalahoa Hwy, Kamuela, HI 96743, USA\\
$^{108}$ Aix-Marseille Universit\'e, CNRS, CNES, LAM, Marseille, France\\
$^{109}$ Department of Astronomy \& Physics and Institute for Computational Astrophysics, Saint Mary's University, 923 Robie Street, Halifax, Nova Scotia, B3H 3C3, Canada\\
$^{110}$ Dipartimento di Fisica, Universit\'a degli studi di Genova, and INFN-Sezione di Genova, via Dodecaneso 33, 16146, Genova, Italy\\
$^{111}$ Ruhr University Bochum, Faculty of Physics and Astronomy, Astronomical Institute (AIRUB), German Centre for Cosmological Lensing (GCCL), 44780 Bochum, Germany\\
$^{112}$ Department of Physics and Astronomy, Vesilinnantie 5, 20014 University of Turku, Finland\\
$^{113}$ AIM, CEA, CNRS, Universit\'{e} Paris-Saclay, Universit\'{e} de Paris, 91191 Gif-sur-Yvette, France\\
$^{114}$ Centre de Calcul de l'IN2P3/CNRS, 21 avenue Pierre de Coubertin 69627 Villeurbanne Cedex, France\\
$^{115}$ Dipartimento di Fisica, Sapienza Universit\`a di Roma, Piazzale Aldo Moro 2, 00185 Roma, Italy\\
$^{116}$ University of Applied Sciences and Arts of Northwestern Switzerland, School of Engineering, 5210 Windisch, Switzerland\\
$^{117}$ INFN-Sezione di Roma, Piazzale Aldo Moro, 2 - c/o Dipartimento di Fisica, Edificio G. Marconi, 00185 Roma, Italy\\
$^{118}$ Centro de Astrof\'{\i}sica da Universidade do Porto, Rua das Estrelas, 4150-762 Porto, Portugal\\
$^{119}$ Zentrum f\"ur Astronomie, Universit\"at Heidelberg, Philosophenweg 12, 69120 Heidelberg, Germany\\
$^{120}$ Dipartimento di Fisica - Sezione di Astronomia, Universit\'a di Trieste, Via Tiepolo 11, 34131 Trieste, Italy\\
$^{121}$ Department of Mathematics and Physics E. De Giorgi, University of Salento, Via per Arnesano, CP-I93, 73100, Lecce, Italy\\
$^{122}$ INAF-Sezione di Lecce, c/o Dipartimento Matematica e Fisica, Via per Arnesano, 73100, Lecce, Italy\\
$^{123}$ INFN, Sezione di Lecce, Via per Arnesano, CP-193, 73100, Lecce, Italy\\
$^{124}$ Institute for Computational Science, University of Zurich, Winterthurerstrasse 190, 8057 Zurich, Switzerland\\
$^{125}$ Universit\'e St Joseph; Faculty of Sciences, Beirut, Lebanon\\
$^{126}$ Department of Astrophysical Sciences, Peyton Hall, Princeton University, Princeton, NJ 08544, USA}

   \date{Received; accepted}

% \abstract{}{}{}{}{} 
% 5 {} token are mandatory
 
  \abstract{
   Material outgassing in a vacuum leads to molecular contamination, a well-known problem in spaceflight. Water is the most common contaminant in cryogenic spacecraft, altering numerous properties of optical systems. Too much ice means that \Euclid's calibration requirements cannot be met anymore. \Euclid must then be thermally decontaminated, which is a month-long risky operation. We need to understand how ice affects our data to build adequate calibration and survey plans. A comprehensive analysis in the context of an astrophysical space survey has not been done before.
   
   In this paper we look at other spacecraft with well-documented outgassing records. We then review the formation of thin ice films, and find that for \Euclid a mix of amorphous and crystalline ices is expected. Their surface topography -- and thus optical properties -- depend on the competing energetic needs of the substrate-water and the water-water interfaces, and they are hard to predict with current theories. We illustrate that with scanning-tunnelling and atomic-force microscope images of thin ice films.

   Sophisticated tools exist to compute contamination rates, and we must understand their underlying physical principles and uncertainties. We find considerable knowledge errors on the diffusion and sublimation coefficients, limiting the accuracy of outgassing estimates. We developed a water transport model to compute contamination rates in \Euclid, and find agreement with industry estimates within the uncertainties. Tests of the \Euclid flight hardware in space simulators did not pick up significant contamination signals, but they were also not geared towards this purpose; our in-flight calibration observations will be much more sensitive. 
   
   To derive a calibration and decontamination strategy, we need to understand the link between the amount of ice in the optics and its effect on the data. There is little research about this, possibly because other spacecraft can decontaminate more easily, quenching the need for a deeper understanding. In our second paper, we quantify the impact of iced optics on \Euclid's data. 
   %Accurate knowledge, however, can only be obtained from self-calibration data collected in-flight.
   }


   \titlerunning{Physics of ice formation in spacecraft}
   \authorrunning{Euclid Collaboration, M. Schirmer et al.}
   
   \maketitle
%
%-------------------------------------------------------------------

% kimoto2018: modeling of contamination, software, exposure probes on ISS
% luey2018: formation of thin films and droplets (destroyed by VUV)
% pirich2010: hydrophobic materials, van der waals description, wetting, coating
% odell2013: chandra organic contamination, evaporation rates; contamination modeling similar to mine
% plucinsky2018: chandra contamination increase
% lin 2018: water monolayers restructure to optimise interaction with subsequent water layers
% he2009: watering of TiO2
% guo2018: 

% https://home.fnal.gov/~mlwong/#degas  outgassing parameters

\section{Introduction}
\Euclid will survey 15\,000\,deg$^2$ of extragalactic sky \citep{scaramella2022} during its nominal six-year mission \citep{laureijs2011,racca2016}. 
%Using weak gravitational lensing and galaxy clustering experiments, its data will help to determine the expansion history of the Universe, and to test the validity of General Relativity at cosmological scales. 
To achieve its cosmology science goals with measurements of weak lensing and galaxy clustering, \Euclid must maintain pristine image quality, and a relative spectrophotometric flux accuracy of about $1\%$ in its optical and near-infrared (NIR) channels. \Euclid will observe from the Sun-Earth Lagrange point L2, which offers exceptional thermal stability and a well-known space environment.

Yet, even at L2 \Euclid will degrade over time due to space weathering. Radiation damage, dust, and meteoroid\footnote{The IAU discouraged the use of the term `micrometeoroid' beginning in 2017. Dust particles are smaller than 30\,\micron, and meteoroids are larger.} impacts will degrade protective thermal blankets and their efficacy \citep{engelhart2017,plis2019}, which can change the electronical and optical performance. Detectors directly suffer from radiation that increases the charge transfer inefficiency of charge-coupled devices \citep[CCDs;][]{massey2014}, and they may decrease the quantum efficiency of some photo-diode architectures \citep{sun2020,crouzet2020}. Dust and meteoroids increase the scattering and transmission loss of optical elements through surface pitting \citep{rodmann2019,mcelwain2023}; ionising radiation has a similar effect on optical surfaces, although on smaller physical scales \citep{roussel2016,simonetto2020}. These environmental factors are well known at L2. \Euclid's calibration program is well suited to account and correct for them, yielding accurate and consistent survey data. Atomic oxygen, the prime cause for spacecraft degradation in low-Earth orbits \citep{banks2003,palusinski2009,samwel2014}, is fortunately not a problem at L2.

However, space weathering is not the only adversary. Ongoing contamination also degrades the performance of optics, solar panels, thermal control, and other sub-systems \citep[e.g.][]{green2001,zhao2009,smith2012b,hui2022}. We distinguish between particulate and molecular contamination, with the latter being composed of volatile (for example H$_2$O and CO) and non-volatile substances, such as polymers. In this paper, we mostly focus on molecular contamination by water ice. In terms of prevention and minimisation of contamination, \Euclid is the best-designed spacecraft by the European Space Agency (ESA) to date. Water from material outgassing is expected to be the only relevant source of contamination, possibly forming thin ice films on optical surfaces throughout the mission.

On ground, contamination is an inherent part of construction and launch, and subject to contamination control plans \citep{kimoto2017,luey2018,patel2019,abeel2022}. In the vacuum of space, contamination is driven by material outgassing \citep{chiggiato2020}. Quartz crystal microbalances (QCMs) can be used to detect surface contamination down to a few $10^{-9}$\,g\,cm$^{-2}$ \citep{dirri2016}. In the case of water, this corresponds to a molecular monolayer with a $10\%$ filling factor. Solar System exploration missions require additional decontamination prior to launch to preserve the pristine states of the bodies they visit, and those of any samples returned to Earth \citep{willson2018,chan2020}.

Even though spacecraft materials can be degassed (baked out) to reduce outgassing, water and other trace materials are recaptured until launch, on timescales of days \citep{scialdone1993} and down to seconds \citep{postberg2009}. The outgassing rate depends, among other factors, on the material's molecular structure, chemical composition, surface finish, coatings, temperature, and mass and mobility of the dissolved contaminants. Outgassing rates for spacecraft materials are usually determined at room temperature, and must be extrapolated to cryogenic conditions. This extrapolation is highly uncertain due to the considerable dependence of diffusion and sublimation coefficients on temperature. Nano-scale restructuring processes in the materials during cool-down also play a role. Accurate contamination forecasts are therefore hard and require considerable effort well beyond the scope of this paper \citep[e.g.][for the \textit{James Webb} Space Telescope; JWST]{brieda2022}.

To counter contamination, temperatures in many spacecraft can be increased locally -- for example for a single lens -- or globally to sublime volatile contaminants. A global decontamination, however, implies a major thermal shock to the spacecraft; it may alter electronic and optical properties, and may even lead to additional contamination \citep[e.g.][]{haemmerle2006,liebing2018}. In the case of \Euclid, on-board heating power is insufficient for a full decontamination; partial Sun exposure of the external telescope baffle is required, implying further risks. A full decontamination cycle for \Euclid lasts about one month, including warm-up, cool-down, and recalibration, and only $1-2$ days are spent at maximum temperature to allow the sublimates to find their way out of the spacecraft. Given a mission duration of six years, this is a very costly procedure.

Volatile and non-volatile molecular contamination has caused throughput losses of 20\% and more in some Earth-observation satellites, posing a substantial  challenge for accurate and consistent long-term environmental and climate monitoring. To this end, the Global Space-Based Intercalibration System \citep[GSICS;][]{goldberg2011} has established terrestrial and bright celestial targets as a reference, used by numerous Earth-observation satellites for cross-calibration and correction. Yet surprisingly, little is known about the effect of iced optics on astrophysical observations. Perhaps this is because local decontamination comes as an easy fix in many spacecraft, readily and frequently applied whenever necessary, or because their calibration requirements are more relaxed; we give examples in Sect.~\ref{sec:spacecraftlessons} and links to to other works in Appendix~\ref{sec:apdx-library}. \Euclid, however, cannot heat individual optical elements alone, nor does it carry internal QCMs to monitor contamination directly. To maintain a spectrophotometric accuracy of 1\% throughout its lifetime, \Euclid has to rely on its own survey and self-calibration data. In this way, we can detect and correct for water ice until a decontamination is required. 

In this context we need to understand (1) the physical properties of thin ice films, (2) their surface topography, (3) their formation on optical substrates, and (4) their temporal evolution in space conditions. We also need to investigate the outgassing and sublimation fluxes in \Euclid, and how accurately they can be known in advance. These points are addressed in the present paper. We present a comprehensive analysis of ice contamination in spacecraft from the bottom-up perspective. This has allowed us to capture, understand, and counter \Euclid's response to ice contamination. In Sect.~\ref{sec:spacecraftlessons} we summarise the molecular contamination experienced by other spacecraft, building a picture of what \Euclid might encounter. In Sect.~\ref{sec:ice} we review the types of water ice that exist in a vacuum at cryogenic temperatures, how they transform into each other, and how their structure depends on the wetting properties of the substrates. In Sect.~\ref{contamination} we review the physics of diffusion, sublimation, and adsorption of water molecules. We also built a simple transport model to estimate the water exchange between surfaces, and thus the contamination rate in \Euclid's payload module (PLM). In Sect.~\ref{tvac} we present results about outgassing from \Euclid's thermal vacuum tests, and we conclude in Sect.~\ref{sec:conclusions}. 

In our second paper we investigate the optical properties of thin ice films and their impact on the spectrophotometric data taken by \Euclid. Specifically, we look at absorption, interference, scattering, polarisation, apodisation, and phase shifts, with each uniquely influencing \Euclid's spectrophotometric data. We have developed strategies to detect, monitor, and -- if possible -- correct for these effects. Only then are we in a position to determine how much ice \Euclid can tolerate on its optics to achieve its cosmological science, and when a decontamination is in order.

\section{Lessons learnt from other spacecraft}\label{sec:spacecraftlessons}
Material outgassing \citep{chiggiato2020} has troubled spacecraft already in the \textit{Mercury}, \textit{Gemini} and \textit{Apollo} programs \citep{leger1972}. Numerous experiments were dedicated to it, such as on the \textit{Mir} space station \citep{wilkes1999}, the Midcourse Space Experiment \citep[MSX,][]{uy1998}, and the International Space Station \citep{palusinski2009}. 

Astrophysical spacecraft have added further insight into contamination and its impressively broad spectrum of effects. Solar System exploration missions are particularly useful, often carrying pressure sensors and mass spectrometers to analyse the interplanetary gas and dust, and thus also the spacecraft's own outgassing constituents. In this section, we summarise the lessons learnt from some missions with a well-documented outgassing record. These are of great importance for our preparation of suitable calibration and decontamination plans.
Appendix~\ref{sec:apdx-library} has a list of references and short summaries for a larger number of astrophysical and Earth-observation satellites.

\subsection{Multi-layer insulation (MLI) thermal blankets\label{sec:mli}}
Spacecraft have both hot and cold sides, in particular in the inner Solar System, and are wrapped in MLI \citep{cepeda2021} blankets to ensure stable operating temperatures. Further thermal shielding may be needed internally to accommodate instrument needs, for example in \Euclid the Near Infrared Spectrometer and Photometer \citep[NISP,][]{maciaszek2016} has its own blanket.

\begin{figure}[t]
\centering
\includegraphics[angle=0,width=1.0\hsize]{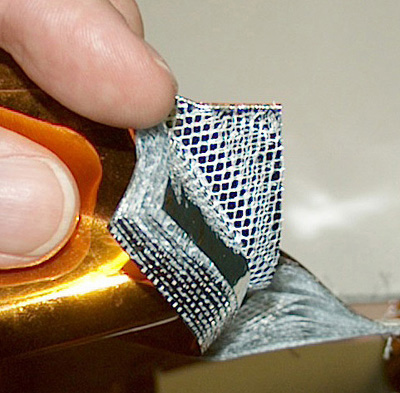}
\caption{Structure of a MLI thermal blanket, the main source of water contamination in spacecraft. Figure credit: John Rossie of Aerospace Educational Development Program (AEDP), \href{https://creativecommons.org/licenses/by-sa/2.5/deed.en}{CC BY-SA 2.5 license}}
\label{fig:mli}
\end{figure}

MLI consist of multiple -- often ten or more -- thin sheets of a high-performance polymer such as Kapton -- a polyimide -- coated with aluminium or gold (Fig.~\ref{fig:mli}). Outer layers may be carbon charged to suppress optical straylight (`black Kapton'; used for NISP). The individual MLI sheets are physically separated by a thin netting to minimise contact and thus thermal conductivity. To avoid rupture due to the rapid depressurisation during launch, the MLI may have venting holes or is perforated.

Similar to other polymers, Kapton -- in particular its amorphous versions -- can trap large amounts of water due to its high gas solubility \citep[e.g.][]{yang1985,sharma2018}; the dissolved water then also has great mobility \citep{chiggiato2020}. After degassing at $125^\circ$\,C for 24\,h in a vacuum, Kapton quickly recaptures $0.6$--$0.7$\% of its initial total mass in terms of water, during 24\,h at $20^\circ$\,C and in 55\% relative humidity (see the National Aeronautics and Space Administration's outgassing database\footnote{\url{https://outgassing.nasa.gov/}}). Further water intake appears to be stopped after this period \citep{scialdone1993}. Due to its common application in spacecraft, MLI is arguably the most important source of water contamination; it may also release other contaminants due to space weathering \citep{chen2016}.  Venting perforations -- if present -- facilitate contamination further, and the MLI may not deplete even after a decade in space (see below).

For completeness, we note that MLI is not the only possible carrier of water and other contaminants in spacecraft. Noteworthy are honeycomb structures \citep{epstein1993}, often comprising an aluminium core with carbon-fibre reinforced polymers (CFRP) that -- depending on their design -- might trap a considerable volume of water.

While water is the most frequent contaminant, other substances such as carbonates may be more troublesome for specific instruments. For \Euclid, water is expected to cause $90$--$95$\% of the overall transmission loss due to molecular and particulate contamination. We concentrate on water from Sect.~\ref{sec:ice} onwards, with a short excursion in Sect.~\ref{sec:hydrazine} where we address contamination from \Euclid's hydrazine thrusters.

\subsection{\textit{Hubble} Space Telescope\label{sec:hubble}}
In its early years, the \HST (HST) carried the Wide Field and Planetary Camera WFPC1, and its successor WFPC2. Both cameras suffered greatly from contamination in the UV \citep{mackenty1993,holtzmann1995}. Photopolymerisation was the cause for the heavy non-volatile contamination of WFPC1 \citep{tveekrem1996,lallo2012}, and the reservoirs of these contaminants depleted within 3 years. WFPC2 was contaminated mostly by water, resulting in typical flux losses of 1\%\,day$^{-1}$ at wavelengths 170--215\,nm. It was thermally decontaminated on average every 28 days between 1993 until at least 2001 \citep{baggett2001}. The contamination rate slowly decreased by a factor of 2 during this time,  and later on WFPC2 was decontaminated every 49 days \citep{gonzaga2010}. 

WFPC2 contamination estimates at wavelengths $\lambda>600$\,nm have poor signal-to-noise ratio (S/N), since WFPC2's UV science cases required decontamination before flux losses became evident at longer wavelengths. In the F555W filter -- corresponding to the blue end of \Euclid's IE~passband -- the mean flux loss between 1993 and 1998 was $1.2\pm0.3$\%\,month$^{-1}$ \citep{baggett1998}. More complex wavelength dependencies were found at longer wavelengths, partially attributed to different contaminants and their intrinsic diffusion-sublimation timescales. 

The Wide-Field Camera 3 (WFC3), installed in 2009, has a throughput loss of up to 0.3\%\,year$^{-1}$ in the ultraviolet and visible (UVIS) channel, not attributed to contamination \citep{shanahan2017}. The infrared (IR) channel loses about 0.1\%\,year$^{-1}$, likely due to photopolymerisation of contaminants \citep{bohlin2019}. More details about HST contamination and control can be found in \cite{clampin1992}, \cite{baggett1996}, and \cite{baggett1998}.

\subsection{ACIS / Chandra\label{sec:chandra}}
The \Chandra X-ray observatory has been operated since 1999. Since the detectors in the Advanced CCD Imaging Spectrometer (ACIS) are sensitive to optical wavelengths, an optical blocking filter (OBF) is used. \cite{plucinsky2018} show that the optical thickness at X-ray wavelengths has been monotonically increasing -- and slowly stabilising -- during the first seven to eight years of the mission, due to contamination of the OBF. In 2010, a phase of increasingly rapid contamination began that is still ongoing at different speeds for different atomic species \citep{plucinsky2020}.

\cite{plucinsky2018} argue that the initial stabilisation could be due to depletion of the contaminants' reservoirs, while the observed acceleration beginning a decade later came as a surprise. Plausibly, radiation damage \citep{engelhart2017,plis2019} or mechanical dust and meteoroid breakdown of the MLI led to an increase in internal temperatures, activating outgassing sources that were dormant previously. The steep temperature dependence of sublimation and diffusion (Sect.~\ref{contamination}) supports this scenario. The slow-down in contamination since 2017 can be explained by the near depletion of the contaminants, by an increased sublimation from the OBF due to higher temperatures, or both.

The atomic composition of the contaminants is available from their X-ray absorption edges. The dominant species is carbon, followed by oxygen and fluorine. Their deposition rate and spatial distribution has changed over time, indicating that multiple contamination sources are at play. Contamination has been active in \Chandra for almost two decades.

Similar contamination effects have been observed in the X-ray Multi-Mirror Mission's (XMM-\textit{Newton}) European Photon Imaging Camera Metal Oxide Semi-conductor cameras (EPIC-MOS), and also the Reflection Grating Spectrometer \citep[RGS;][]{plucinsky2012}. More details are given in the official calibration release documents\footnote{\url{https://xmmweb.esac.esa.int/docs/documents/CAL-SRN-0390-2-2.pdf}\\
\url{https://xmmweb.esac.esa.int/docs/documents/CAL-SRN-0305-1-0.pdf}\\
\url{https://xmmweb.esac.esa.int/docs/documents/CAL-SRN-0314-1-0.pdf}}.

\subsection{Cassini \label{sec:cassini}}
\subsubsection{Cosmic Dust Analyzer (CDA)\label{sec:cassini_cda}}
\Cassini launched in 1997 and reached Saturn in 2004. It carried the Cosmic Dust Analyzer (CDA), which measured mass, speed, direction, and chemical composition of cations. The latter were extracted from the gas and plasma cloud created by the impact of particles on a rhodium target plate, liberating contaminants as well \citep{postberg2009}.

During \Cassini's cruise phase, the CDA was contaminated by rocket exhaust fumes, outgassing, Solar wind, and the interplanetary medium. Beginning in 2000, after \Cassini's last inner Solar System fly-bys, the CDA was decontaminated by heating the target plate to 370\,K for 8\,h every few months. This removed volatile contaminants such as hydrocarbons and water ice.

\cite{postberg2009} identified H$^+$ and C$^+$ as the dominant contaminants with O$^+$ at lower levels, but they could not unambiguously locate their origin. Direct hydrocarbon contamination of the target plate has been considered unlikely. Plausibly, hydrocarbons elsewhere in the spacecraft were photolysed by the UV background, and the broken-down constituents formed an amorphous, non-volatile carbon-rich layer on the target plate. This particular contaminant likely formed prior to 2000 while \Cassini was still in the inner Solar System. Any halogen contaminants, such as Cl$^-$ and F$^-$, remained undetected since they were propelled away from the detector. Contamination in the CDA mass spectra was taken into account until the end of the mission in 2017 \citep[e.g.][]{altobelli2016}.

\subsubsection{Narrow Angle Camera (NAC)\label{sec:cassini_nac}}
\begin{figure}[t]
\centering
\includegraphics[angle=0,width=1.0\hsize]{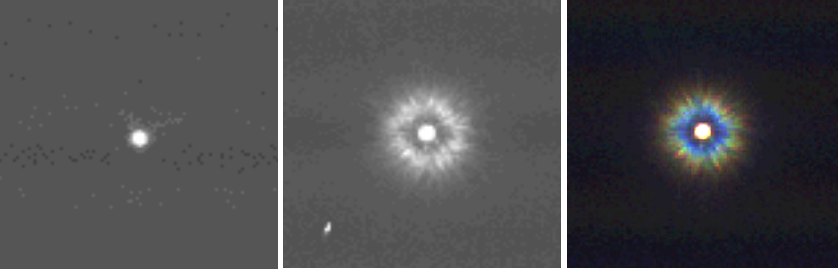}
\caption{Effect of ice on the point-spread function (PSF). \textit{Left panel}: \Cassini\;/ NAC image of the star Maia (Pleiades) taken in the broadband CL1/GRN filter ($\lambda_{\rm eff}=568$\,nm) before the contamination event. \textit{Middle panel}: Bright star $\alpha$ PsA in the same filter, after the contamination event. \textit{Right panel}: Colour image of Maia in filter combinations UV2/UV3 ($\lambda_{\rm eff}=316$\,nm; blue), CL1/GRN (green), and IR2/IR1 ($\lambda_{\rm eff}=827$\,nm; red), showing the chromaticity of molecular scattering. Figure credit: Adapted from \cite{haemmerle2006}.}
\label{fig:cassini_nac}
\end{figure}

\Cassini's NAC was decontaminated at $30^\circ$\,C every six months for 14\,h during the cruise phase. Until the Jupiter fly-by in late 2000, the NAC detector was kept warm at $0^\circ$\,C to minimise radiation damage by means of continuous annealing \citep{dale1991,holmes1991,bassler2010}. Contamination was absent in the Jupiter images taken with a detector temperature of $-90^\circ$\,C, but in 2001 a considerable haze appeared (Fig.~\ref{fig:cassini_nac}). It contained about 30\% of the total flux at 827\,nm and 80\% of the flux at 316\,nm.

This was surprising because the haze manifested within a few days after a decontamination. The main difference to the earlier 12 decontamination cycles was that the latest one heated the NAC from $-90^\circ$\,C to $+30^\circ$\,C, whereas all prior cycles went from $0^\circ$\,C to $+30^\circ$\,C. We know from Earth-orbiting satellites that shadow passages can release considerable amounts of water and particulates (see also Sect.~\ref{sec:msx}), due to rapidly changing temperatures and related mechanical stresses; this is known as `orbital thermo-cycling'. \cite{haemmerle2006} argue that a similar effect caused the NAC contamination, concluding that a decontamination can cause contamination if executed too quickly.

To recover NAC it had to be decontaminated twice: Once during a careful slow heating for seven days to $-7^\circ$\,C, which removed most of the haze that was likely due to water vapour. A remaining haze in the UV images was cleared by a another seven day decontamination to $+4^\circ$\,C, probably due to very small particulates or molecular contamination other than water.
Similarly, reoccurring contamination events were observed with the optical navigation camera onboard STARDUST \citep{bhaskaran2004}.

\subsection{XMM-\textit{Newton} Optical Monitor\label{xmm}}
Similar to \textit{Chandra}, the X-ray Multi-mirror Mission (XMM-\textit{Newton}) has experienced considerable molecular contamination of its X-ray imaging and spectroscopy cameras \citep{schartel2022}. The most likely contaminants are hydrocarbons, and other contaminants are suspected as well. Their origin is not well understood, and contamination has continuously increased over 22 years since launch.

Of particular interest to us is the Optical Monitor (OM), observing in the 180--700\,nm range \citep{mason2001}. The in-orbit commissioning of the OM showed a chromatic throughput loss of 16\% to 56\% compared to pre-launch expectations, with the largest losses occurring in the UV \citep{kirsch2005,schartel2022}. This contamination is attributed to non-volatile hydrocarbons, as the OM detector is kept at 300\,K, and the entire optics at 290\,K \citep{stramaccioni2000}; surface contamination by water does not persist in a vacuum at these temperatures (Sect.~\ref{sec:theoretical_sublim_rates}). 

Contamination of the OM has increased since, in parallel to an expected degradation of the detector's photocathode, which causes additional throughput losses up to 2.8\% year$^{-1}$ \citep{kirsch2005}. Notably, the OM's point-spread function (PSF) appears unaffected\footnote{\url{https://xmmweb.esac.esa.int/docs/documents/CAL-TN-0019.pdf}} by the increasing contamination, and a chromatic aureole from scattering as in the contaminated \textit{Cassini} NAC images (Fig.~\ref{fig:cassini_nac}) seems absent. Therefore, absorption by organic non-volatile contaminants is the most likely explanation for the observed throughput loss. The UV/Optical telescope (UVOT) onboard the \textit{Swift} Gamma-ray observatory inherited the OM design with improved contamination control \citep{roming2005}, and has shown little impact from contamination since \citep{poole2008,breeveld2010,kuin2015}.

\subsection{Genesis\label{sec:genesis}}
Genesis was a sample return mission probing the Solar wind, exposing ultra-clean sample containers for 850 days at Lagrange point L1. The containers were purged with dry nitrogen from assembly until launch to minimise on-ground contamination. Upon their recovery, the containers showed pervasive stains from material outgassing, composed of H, C, N, O, Si, and F. The root contaminants were not identified, but plausibly contained hydrocarbon, siloxane, and fluorocarbon components that were either vacuum pyrolysed, or polymerised by the UV background, or both \citep{burnett2005,calaway2006}; for the effect of UV-photofixation of contaminants, see also \cite{roussel2016}. As for the possible sources, \cite{burnett2005} list among others seals and locking elements, the electroplated gold concentrator, sealants and greases, residual films from pre-flight storage containers or processing, and residue from anodisation processing.

\subsection{Midcourse Space Experiment (MSX)\label{sec:msx}}
MSX was launched in 1996 into a Sun-synchronous orbit at 903\,km altitude and inclination of \ang{99;;}, carrying a total of ten contamination monitoring instruments; among others a neutral mass spectrometer (NMS) and a total pressure sensor (TPS) to analyse its gaseous surroundings, and five QCMs to investigate film depositions on external and internal surfaces \citep{uy1998}. MSX was operated for 12 years.

The NMS data show a pressure decrease with time $t$ as $t^{-1}$ for the first few days, then slowing down to approximately $t^{-0.5}$ over the next six months \citep{uy2003}, which corresponds to a $1/{\rm e}$ decay time of $t_{\rm e}=45$\,days. The TPS data shows a $t^{-0.6}$ dependence ($t_{\rm e}=30$\,days) over the first six months. This pressure evolution is attributed to the sublimation of superficial water ice, followed by diffusion and sublimation of absorbed water.

After the end of its initial ten month cryogenic phase, the MSX was inclined on a yearly basis by \ang{30} towards the Sun to heat its baffle and primary mirror \citep{uy1998}. Even after six years, these Sun exposure tests were always accompanied by a $100$-fold increase in TPS water vapour pressure from the sudden illumination of MLI that otherwise remained in the shadow; the pressure peaks even increased with every repetition of this test. \cite{uy2003} conclude that the MLI acts as a deep water reservoir and continuous source of contamination over many years, and that it is difficult to deplete. The MLI was also found to react very quickly to even small changes in the solar illumination angle. \cite{uy2003} and \cite{wood2003} also report numerous high-pressure transients unrelated to changes in solar exposure. These could be caused by rupturing, meteoroid impacts, and stress-release events due to orbital thermo-cycling, and are evidenced by an increasing particle density in the spacecraft's local environment (orbital degradation). 

The QCM results are described by \cite{wood2003}. During the initial, ten month long cryogenic phase, the contamination layers grew up to 16\,nm thick, depending on which parts of the spacecraft were in the QCMs' field of view. 
The internal QCM showed the highest contamination, mostly from Ar -- used as a cryogen -- and O, whereas H$_2$O and CO$_2$ were not detected. During the baffle's Sun-exposure tests following the cryogenic phase, up to 20\,nm of water were deposited on the internal QCM, indicating that the cold baffles had trapped considerable amounts of water. The water began to evaporate noticeably once the QCM was heated to 150\,K, and was gone once 165\,K were exceeded. Of the external QCMs, the ones facing the solar panels showed the highest rate of contamination during the first two years of the mission, followed by incomplete sublimation over the next three years, indicating the presence of non-volatile contaminants.

\begin{figure}[t]
\centering
\includegraphics[angle=0,width=1.0\hsize]{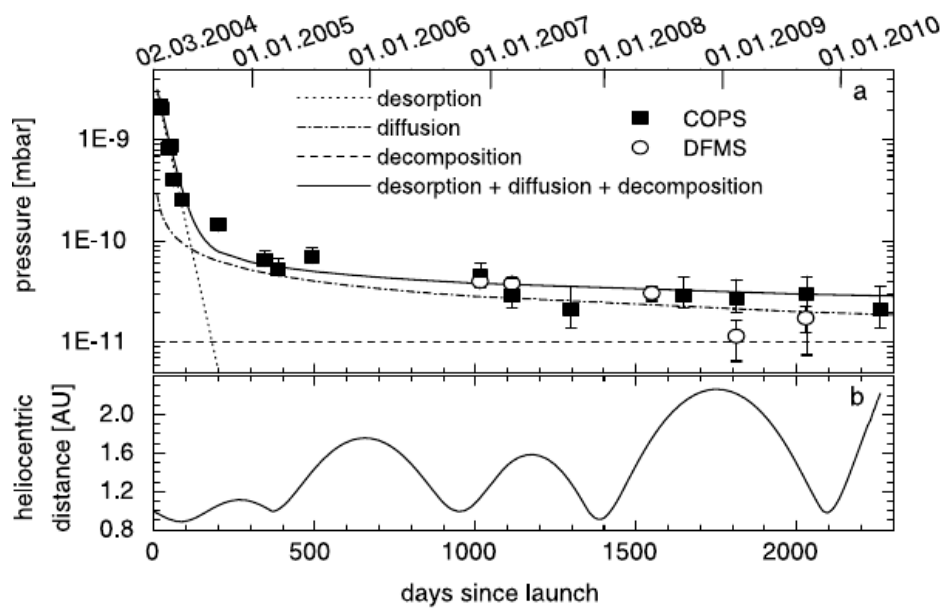}
\caption{Pressure around Rosetta due to water outgassing, showing that spacecraft travel for many years in their own gas cloud. Initially, desorption from surfaces is the dominant source, and in this case detected for up to 200 days after launch. Diffusion-sublimation then supports the cloud for years after, with a pedestal from decomposition due to UV- and particle radiation. The outgassing rate appears fairly independent of heliocentric distance. Typical interplanetary pressure is a few $10^{-12}$\,mbar and below. Figure reproduced from \cite{schlaeppi2010}.}
\label{fig:rosetta_pressure}
\end{figure}

\subsection{ROSINA / Rosetta\label{sec:rosetta}}
Rosetta carried the Rosetta Orbiter Spectrometer for Ion and Neutral Analysis (ROSINA), consisting of two mass spectrometers (RTOF and DFMS), and the Comet Pressure Sensor (COPS). Rosetta was launched in 2004 and arrived at comet P67 in 2014. ROSINA was active during extended periods of the cruise phase and two asteroid fly-bys, in particular also to understand contamination by outgassing \citep{schlaeppi2010}. 

The initial desorption of water from Rosetta's surfaces had a $1/{\rm e}$ decay time of 30 days and could be detected for the first 200 days of the mission (Fig.~\ref{fig:rosetta_pressure}). Once this source depleted, diffusion-sublimation became the dominant source in both DFMS and COPS data. After three years, the pressure around Rosetta had stabilised at $3\times10^{-11}$\,mbar. For comparison, the typical pressure in interplanetary space at these heliocentric distances is considered to be a few $10^{-12}$\,mbar \citep{postberg2009} and below. The mass spectrometers did not have any direct line of sight to structural parts of the spacecraft, whereas the pressure sensor had a nearly full solid-angle field of view. \cite{schlaeppi2010} show that the pressure sensors and mass spectrometers reacted mostly to return flux from self-scattering. In other words, Rosetta did travel in its own gas cloud, dense enough such that backscattering caused contamination elsewhere on the spacecraft. 

Similar to MSX, ROSINA found the gas pressure to be highly dependent on the spacecraft's Sun attitude \citep{schlaeppi2010}. This was noticed during the asteroid fly-bys, when Rosetta was reoriented to keep the target in sight and to protect some instruments from direct Sun exposure. The sudden illumination of structural parts that had been in the shadow for years resulted in the pressure exceeding $10^{-8}$\,mbar within a few tens of seconds after a reorientation. Likewise, the chemical composition of the gas phase changed, an effect that was also observed after the switch-on of previously dormant instruments. 

Outgassing from suddenly exposed, previously unilluminated components can cause a considerable acceleration of the spacecraft. For example, when OSIRIS-REx exposed its sample-return capsule to the Sun on its outbound journey, the acceleration exceeded that by Solar radiation pressure by one order of magnitude \citep{sandford2020}.

\cite{schlaeppi2010} report 146 different chemical constituents in the ROSINA outgassing data, from hydrocarbons, PAHs, C-O, N-O and C-O compounds to S, F, and Cl. The dominant species detected by DFMS are H$_2$O, followed by CO, N and CO$_2$. Hydrocarbon compounds may originate from polycarbonates (structural parts) and solvents, nitrogen-bearing compounds from adhesives, epoxies, coatings, and structural parts. Halogens point at brazing and lubricants, structure and tapes. Curiously, the RTOF spectra were dominated by F followed by H$_2$O. The high fluorine detection has been explained by a F-bearing lubricant used in the antenna, which is Sun-lit and closer to RTOF than to DFMS -- neither of which has direct lines of sight to the spacecraft. Again, this shows that contaminants evaporated into space can re-contaminate the spacecraft elsewhere through backscattering \citep[see also][]{bieler2016}. \cite{schlaeppi2010} estimate that several hundred grams of nonmetallic material and water outgas every year from Rosetta.

\begin{figure}[t]
\centering
\includegraphics[angle=0,width=1.0\hsize]{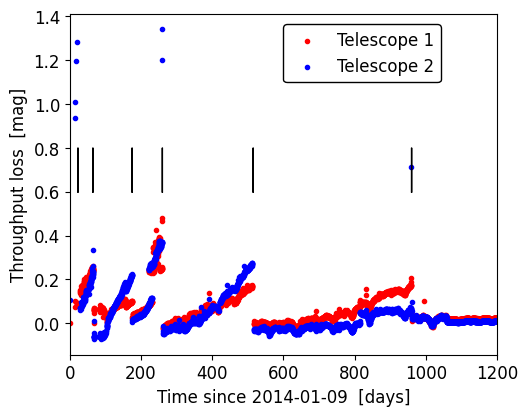}
\caption{Throughput loss for Gaia's telescopes since the beginning of operations. Initially, a rapid loss of $0.06$\,mag\,day$^{-1}$ was observed. A total of six decontaminations (indicated by vertical lines) were required over 2.6 years to reach a nearly stable state. Minor discontinuities in the data are artefacts due to an incremental calibration strategy.}
\label{fig:gaia}
\end{figure}

\subsection{Gaia\label{sec:gaia}}
Gaia is similar to \Euclid, in the sense that it is a wide-field astrophysical survey mission. Its mirrors and telescope structure are made of silicon carbide \citep[SiC;][]{bougoin2011}, as are \Euclid's \citep{bougoin2019}. SiC is known for its high strength, hardness, thermal conductivity, and low thermal expansion. Gaia had an industry forecast of very low water contamination. However, water heavily contaminated the optical system, leading to early and rapid transmission loss that required prompt decontamination (Fig.~\ref{fig:gaia}). A total of six decontaminations were needed over 2.6 years to reach a quasi-stable state \citep[see also][]{gaia2016}. As of now, no clear consensus has been achieved about the nature and origin of the contamination. Possibly, there is a contamination path from the service module (SVM) to the PLM, even though the two are separated by a single-layer insulation (SLI, as is the case for \Euclid, see Figs.~\ref{fig:euclid_svm2} and \ref{fig:euclid_mating_plm} in the Appendix). Contamination is spatially and temporally variable across Gaia's focal planes, and it appears to have switched between Gaia's two mirror systems \citep{riello2021}. We note that the Gaia PLM is fully covered in MLI, very close to the optical surfaces\footnote{\href{https://sci.esa.int/web/gaia/-/51009-gaia-payload-module-undergoing-acceptance-vibration-testing}{Photo of the MLI wrapping the Gaia optics and structure.}}.

Gaia carries internal laser interferometers to monitor its optical alignment. One of the most important lessons for \Euclid is that Gaia's SiC structure did not exactly resume its previous alignment after a decontamination. Moreover, slow and continuous focus drifts are seen over years after the last decontamination \citep{mora2016}. This implies that a decontamination of \Euclid requires a careful check of the PSF calibration.

Being $10$--$20$\,K warmer than \Euclid, water in Gaia's MLI is more mobile and the outgassing rate  considerably higher (see Sect.~\ref{contamination}), but it is not at all clear whether this can explain Gaia's initial high transmission loss. Higher temperatures also mean higher sublimation fluxes, beneficial if the ice is located already on optical surfaces, but detrimental if located on -- or still in -- other surfaces from where it can contaminate optics. Given Gaia's completely different design, we cannot conclude whether \Euclid's lower temperature puts it at an advantage or disadvantage compared to Gaia, and on what timescales. \Euclid's design benefited considerably from the Gaia experience.

\subsection{Take-home points}
The main contamination lessons are: 
(i) water is the most common contaminant for spacecraft operating in or near cryogenic conditions; it is found on -- and in -- numerous materials, with MLI being the most important reservoir due to its large area and high solubility of water in it;
(ii) contamination reservoirs deplete very slowly, and in the worst case will be active during \Euclid's entire life;
(iii) contamination rates, chemical composition, and location are time variable, given the depletion of some reservoirs and the activation of others, for example due to temperature changes;
(iv) spacecraft travel in their own gas cloud with sufficient gas pressure for backscattering, that is molecules evaporating into outer space can recontaminate the spacecraft elsewhere;
(v) the chemical composition of the gas cloud is spatially variable, with water being dominant on the shadow side, and decomposed substances at the spacecraft's Sun-illuminated side;
(vi) the pressure and chemical composition of the gas cloud respond within seconds to small changes in the spacecraft's Sun attitude and to instrument operations such as a switch-on;
(vii) water re-absorption on ground is both hard to avoid despite cleaning and degassing efforts, and hard to track for estimates of the absolute amount of water re-absorbed;
(viii) hydrocarbons and non-volatile organic compounds can considerably reduce the optical throughput by means of absorption.

\begin{figure}[t]
\begin{center}
\includegraphics[angle=0,width=0.8\hsize]{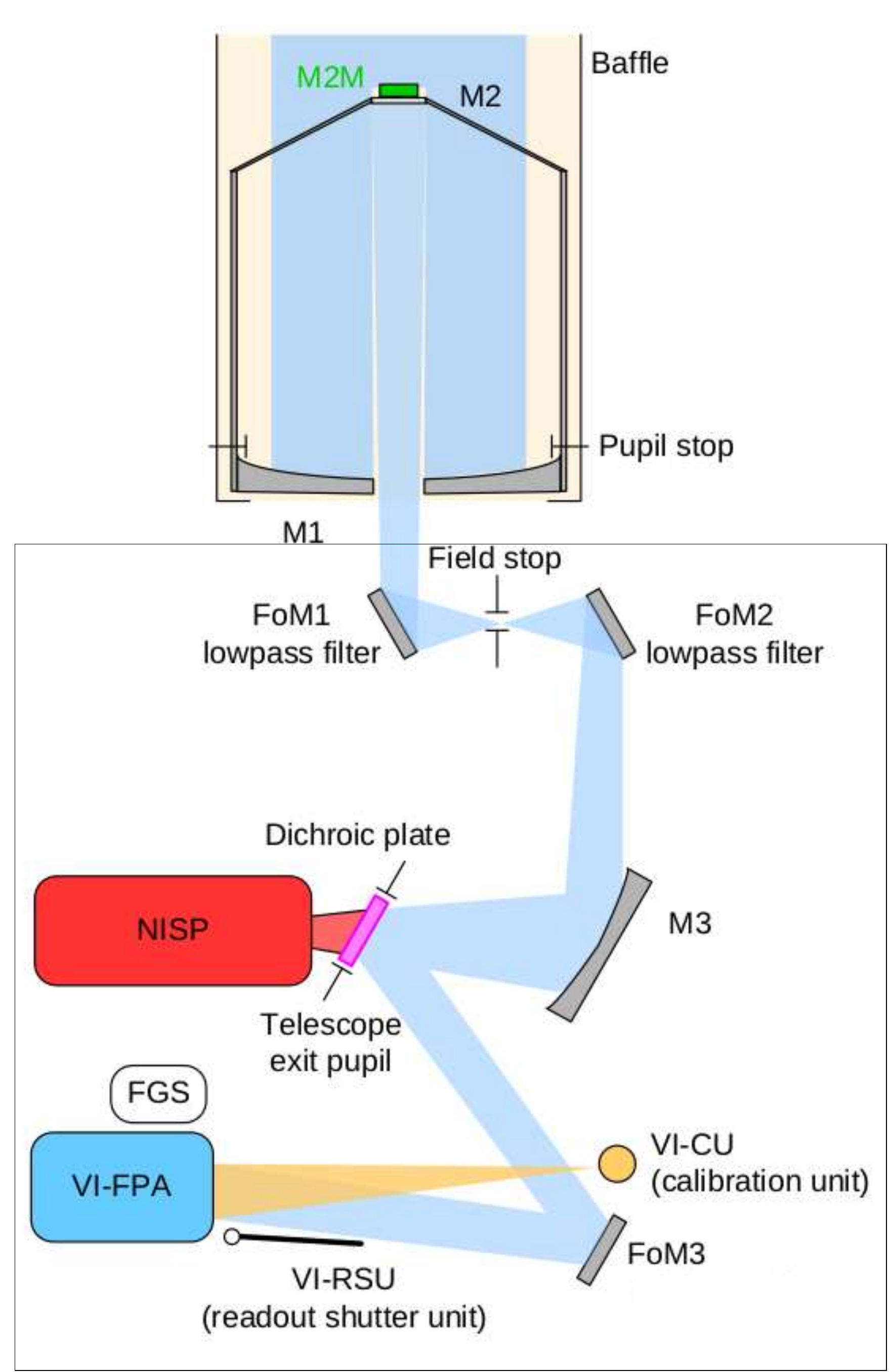}
\end{center}
\caption{Schematic view of optical surfaces in \Euclid. The telescope cavity contains M1, M2 and the baffle, and the instrument cavity (box) the remainder of the PLM. Figure credit: D. Filleul,  Airbus Defence and Space. A high-resolution 3D rendering of the instrument cavity and a photograph of the real flight hardware are shown in Figs.~\ref{fig:euclid_CAD_annotated} and \ref{fig:euclid_instcavity}.}
\label{fig:PLM}
\end{figure}

\begin{table}
\caption{Temperatures of PLM elements for nominal operating conditions, a ‘warm’ comparison case, and for decontamination mode.}
\smallskip
\label{euclid_temperatures}
\smallskip
\begin{tabular}{lllll}
\hline
\rowcolor{gray!20}
Common path & $T_{\rm nominal}$ & $T_{\rm warm}$ & $T_{\rm decont.}$\\
\hline
M1   & 117\,K & 123\,K & 220\,K \\
M2   & 104\,K & 111\,K & 289\,K \\
FoM1 & 123\,K & 128\,K & 220\,K \\
FoM2 & 122\,K & 126\,K & 221\,K \\
M3   & 122\,K & 129\,K & 220\,K \\
Dichroic & 122\,K & 126\,K & 221\,K \\
\hline
\rowcolor{gray!20}
NISP path & & & \\
\hline
Corrector lens & 130\,K & 131\,K & 218\,K \\
Filter / Grism &  133\,K & 133\,K & 206\,K \\
Camera lenses &  132\,K & 132\,K & 204\,K \\
Detector & 95\,K & 95\,K & 200\,K \\
\hline
\rowcolor{gray!20}
VIS path & & & \\
\hline
FoM3     &  118\,K & 123\,K & 220\,K \\
Detector &  152\,K & 156\,K & 270\,K \\
\hline
\rowcolor{gray!20}
Structural & & & \\
\hline
Baffle & 100\,K & 108\,K & 205\,K \\
PLM baseplate & 120\,K & 125\,K & 207\,K \\
\hline
\end{tabular}
\footnotesize
\tablefoot{The order of the components is as they appear in the optical paths towards NISP and VIS (see Fig.~\ref{fig:PLM}). Here `M' stands for mirrors with optical power, `FoM' for flat folding mirrors, and `L' for lenses inside NISP. Temperatures were taken from the PLM critical design review thermal analysis report. Accurate operational values will only be known after launch and may deviate by a few kelvin from the ones tabulated here. We note that the VIS detectors are considerably warmer than their surrounding, whereas the NISP detectors are colder, which determines their contamination experience.}
\end{table}

\subsection{Pertinent technical details about \Euclid\label{sec:pertinentdetails}}
For better understanding of the remainder of this paper, we provide here some technical details of \Euclid's PLM. A schematic layout of the optical configuration -- a three-mirror anastigmat Korsch design \citep{korsch1977} -- is shown in Fig.~\ref{fig:PLM}; more details are given in \cite{venancio2014}. Mirrors M1, M2, and M3 are powered mirrors, whereas FoM1 to FoM3 are flat. The dichroic plate separates the near-infrared from the optical wavelengths for simultaneous observations with VIS and NISP.

The silver coatings on mirrors M1, M2, M3 and FoM3 have additional layers for chemical and physical protection. The designs of these protective layers were not disclosed to us. Usually, they are complex, see for example \cite{sheik2008} for the \textit{Kepler} Space Telescope, and also 
\cite{garoli2010}. The top-most layer is of great importance for the formation and structure of ice films, as we discuss next in Sect.~\ref{sec:ice}. The entire layer stack is relevant for the optical properties of contaminating ice films, which we will show in our second paper.

The folding mirrors FoM1 and FoM2 have a high-performance dielectric coating stack including layers of gold, to provide a wavelength cut-off below $0.42$\,\micron. More details about the stacks were not disclosed by industry. The dichroic element and the NISP filters have alternating layers of Nb$_2$O$_5$ and SiO$_2$. The coatings on the fused silica NISP lenses might include TiO$_2$. Jointly, the mirrors and the dichroic plate provide a complex chromatic selection function that defines the passbands -- and out-of-band blocking -- for the VIS and NISP instruments \citep[for details, see][]{schirmer2022}.

Relevant for ice formation are also the in-flight temperatures of the optical and structural elements in the PLM. An estimate of the expected temperatures is given in Table~\ref{euclid_temperatures}. Exact values are difficult to predict from thermal modelling, and the actual temperatures might deviate by a few kelvin. Small changes in temperature may have a large impact on contamination, as we show in Sect.~\ref{contamination}. To this end, we use a `warm' case for comparison. The warm case is not realistic; it is a part of the thermal analysis, showing that \Euclid's temperature control systems can keep the spacecraft within operational limits even in unusual conditions. 

\section{Water ice types in spacecraft conditions\label{sec:ice}}
The rest of this paper focuses on the effects of water, the most common -- and for \Euclid\xspace-- most important contaminant. Water shows complex behaviour in its solid and liquid phases. This is attributed to four hydrogen bonds available to a water molecule to connect to its neighbours, and the two lone electron pairs of oxygen forcing the molecule into its bent shape. A water molecule is 0.28\,nm\;in size. Depending on temperature and pressure, water can form at least 20 different types of ice \citep{gasser2021,rosufinsen2023}.

%\begin{figure}[t]
%\centering
%\includegraphics[angle=-90,width=0.8\hsize]{molecule.pdf}
%\caption{Covalent and hydrogen bonds in and between water molecules. A water molecule is about 2.8\,\AA\;in size. Figure credit: Wikipedista/Qwerter (public domain).}
%\label{molecule}
%\end{figure}

The formation and structure of thin water ice films on nanometer and micrometer scales has been very actively researched  \citep[see][for a review]{salzmann2019}. However, to the best of our knowledge, this has never been studied in the context of contamination of astrophysical observatories. Given \Euclid's extraordinary calibration requirements, we need to understand ice evolution at a molecular level and how the numerous related physical processes lead to measurable effects in \Euclid data.

In Sects.~\ref{sec:amorphous_ice} to \ref{sec:irradiation}, we introduce the various types of ice forming in spacecraft, that is in a high vacuum and for very low deposition rates. In laboratory experiments, thin ice films are usually deposited with $0.01$--$100$\,nm\,min$^{-1}$. Even the lowest rate of 0.01\,nm\,min$^{-1}$ is 2--4 orders of magnitude (or more) higher than what \Euclid might experience in flight (Sect.~\ref{sec:contamination_forecast}). Yet for example $0.1$\,nm\,min$^{-1}$ are well applicable, since the latent heat released by adsorption of water molecules is rapidly dissipated in bulk ice \citep{brown1996a}, and eventually in the substrate before the next molecules are deposited. The thickness of laboratory ice films ranges from a few \AA\;-- that is incomplete monolayers -- to several \micron. In Sects.~\ref{sec:surface_properties} and \ref{sec:surface_RMS_estimates} we review how the surface topography of the ice depends on the substrate.

Studies of molecular contamination in the material sciences and by industry usually parameterise thin-film deposits in units of surface density; likewise for deposition, condensation, and sublimation fluxes. For our purposes, we parameterise ice films in terms of their thickness, which is more directly linked to their optical properties that we study in our second paper. For practical purposes we approximate that $1\,{\rm nm}\propto1\times10^{-7}$\,g\,cm$^{-2}$.

The scanning tunnelling microscope (STM) and atomic force microscope (AFM) data of ice surfaces shown in this section are available upon informal request. The surface-height profiles are encoded in ASCII x,y,z format.

\begin{figure}[t]
\centering
\includegraphics[angle=0,width=1.0\hsize]{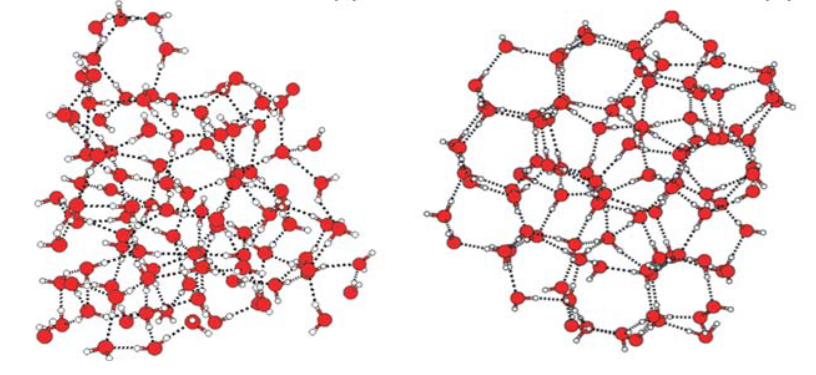}
\caption{Typical structure of high-density (\textit{left panel}) and low-density (\textit{right panel}) amorphous ice. Red dots represent the oxygen atoms, and small white circles the hydrogen atoms. Hydrogen bonds are indicated by dashed lines. Coherent structures are absent. Figure adapted from \citet{belosludov2008}; see also \cite{he2019}.}
\label{fig:amorphous_ice}
\end{figure}

\subsection{\label{sec:amorphous_ice}Amorphous ice (\texorpdfstring{$T\lesssim120$\,K}{T<120 K})}
Amorphous or non-crystalline ice, also called amorphous solid water or vitreous ice, is characterised by the absence of coherent crystal structures down to scales of individual water molecules. In a vacuum, it exists in three states, with the two coldest ones being highly porous at a molecular level (Fig.~\ref{fig:amorphous_ice}). For reviews about amorphous ices see for example \cite{limmer2014}, \cite{he2019}, and \cite{cao2021}.

\subsubsection{High-density amorphous ice \texorpdfstring{\Iah}{} (\texorpdfstring{$T<30$\,K}{T<30 K})}
High-density amorphous ice \Iah\footnote{The 19 known types of ice are labelled with Roman Numerals \textrm{I} to \textrm{XIX}. Ice types in spacecraft are all variants of type \textrm{I}.} forms when water vapour is deposited at temperatures below $T=30$\,K \citep{jenniskens1994}. It has a typical density of $1.15$\,g$\,$cm$^{-3}$ \citep{cao2021} and has the least structured state of all ice types. Between $30$--$70$\,K, one of the hydrogen bonds in ice \Iah breaks, irreversibly transforming ice \Iah into low-density, amorphous ice \Ial on timescales of a day \citep{schrivermazzuoli2000}.

Ice \Iah will not be found in \Euclid because temperatures are above 80\,K (see Table \ref{euclid_temperatures}). It may be present in other spacecraft such as the \textit{James Webb} Space Telescope, where temperatures reach below 40\,K \citep{lightsey2012,wright2015}.

\begin{figure}[t]
\centering
\includegraphics[angle=0,width=1.0\hsize]{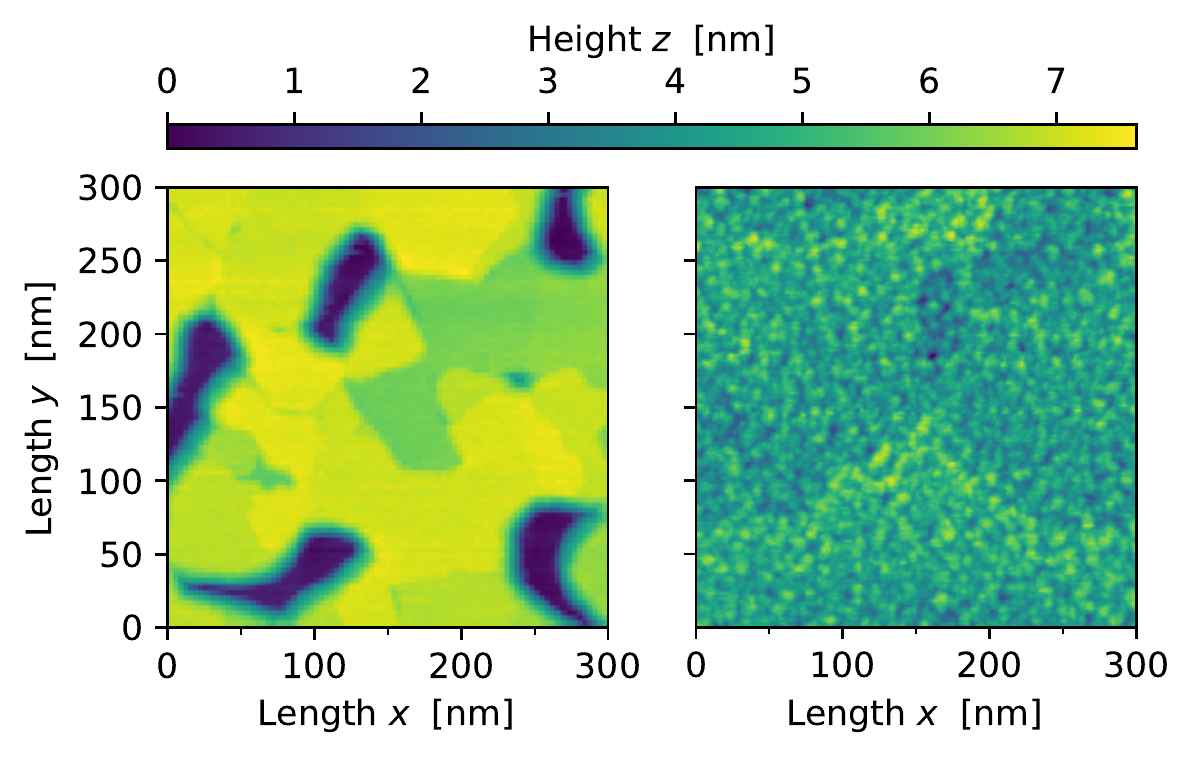}
\caption{Comparison of crystalline and amorphous ice topography. \textit{Left panel}: STM image of a polycrystalline ice film, average thickness 6\,nm, grown at 145\,K on Pt(111). Surface steps of bilayer height (0.37\,nm) are easily resolved. \textit{Right panel}: Same, for a 6\,nm thick amorphous ice film grown at 100\,K on Pt(111), revealing high surface roughness at nanometer scales. Two surface steps are visible in the otherwise atomically flat Pt(111) substrate, replicated by the amorphous ice film. Data originally taken by \citet{thuermer2008}.}
\label{fig:ice_STM}
\end{figure}

\subsubsection{Low density amorphous ice \texorpdfstring{\Ial}{} (\texorpdfstring{$30\,K\lesssim T\lesssim 120$\,K}{30 K < T < 120 K})\label{sec:lowdensity_am_ice}}
Low density amorphous ice \Ial is created by vapour deposition between 30\,K and $110$ to $120$\,K, the upper limit depending on the deposition rate (Sect.~\ref{sec:deposition_rates}).
The density of ice \Ial is $0.94$\,g\,cm$^{-3}$, neglecting variations in porosity. Porosity itself is parameterised by the internal surface area per mass, and for \Ial is typically  $150$--$500$\,m$^2$\,g$^{-1}$ \citep{mitlin2002}. Ice \Ial can be thought of as an open network of water molecules, where all pores are directly connected to the top surface \citep{he2019}, independent of the thickness of the ice. The top surface of ice \Ial is very rough at the nanometer scale when compared to crystalline ice (Fig.~\ref{fig:ice_STM}). The large surface area of amorphous ice facilitates astrochemical processes \citep{watanabe2008,gudipati2013}.

Amorphous ice is distinguished from crystalline ice by its large surface area and by the high internal vapour pressure at highly curved surface elements \citep{nachbar2018a,nachbar2018b}. This enhances the sublimation flux by factors 2--100 compared to crystalline ice at the same temperature (Sect.~\ref{sec:sublimation_amorphous}). Yet the absolute sublimation flux at temperatures where ice \Ial can form is very low (Sect.~\ref{sec:theoretical_sublim_rates}). 

In \Euclid, ice \Ial can occur on the NISP detectors (95\,K), the external baffle (100\,K), and the secondary mirror (M2; 104\,K). It will remain amorphous during the mission (Fig.~\ref{fig:crystal_timescale_euclid}). On the NISP detector, ice \Ial would modulate the quantum efficiency through interference effects \citep{holmes2016} and possibly severely affect the pixel response non-uniformity (PRNU); we address these effects in our second paper. Elsewhere in the PLM at $T\gtrsim120$\,K (Table \ref{euclid_temperatures}), ice \Ial would crystallise within a few days or weeks. However, these parts of the PLM are usually not cold enough to form amorphous ice in the first place.

\subsubsection{Restrained amorphous ice \texorpdfstring{\Iar}{} and onset of crystallisation (\texorpdfstring{$120\,K\lesssim T\lesssim 160$\,K}{120 K < T < 160 K})\label{sec:ice_restrained}}
When amorphous ice is heated to $120$--$140$\,K, or water vapour deposited at these temperatures at a high rate, surface reorganisation starts to collapse the internal pores \citep{hessinger1996} and reduces the number of `dangling bonds', that is unsatisfied OH bonds. The resulting modified state is called restrained amorphous ice \Iar; the transformation cannot be reversed by means of cooling. At these temperatures, nanocrystals begin to form in the amorphous phase through nucleation, and grow into crystalline clusters \citep{kouchi1994,nachbar2018a}. For 3D simulations of the transition process from amorphous ice to crystalline ice see \cite{he2019}.

\begin{figure}[t]
\centering
\includegraphics[angle=0,width=1.0\hsize]{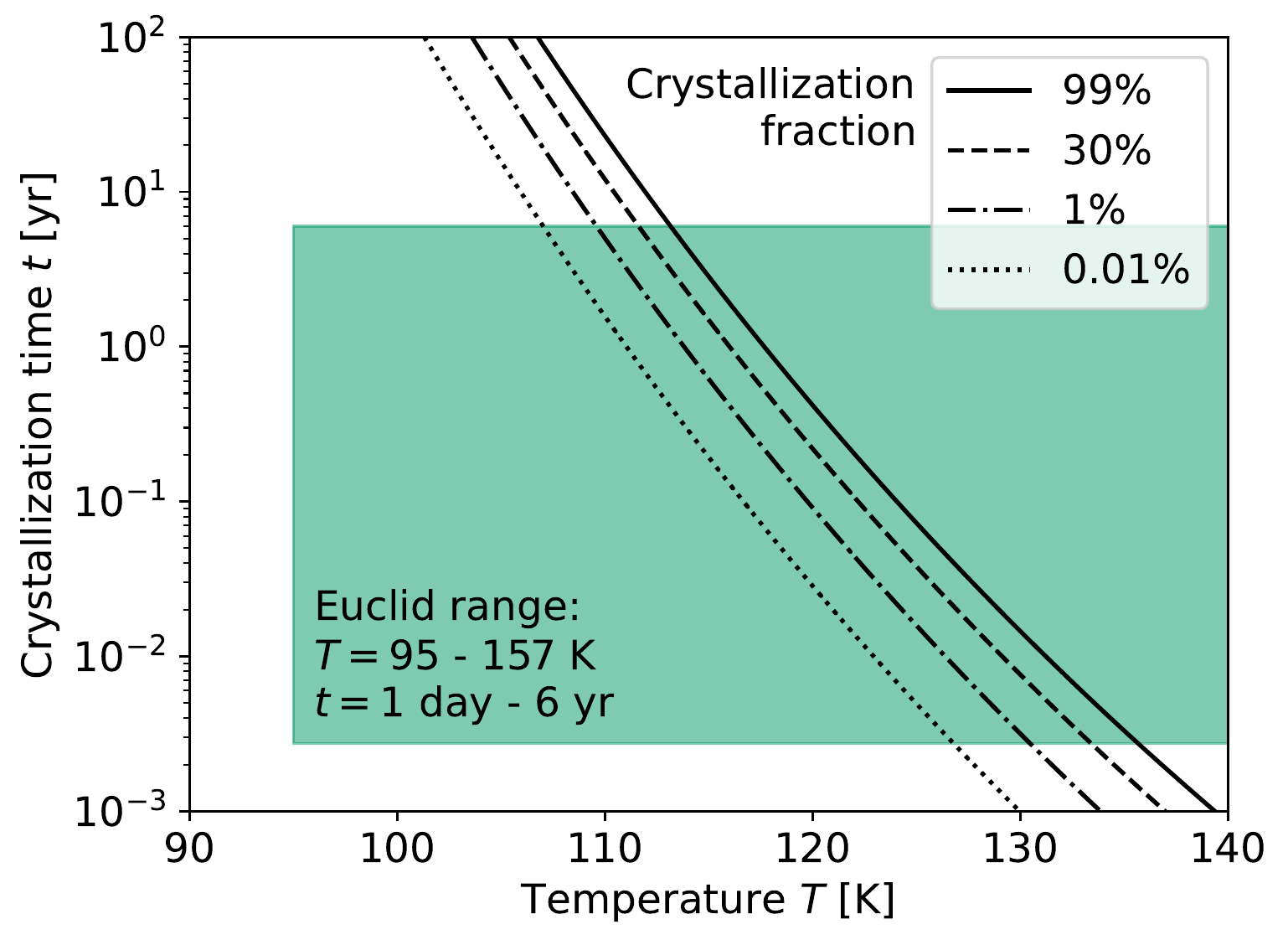}
\caption{Annealing time for amorphous ice \Ial to reach different fractions of crystallisation, using the \cite{kouchi1994} formalism that is based on kinetic theory of crystallisation. The shaded box shows the relevant time and temperate ranges for \Euclid. The crystallisation speed can be greatly accelerated  in the case of epitaxial growth on suitable substrates \citep{dohnalek2000}.}
\label{fig:crystal_timescale_euclid}
\end{figure}

Amorphous ice is meta-stable with respect to crystallisation (Fig.~\ref{fig:crystal_timescale_euclid}). Even at temperatures as low as $80$\,K it will eventually anneal into stacking disordered ice (Sect.~\ref{sec:cubic_ice}), albeit on geological timescales. Depending on the heating rate and deposition speed, crystallisation in laboratory experiments is observed mostly between $120$--$160$\,K \citep{laspisa2001,mitlin2002,mastrapa2013,he2022}. Amorphous constituents in the crystalline phase are uncommon above $160$\,K \citep{kuhs2012}, and do not survive $175$--$180$\,K for more than a few hours. Crystallisation cannot be reversed by cooling. Annealing of amorphous ice does not necessarily result in the same crystalline structures as depositing water at higher temperatures when crystalline ice forms directly \citep{hessinger1996a}.

In \Euclid, ice \Iar will occur only intermittently when heating cold surfaces covered with ice \Ial to their decontamination temperature (Table \ref{euclid_temperatures}). Otherwise, it would crystallise on timescales of days to a few months.

\subsection{Crystalline ice\label{sec:crystalline_ice}}
\begin{figure}[t]
\centering
\includegraphics[angle=0,width=1.0\hsize]{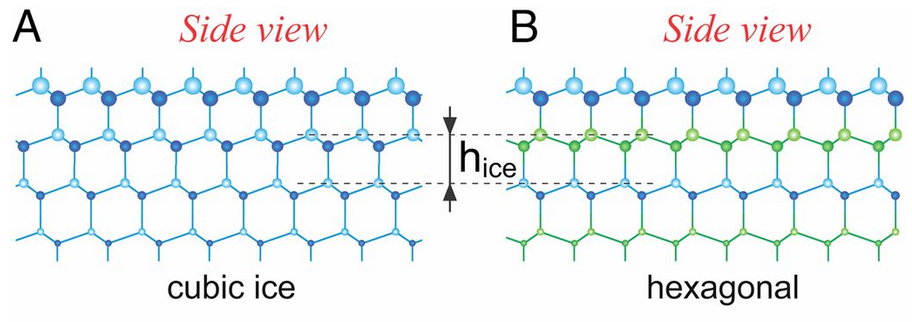}
\caption{Side views of four stacked corrugated bilayers of ice. The dots are the oxygen atoms, connected by hydrogen bonds. Thicker dots are higher up in the stack. The \textit{left panel} shows cubic ice \Ic. The \textit{right panel} shows hexagonal ice \Ih, where every second bilayer is rotated by $180^\circ$ around its surface normal axis; $h_{\rm ice}$ refers to the bilayer height. Figure credit: \citet{thuermer2013}.}
\label{fig:ice_crystalline_structure}
\end{figure}

In crystalline water ice, the oxygen atoms of six water molecules connect via hydrogen bonds to form corrugated hexagons. These hexagons merge into extended, 2-dimensional corrugated bilayers, which can be stacked in two ways: without rotation, forming cubic ice \Ic, and by rotating every other bilayer by $180^\circ$, forming  hexagonal ice \Ih (Fig.~\ref{fig:ice_crystalline_structure}). The hexagonal stacking order is energetically preferred over the cubic stacking order.

\subsubsection{\label{sec:cubic_ice}Stacking disordered ice \texorpdfstring{\Isd}{} (\texorpdfstring{$120\,K\lesssim T\lesssim 160$\,K}{120 K < T < 160 K})}
Cubic ice \Ic was first described by \citet{koenig1943} and wrongly thought to exist at a macroscopic scale at $120$--$160$\,K. It is now known that at these temperatures the ice consists of cubic and hexagonal layers, interlaced in a complex non-random fashion  described as `stacking disordered ice' \Isd \citep{kuhs2012}. Pure cubic ice exists essentially only in nanocrystals and in ice films a few nanometer thick \citep{kuhs2012,thuermer2013,malkin2015,nachbar2018a}. At a macroscopic level, pure cubic ice was created only recently by \cite{delrosso2020}.

Stacking disordered ice \Isd is meta-stable and forms via vapour deposition between $120$--$185$\,K. There is a large number of crystal defects and stacking faults in ice \Isd, requiring specific energies to be healed \citep{hondoh2015}: At $T=130$\,K, the least stable defects heal in about one week, whereas the timescale of most other defects exceeds one year. At $140$\,K, simple defects heal in one day, and within 1\,h at $150$\,K. The transformation from ice \Isd to ice \Ih speeds up considerably at $175$\,K and above \citep{kuhs2012,hondoh2015,delrosso2020}. Cubic sequences disappear within $1$\,h when ice \Isd is heated to $210$\,K, and they are essentially absent above $240$\,K.

In \Euclid, all mirrors are at or below $T=120$\,K (Table \ref{euclid_temperatures}). The transformation of any ice \Isd deposits to ice \Ih is therefore negligible on mission timescales.

%Therefore, once an ice layer has formed, it will maintain its structure and surface roughness unless modified by sublimation or further deposition. Warmer surfaces such as the NISP optics (132\,K) and the VIS focal plane (155\,K) are not expected to contaminate with water ice, as their sublimation fluxes are higher than the deposition rates (See Sect.~XX).

%The surface of ice \Isd is rougher than that of hexagonal ice at a molecular scale \citep{kouchi1994,kuhs2012}, but not as rough as that of amorphous ice. 

%\begin{figure}[t]
%\centering
%\includegraphics[angle=0,width=1.0\hsize]{cubic_ice_transformation.png}
%\caption{Decrease of the fraction of cubic stacking sequences in ice \Isd, as a function of time after vapour deposition at different temperatures. Figure credit: \citet{kuhs2012}.}
%\label{fig:cubic_ice_transformation}
%\end{figure}

\begin{figure*}[t]
\centering
\includegraphics[angle=0,width=1.0\hsize]{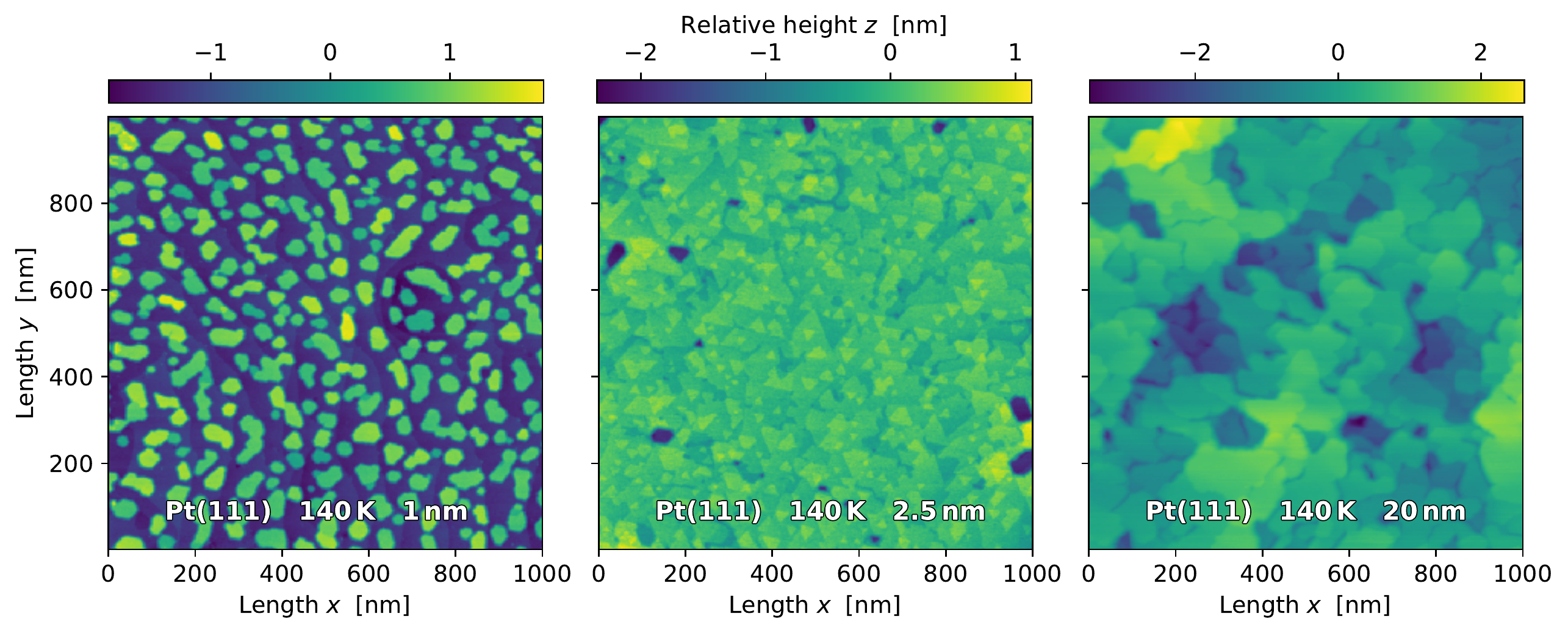}
\caption{Growth of crystalline ice on atomically flat Pt(111). The temperature and the mean film thickness are indicated. Shown is the relative height of each film, as the absolute height is difficult to assess for the thickest film that does not expose the substrate anymore. \textit{Left panel}: $2$--$3$\,nm ($7$--$10$ layers) high, flat-top crystallites appear in the wetting monolayer (dark blue), imaged with an STM. \textit{Middle panel:} Further deposition causes the crystallites to grow laterally and overlap each other (STM). \textit{Right panel:} A thick ice film that would not conduct sufficient electricity anymore from the substrate to the tip of an STM; an AFM was used instead. More details can be found in \cite{thuermer2008}, \cite{thuermer2013}, and \cite{thuermer2014}, who also took these data. For comparison, \Euclid's SiC mirrors have a typical surface roughness of $0.9$--$1.1$\,nm.}
\label{fig:crystal_growth}
\end{figure*}

\subsubsection{\label{sec:hexagonal_ice}Hexagonal ice \texorpdfstring{\Ih}{} (\texorpdfstring{$T\gtrsim120-185$\,K}{T > 120--185 K})}
Hexagonal ice \Ih forms from ice \Isd upon heating (Sect.~\ref{sec:cubic_ice}), or via vapour deposition at high rates ($\gtrsim 1$\,nm\,s$^{-1}$) at $T>185$\,K. It can also form by slow ($0.1$\,nm\,min$^{-1}$) vapour deposition at temperatures as low as $120$\,K in ultra-high vacuum \citep[$p\sim3\times10^{-11}$\,mbar; see][]{thuermer2013}. Once formed, ice \Ih is stable against cooling at least down to $T=5$\,K. \cite{rosufinsen2023} show that ice \Ih can be mechanically transformed into a previously unknown, medium-density amorphous ice; we do not consider this further as this process does not happen in \Euclid. On Earth, all naturally occurring ice is hexagonal, apart from very cold high-altitude cirrus clouds, where ice \Ic may be found. 

The physical properties of ices \Ih, \Isd, and \Ic are similar \citep{bertie1969,kuhs2012,mastrapa2013} for the purposes of the current paper, so we do not distinguish between them. However, the optical properties do show smaller differences in the refractive index \citep{he2022} that could be relevant for modelling effects in the data (see our second paper).

\subsection{Deposition rate and crystallinity\label{sec:deposition_rates}}
Whether vapour deposition initially leads to amorphous or crystalline ice depends on temperature, film thickness, and deposition rate. The latent heat released upon adsorption facilitates surface diffusion of water molecules and thus their settlement into energetically preferred configurations. With very high deposition rates, ice \Ial is formed initially, but dissipation of the latent heat is impeded by the low thermal conductivity of \Ial \citep{cuppen2022}, and crystallisation occurs. See also \cite{he2022}, \cite{cao2021}, \cite{watanabe2008}, \cite{laspisa2001}, and \cite{kouchi1994}.

For low deposition rates and $T\gtrsim120$\,K, water molecules can settle into crystalline structures before being disturbed by other incoming molecules \citep{kouchi1994,thuermer2008}. At $105$--$120$\,K, ice films may be amorphous, crystalline, or a mixture of both. At $100$\,K and below, they are always amorphous even when grown very slowly \citep[$0.1$\,nm\,min$^{-1}$;][]{laspisa2001,thuermer2008}.

In \Euclid, deposition rates are anticipated to be very low. We expect crystalline ice at $T\gtrsim120$\,K, amorphous ice at $T\lesssim110$\,K, and a mixture for the range $T\sim110$--$120$\,K.

\subsection{\label{sec:irradiation}Amorphisation through irradiation}
Crystalline ice can be amorphised by proton, heavy ion, and UV irradiation, which dissociate (photolyse) water molecules \citep{raut2008,fama2010,rothard2017}. The freed hydrogen atoms diffuse through the crystal and recombine with the fragments of other dissociated molecules, thus breaking down the crystalline structure. Irradiation experiments have shown that amorphisation processes become effective only at $70$\,K and below \citep{kouchi1990,mastrapa2006}. Typical timescales range between one year to several $10^5$ years, depending on environment and ice thickness \citep[see also][]{dartois2013,dartois2015}.

Temperatures in the \Euclid PLM are above 80\,K. At L2, irradiation-induced compaction and amorphisation of crystalline ice is negligible.

\subsection{\label{sec:surface_properties}Wetting of surfaces and growth of ice films}
So far we have reviewed ice types alone. We now
shift our focus to the substrate-water interface and its important influence on ice films growing on a substrate.

\subsubsection{Energetic needs of the substrate-water interface}
In general, surface atoms of a clean solid do not have all their bonding requirements fulfilled. Eventually, molecules in the surrounding gas phase are adsorbed due to van der Waals forces, covalent binding or electrostatic attraction, releasing latent heat in the process.

When water molecules adsorb on a substrate (`wetting'), they settle into energetically preferred locations determined by the surface's topography and electronic configuration. Above 40\,K, water molecules have enough energy for surface diffusion and form hydrogen bonds with neighbouring water molecules. The topography of these superficial water structures depends on the energetic needs of the substrate material; it can vary widely between 1D filaments, isolated clusters surrounded by `dry' substrate, and 2D contiguous films (wetting monolayers). At higher temperatures, water molecules may partially dissociate forming a mix of H, OH, and H$_2$O. For a comprehensive introduction and review see \cite{hodgson2009} and \cite{bjoerneholm2016}.

\begin{figure}[t]
\centering
\includegraphics[angle=0,width=1.0\hsize]{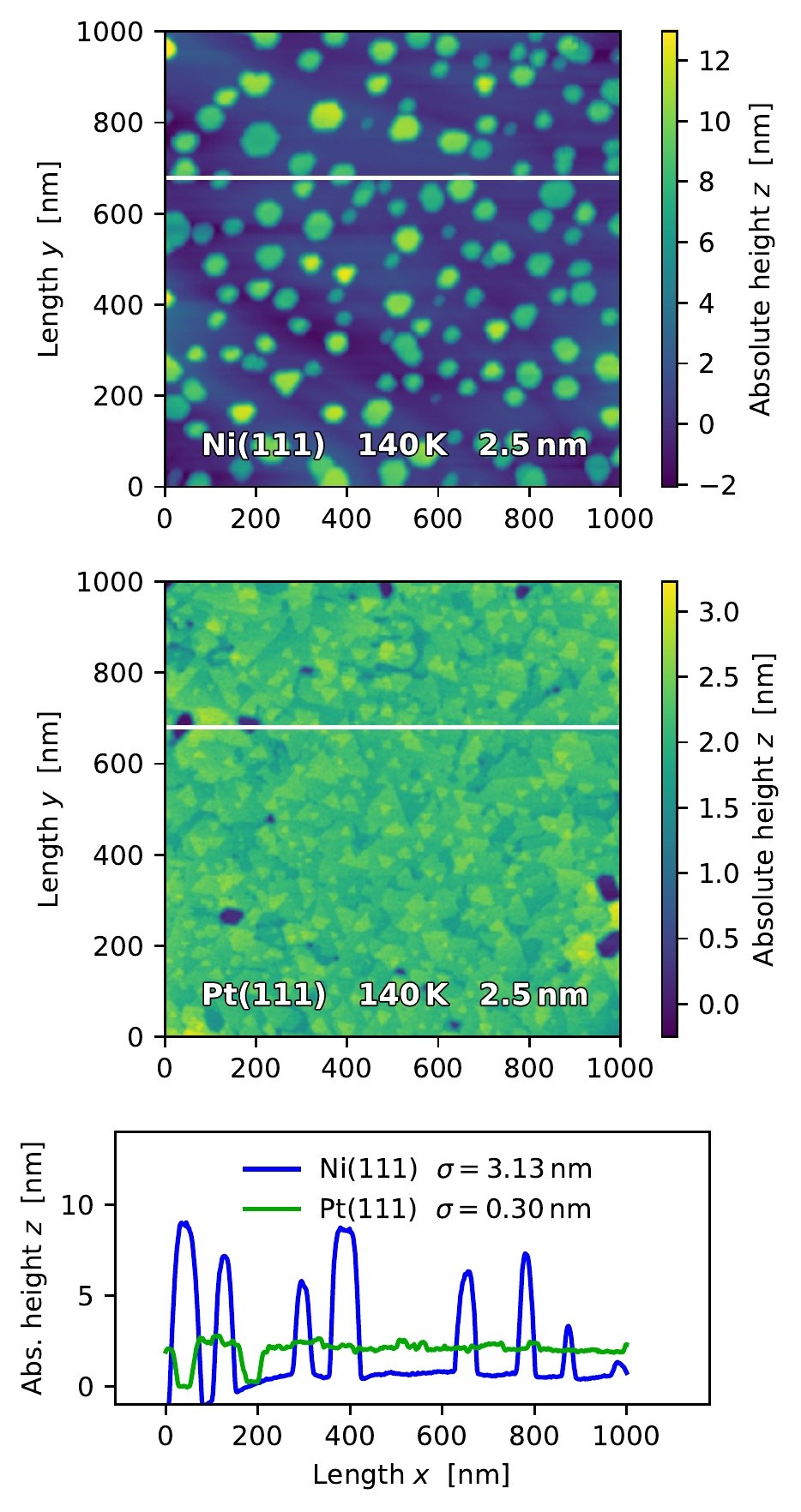}
\caption{Effect of the substrate on ice topography, for an average amount of 2.5\,nm ice. \textit{Top panel}: On Ni(111), a wetting bilayer (dark blue) is formed, in which isolated crystallites grow in height that cover 15\% of the surface area. \textit{Middle panel:} On Pt(111), crystallites quickly overgrow each other, forming a contiguous film. The wetting monolayer (dark spots) is still exposed in a few places. \textit{Bottom panel}: Height profiles measured along the horizontal lines shown in the upper panels. The standard deviation of the height distribution for Ni(111) is ten times that of Pt(111). The profile of the Ni(111) crystallites is convolved with the width of the STM's scanning tip; in reality, the walls of the crystallites are more vertical. To directly compare the height profiles, we plot the absolute height above a substrate mean reference, whereas in Fig.~\ref{fig:crystal_growth} we show the relative heights. The data for these plots were taken by \citet{thuermer2014}, who also inspired this figure.}
\label{fig:pt_ni_wetting}
\end{figure}

Once enough water is deposited for more than a monolayer (a wetting layer with a thickness of one molecule), the energetic constraints of the substrate-water interface need to be balanced with those of the water-water interface \citep{thuermer2014,lin2018,maier2018}. This results in a complex restructuring of the water-substrate interface that depends on the substrate's lattice constant, structure, electronic needs, and how water molecules in direct contact with the substrate orient themselves. The effects may reach just a few layers into the ice, or well beyond 100 layers (25--40\,nm thickness). Density functional theory can predict these structures for a given substrate, yet the case of water remains difficult \citep[][]{tamijani2020}. 

\subsubsection{Influence of the substrate on ice film topography}
We now compare wetting layers on two atomically flat, close-packed, and monocrystalline surfaces. We choose Pt(111) and Ni(111), two well-studied surfaces that illustrate the strong influence of the substrate on the growing ice films; the (111) tuple is the Miller index, describing the orientation of the atomic lattice exposed at the surface. 

On Pt(111), a contiguous monolayer is formed at first. Further deposition of water results in $50$--$150$\,nm wide crystallites surrounded by the monolayer. The crystallites have flat-top surfaces and heights of $2$--$3$\,nm ($7$--$10$ layers). Further deposition makes the crystallites grow mostly laterally and coalesce with their neighbours, maintaining an intact wetting layer in between. Eventually, all crystallites have merged, forming a contiguous polycrystalline film  (Fig.~\ref{fig:crystal_growth}). Therein, crystallites overgrow each other, leading to the preferential formation of hexagonal ice \Ih at temperatures as low as $115$--$140$\,K \citep[see Fig.~\ref{fig:crystal_growth}, and][]{thuermer2013}. 
%The surface RMS 
%$\sigma_{\rm ice}$ of contiguous ice on Pt(111) with a thickness $z$ of 2.5, 9, 20 and 40\,nm is about 0.25, 0.4, 0.8 and 1.0\,nm, respectively, roughly scaling as
%\begin{equation}
%    \sigma_{\rm ice} = 0.16\,\left(\frac{z}{{\rm nm}}\right)^{\;0.5}\;.
%    \label{ice_scaling}
%\end{equation}
%How well this trend continues for thicker films is unclear at this point, due to a lack of high-resolution microscopy data for ice films thicker than 40\,nm. In this work we use Eq.~(\ref{ice_scaling}) as a \textit{lower limit} to estimate photon scattering losses (Sect.~XX).
%UPDATE after fixing scattering section
 
On Ni(111), instead of a monolayer, the wetting layer is two molecules thick (bilayer). The emerging crystallites are much taller than those on Pt(111) and have smaller diameters of $30-60$\,nm. At a mean film thickness of 2.5\,nm -- when on Pt(111) a continuous film has formed -- the crystallites on Ni(111) are still well isolated, covering just 15\% of the surface (Fig.~\ref{fig:pt_ni_wetting}). This is attributed to a larger driving force for dewetting, presumably due to a lower surface energy of the wetting bilayer, or due to an increased energy of the interface between the crystallites and Ni(111). There are no high-resolution microscopy data for thicker films of ice on Ni(111) available at this point. However, based on comparison with yet another close-packed metal surface, Ru(0001), we predict with some confidence that the trend of ice films on Ni(111) being much rougher than those of equal mean thickness on Pt(111), will persist up to at least 100 molecular layers, if not indefinitely. The gas adsorption experiments by \cite{haq2007} for Ru(0001) have revealed that although the crystallites cover already 50\% of the surface at a mean film thickness of 2.5\,nm, it takes about 90 layers for the ice to fully coalesce. We thus infer that ice on Ni(111) will not coalesce for thicknesses up to at least 100 layers and remain much rougher than on Pt(111).

Quoting \citet{maier2018}: `On metal surfaces, the adsorption energy of water is comparable to the hydrogen bond strength among water molecules. Therefore, the delicate balance between competing water–water and the water–metal interactions leads to a rich variety of structures that form at the interface between water and seemingly simple, flat metal surfaces.'

\citet{thuermer2014} conclude similarly:
`Even for simple atomically flat close-packed metal substrates, the question of how water wets is surprisingly difficult. The delicate balance between optimising water-water bonding and water-metal interaction, the effect of the metal lattice constant, and [...] the possibility of water dissociation, all contribute to a complexity that renders predictions of water layer structure unfeasible.
Density functional theory [...] is not yet able to find the lowest-energy configuration of a water layer on a metal substrate reliably.'

\subsection{Impossibility to predict ice topography for \Euclid's optical surfaces\label{sec:surface_RMS_estimates}}
For \Euclid, the situation is exacerbated, as most coating materials have not been disclosed to us by industry (see Sect.~\ref{sec:pertinentdetails}). Wetting experiments were conducted for crystalline metal oxides such as Al$_2$O$_3$ \citep{tamijani2020} and TiO$_2$ \citep{he2009}, common optical coating materials. However, this does not help us, even if these materials were actually used in \Euclid.

First, the wetting process is highly dependent on the crystal planes (Miller indices) exposed at the surface, which we do not know in general. Second, vapour deposition of metal and semiconductor oxides generally results in amorphous and polycrystalline films \citep{kazmerski2012} that are also not atomically flat. Third, while the topography of a substrate is often replicated in dense optical coating layers \citep{trost2015}, this does not hold for contaminating ice films. There, long-range forces from crystallisation and the substrate-water interface control the topography on nano- and micrometer scales, together with growth spirals over substrate-surface steps \citep[][]{thuermer2008} and shadowing effects during deposition \citep{labello2011}.

% replication of surface structure: trost 2015

\subsection{Conclusions for ice in Euclid}
\paragraph{NISP detectors, M2, and external baffle:} These are the only places where low-density amorphous ice may form. Only if deposition occurs already during cool-down at $T\gtrsim120$\,K, crystalline ice is expected, with a top amorphous layer from further contamination. 
\paragraph{All other optical surfaces:} Polycrystalline ices \Isd and \Ih are expected. Their exact nano-scale crystalline composition is not relevant for \Euclid data. However, long-range forces in polycrystalline ice films determine the surface topography on scales of 100\,nm and above, and may thus have a noticeable impact on optical scattering and wavefront errors. These are difficult to model and predict, and it is not a priori clear how amorphous and crystalline ices manifest in the data. Crystalline ices have a narrow absorption line at 1.65\,\micron~that would be detectable in heavier contamination scenarios (see our second paper).
\paragraph{Internal and external processes:} Annealing and irradiation can break down the nanoscopic structure of ice films. They are highly inefficient at $90$--$120$\,K and can be ignored for \Euclid.
\paragraph{The structure of ices are mostly stable:} Ice films are predominantly modified by sublimation and further deposition. Mechanical surface restructuring will occur from dust and meteoroid pitting on M1 \citep{gruen1985,evans2000}.
\paragraph{The topography of ice films cannot be predicted:} The energetic needs of the substrate-water interface and the water-water interface are very complex. Also, we do not know the composition of the top-most coating layer on most surfaces.

% UPDATE: include later in Sect.~5
%Since forecasts of the RMS and thus scattering are currently not possible, we are planning dedicated %contamination tests. Flight-like witness samples with identical protected silver and dielectric coatings (as on M1 and FoM1) shall be contaminated with various amounts of water. We then measure the bidirectional %reflectance distribution function (BRDF) at different wavelengths and deposition temperatures for %crystalline and amorphous ice. With these tests, we can link an absolute amount of water to an optical %transmission loss, and make a forecast about the frequency of in-flight decontamination procedures. 

\section{\label{contamination}Contamination and decontamination modelling}
A single thermal decontamination cycle for \Euclid takes about 18 days, not counting subsequent recalibrations. Estimates of the contamination rate are thus of great interest for mission planning. Outgassing is driven by bulk diffusion of dissolved molecules in a substrate, followed by their sublimation. Our knowledge uncertainties of these processes limit the accuracy of contamination models; an estimate of a single decontamination per year could quickly become several per year, or none at all. 

Sophisticated codes exist to compute outgassing and contamination rates \citep[e.g.][]{brieda2020,zitouni2020}. They were applied for example to compute the contamination of the JWST during its initial 180 days in flight, accounting for JWST's complex unfolding sequence \citep{brieda2022}. In this section we aim much lower, developing an understanding of the dynamics of molecular contamination to inform our calibration strategy. We break down the contamination process into the underlying basic physics and geometry, and develop a transport model for the water exchange between surfaces in \Euclid. The main result is shown in Table~\ref{contamination_rate_table}, listing estimated contamination rates for the optical surfaces in \Euclid. These are indicative only and highly uncertain. To arrive at these values, we need sublimation and condensation rates (Sects.~\ref{sec:diffusion} and \ref{sec:theoretical_sublim_rates}), the vapour pressure from sublimed ice in \Euclid's cavities (Sect.~\ref{sec:vapour_pressure}), the effect of geometry on the sublimation and condensation rates between two surfaces facing each other (Sect.~\ref{sec:fluxcomputation}), and lastly geometrical models of the PLM to compute the water exchange flux between surfaces (Sect.~\ref{sec:contamination_forecast}). Finally, in Sect.~\ref{decontamination_procedure} we provide an overview of the thermal decontamination procedure.

Of equal interest is the impact of contamination on the data, which ultimately drives how often we have to decontaminate \Euclid. This will be addressed in the second paper.

\subsection{Methodologies\label{sec:methodologies}}
Here as in Sect.~\ref{sec:ice} we make extensive use of literature in the material sciences, outside the astronomical context. For better understanding we summarise  basic measurement principles.

The water update of a material can be determined using dynamic gravimetric vapour-sorption\footnote{The term `sorption' refers to the uptake of a substance by some material at (i) the material's surface (adsorption), and (ii) by integration into the material's atomic structure (absorption), without distinguishing between these processes.} experiments, where a material is exposed to various degrees of relative humidity \citep[e.g.][]{sharma2018}. Fourier-transform infrared (FTIR) spectroscopy is another method to measure the absorption or emission of water \citep[e.g.][]{scherillo2014}. These experiments are typically conducted at room temperature or higher.

The surface- and bulk-diffusion coefficients can be determined from transport models that describe the dynamic mass balance determined by sorption experiments. An alternative is laser-induced thermal desorption (LIDT) coupled with mass spectrometry, possibly using different isotopologues such as H$_2^{\,16}$O and H$_2^{\,18}$O ice as in \cite{brown1996}.

Different methodologies are available to measure sublimation and condensation rates, that is the change of ice-film thickness. The change in mass can be tracked by depositing ice films directly on cryogenic QCMs \citep[e.g.][]{sack1993}. Alternatively, the film thickness is determined directly using interference fringe counts in a reflected laser beam, or using FTIR reflection-absorption spectroscopy, exploiting the very strong absorption line of water ice at 3\,\micron~\citep[as e.g. in][]{ghesquiere2015}; details about water-ice absorption are presented in our second paper.

\subsection{Diffusion\label{sec:diffusion}} 
The first Fick law relates the diffusion flux, $\boldsymbol{j}_{\rm d}$, of absorbed particles to the spatial gradient of their concentration, $c$ \citep[see also][]{chiggiato2020}. In one dimension, 
\begin{equation}
    j_{\rm d}(l,t) = -D\,\frac{\partial c(l,t)}{\partial l}\;,
\end{equation}
where $l$ and $t$ are space and time, respectively. The diffusion coefficient $D$ is described by an Arrhenius-type law,
\begin{equation}
    D = D_0\,{\rm exp}\left(-\frac{E_{\rm d}}{k_{\rm B}\,T}\right)\;.
\end{equation}
Here, $D_0$ is the pre-exponential factor, $k_{\rm B}$ Boltzmann's constant, and $E_{\rm d}$ the diffusion activation energy. The constants $D_0$ and $E_{\rm d}$ depend on mass of the absorbed molecules, their size, and on the nanoscopic structure of the substrate. Using Kapton and ice as examples, we show that $D$ is highly sensitive to these parameters:

In amorphous Kapton, $E_{\rm d}=0.2$\,eV \citep{yang1985} and $D$ may vary by a factor of 3 depending on the orientation of the polymers, the thickness of Kapton, and the presence of aggregates \citep{yang1986}. This, and the absorption of water by Kapton, was further studied by \cite{sharma2018}, who find $E_{\rm d}=0.3$--$0.4$\,eV, and that $D$ can change by a factor of $10$, depending on the addition of aggregates. We note that lowering $E_{\rm d}$ from $0.40$\,eV to $0.39$\,eV at 120\,K -- a typical \Euclid temperature -- increases $j_{\rm d}$ by a factor 2.6. Thus $j_{\rm d}$ is highly susceptible to measurement errors of $E_{\rm d}$ and to the addition of aggregates.

Next, we consider the mobility of dissolved water molecules in ice. In amorphous ice \Ial, the porous structure greatly facilitates diffusion jumps of water molecules, resulting in a low $E_{\rm d}=0.08$--$0.25$\,eV \citep{ghesquiere2015}. The mean-square displacement of a particle due to bulk diffusion is given by 
\begin{equation}
    \langle (\Delta l)^2\rangle = D\,t\,.
    \label{eq:diffusionspeed}
\end{equation}
Accordingly, and using the computations in \cite{ghesquiere2015}, it would take a water molecule $\sim$\,0.5\,s to cross an amorphous ice film of 10\,nm thickness at 120\,K. In crystalline ice \Ih, this would take $120$\,s. \cite{brown1996} report even lower diffusion rates for ice \Ih, finding $E_{\rm d}=0.7$\,eV at $160$\,K. Using the Arrhenius law to compute the respective $D$ at 120\,K, we find that bulk diffusion is essentially incapacitated \citep[see also][]{labello2011} in ice \Ih in \Euclid, at least on mission timescales.

This means that an existing film of amorphous ice on Kapton does not slow down the diffusion flux $j_{\rm d}$ from Kapton at all, nor from any other substrate in \Euclid. Water molecules easily reach the top of the ice surface where they eventually sublime, unless they get more permanently integrated into the bulk amorphous ice. Therefore, amorphous ice films should grow continuously by substrate diffusion from below and by deposition on top.

Contiguous crystalline ice films, on the other hand, act as an effective diffusion barrier with $E_{\rm d}=0.7$\,eV. Considering the lower sublimation energy of water \citep[$E_{\rm sub}=0.45$--$0.53$\,eV,][]{sack1993,feistel2007,shakeel2018}, any water flux emanating from a surface contaminated with crystalline ice is due to sublimation of this ice, and not due to substrate outgassing. Yet, efficient diffusion channels from the substrate to the surface of the bulk ice could still exist, for example along fault lines and domain walls in polycrystalline ice, or if the surface roughness is very high -- such as on Ni(111) -- exposing large areas of thin wetting layers (Sect.~\ref{sec:surface_properties}).

The take-home message is that estimates of the outgassing rates in a spacecraft are highly uncertain at low temperatures: (i) They depend strongly on the substrate's nanoscopic structure and aggregates. (ii) Small temperature changes of a few kelvin result in an order-of magnitude change in $j_{\rm d}$. Temperatures of spacecraft sub-systems are difficult to estimate prior to launch and may change over time due to radiation damage and mechanical erosion of the insulation. (iii) Small measurement errors in $E_{\rm d}$ at the percent level change $j_{\rm d}$ by a factor of a few. (iv) Outgassing databases use measurements at room temperature because of the much higher signal and simpler non-cryogenic experimental setup. Extrapolations down to 120\,K span many orders of magnitude in $j_{\rm d}$ and ignore all restructuring processes at a microscopic level and below that might occur during cool-down from thermal contraction. For example, thermal stress-induced micro-fractures likely caused the sudden contamination of \Cassini\,/ NAC \citep[see Sect.~\ref{sec:cassini_nac}, and][]{haemmerle2006}.

% For \Euclid and its long thermal cycling time, it is therefore paramount to develop suitable correction methods at data level, to increase its resilience against the adverse effects from contamination and to decrease the amount of decontamination cycles and their inherent risks. 

Accurate diffusion and outgassing forecasts for \Euclid are therefore not feasible. However, we can still build a sublimation model, knowing that contiguous crystalline ice films act as effective diffusion barriers. Therefore we adopt a `glacial' scenario, in which all surfaces in \Euclid are already contaminated by crystalline ice films. The model shall be stationary, that is we do not consider self-depletion by sublimation. Such a glacial scenario could be the case immediately after launch, or after a long period without decontamination, that is a worst-case scenario. We recall that amorphous ice deposited at $120$\,K crystallises within a few weeks to months (Fig.~\ref{fig:crystal_timescale_euclid}). 

We use this glacial model to forecast the change of ice thickness and the amount of water escaping into space through the telescope front aperture. Since the model ignores diffusion, it cannot forecast the contamination rate of an initially uncontaminated spacecraft, nor the depletion times of the various water reservoirs.

%\subsection{Scope and limitations of the transport model}
%For our model we need to evaluate three aspects: the sublimation-condensation rates of water ice, the resulting vapour pressure in the telescope and instrument cavities, and the geometry of the exchange surfaces and their temperatures.

\subsection{Sublimation-condensation rates\label{sec:theoretical_sublim_rates}}
\subsubsection{General approach with the Hertz--Knudsen equation}
Deposition and sublimation happen simultaneously, and their rate is commonly described by the Hertz-Knudsen equation from classical kinetic gas theory. In the case of equal temperature $T$ of a substrate and its surrounding gas phase, we have 
\begin{equation}
    j_{\rm s}(T)=\sqrt{\frac{m}{2\pi\, k_{\rm B}\,T}}\;\left[\sigma{\rm_s}\,p_{\rm sat}(T)-\sigma{\rm_c}\,p(T)\right]\;.
    \label{eq:hertzknudsen}
\end{equation}
Here, $j_{\rm s}(T)$ is the sublimation flux (in kg\,m$^{-2}$\,s$^{-1}$), $m$ the mass of the subliming molecule, $p_{\rm sat}(T)$ the equilibrium saturation-vapour pressure for which sublimation and deposition rates are equal, and $p(T)$ the pressure in the gas phase. The sublimation and condensation coefficients, $\sigma_{\rm s}$ and $\sigma_{\rm c}$, are the fractions of molecules that sublime and re-condense (backscatter) upon reaching the surface; they are difficult to determine accurately. \citet{persad2016} derive a quantum-mechanical formulation for $j_{\rm s}(T)$, but its computation requires knowledge of the local curvatures of the substrate-gas interface, which are not known for ices on \Euclid's surfaces.

The back-scattering term $\sigma_{\rm c}\,p(T)$ accounts for sublimed molecules that immediately redeposit again on the surface after collisions with other sublimed water molecules in the vapour phase. This is negligible for \Euclid, where the mean free path length is thousands of kilometres (Sect.~\ref{sec:vapour_pressure}). Subliming molecules hit other surfaces and stick to them (Sect.~\ref{sec:vapour_pressure}) before colliding with other molecules in the gas phase, and thus $\sigma_{\rm c}\,p=0$ in Eq.~(\ref{eq:hertzknudsen}).

%In summary, sublimation pressure is the vapor pressure of a solid at its sublimation point, while saturation vapor pressure is the vapor pressure of a liquid or solid in equilibrium with its own vapor at a given temperature.

%The sublimation point, also known as the triple point, is the temperature and pressure at which a substance can exist in all three phases - solid, liquid, and gas - at the same time. At the sublimation point, the solid phase of a substance can transition directly into a gas phase without going through a liquid phase, and the vapor pressure of the solid and gas phases are equal. The sublimation point is a characteristic physical property of a substance, and varies depending on the substance. For example, the sublimation point of dry ice (solid carbon dioxide) is -78.5°C (-109.3°F) at standard atmospheric pressure, while the sublimation point of water is 0°C (32°F) at a pressure of 0.006 atmospheres.

\begin{figure}[t]
\begin{center}
\includegraphics[angle=0,width=1.0\hsize]{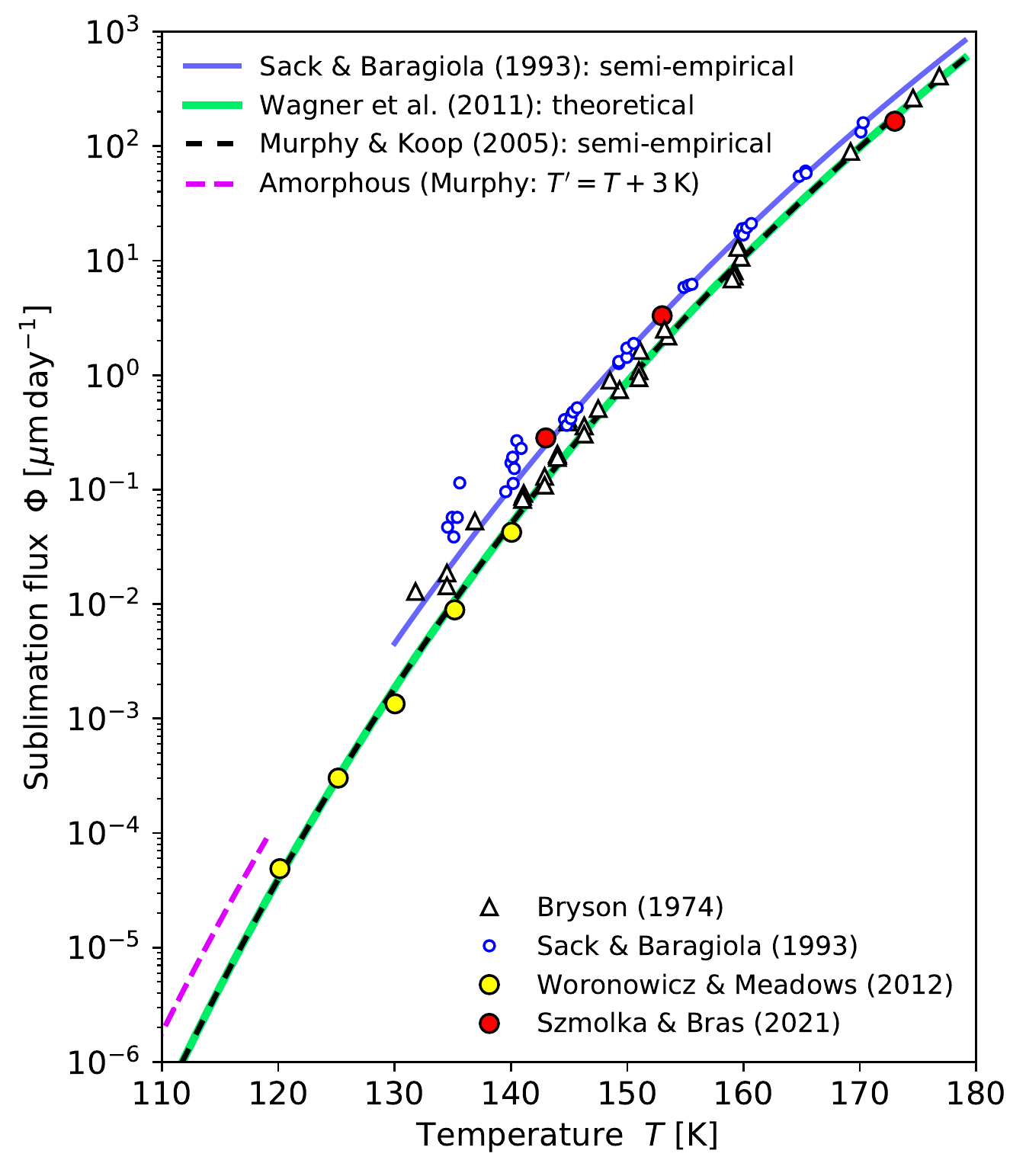}
\end{center}
\caption{Sublimation-flux models for amorphous and crystalline (hexagonal) ice. Overlaid are various measurements. The model for amorphous ice is shown up to 120\,K by the dashed pink line; it is obtained by shifting the \cite{murphy2005} curve by 3\,K to lower temperatures.}
\label{fig:sublimation_rate}
\end{figure}

\subsubsection{Theoretical and empirical estimates for hexagonal ice}
To evaluate Eq.~(\ref{eq:hertzknudsen}) for a vacuum, we can replace $\sigma_{\rm s}\,p_{\rm sat}(T)$ with the sublimation pressure $p_{\rm sub}(T)$. \cite{wagner2011}  derive $p_{\rm sub}(T)$ for a planar surface of monocrystalline ice \Ih based on the thermodynamics of the sublimation zone, valid from 50\,K to $T_{\rm t}=273.16$\,K,
\begin{equation}
    p_{\rm sub}(T) = p_{\rm t}\,\exp\left[\mathcal{T}^{-1}\sum_{i=1}^3\,a_i \mathcal{T}^{b_i}\right]\;,
    \label{eq:wagner}
\end{equation}
where $\mathcal{T}=T/T_{\rm t}$, $p_{\rm t}=611.657$\,Pa, and
\begin{alignat}{2}
a_1 &= -0.212\,144\,006 \times 10^2, &\hspace{0.5cm} b_1 &= 0.333\,333\,333 \times 10^{-2}, \nonumber \\
a_2 &= \;\;0.273\,203\,819 \times 10^2, &\hspace{0.5cm} b_2 &= 0.120\,666\,667 \times 10^1, \nonumber \\
a_3 &= -0.610\,598\,130 \times 10^1, &\hspace{0.5cm} b_3 &= 0.170\,333\,333 \times 10^1. \nonumber
\end{alignat}
Frequently used is \cite{murphy2005}, also based on thermodynamic considerations. We rewrite their result as
\begin{equation}
p_{\rm sub}(T) = \exp\left[c_1+c_2/\mathcal{T}+c_3\,{\rm ln}\,(\mathcal{T}) +c_4\,\mathcal{T}\right]\;{\rm Pa}
\label{eq:murphy}
\end{equation}
with 
%\begin{alignat}{2}
%c_1 &= 29.3577, &\hspace{0.1cm} c_2 &= -20.9521, \nonumber \\
%c_3 &= 3.53068, &\hspace{0.1cm} c_4 &= -1.98951. \nonumber \\
%\end{alignat}
$c_1 = 29.3577$, $c_2 = -20.9521$, $c_3 = 3.53068$, and $c_4 = -1.98951$.
%
%\begin{equation}
%p_{\rm sub}(T) = \exp\left[c_1+c_2/T+c_3\,{\rm ln}\left(\frac{T}{1\,{\rm K}}\right) +c_4\,T\right]\;{\rm Pa}
%\end{equation}
%with 
%$c_1 = 9.550426$, $c_2 = 5723.265$\,K, $c_3 = 3.53068$, and $c_4 = -0.00728332$\,K$^{-1}$.
%
This agrees to within 0.3\% with \cite{wagner2011} in the 90--210\,K range, and hereafter we collectively refer to Eqs.~(\ref{eq:wagner}) and (\ref{eq:murphy}) as the WMK models. 

The resulting sublimation fluxes $j_{\rm s}(T)$ are shown in Fig.~\ref{fig:sublimation_rate}, expressed as a loss rate for the ice film thickness. But how accurate are these theoretical models? The  surface roughness of polycrystalline ice \Ih (Sect.~\ref{sec:ice}) enlarges the effective surface area and increases the sublimation flux. Surface roughness also means larger nano-scale surface curvatures, thus higher internal vapour pressure \citep{andreas2007,nachbar2018a,nachbar2018b} and higher sublimation.

The WMK models are in very good agreement with the sublimation fluxes measured by \cite{woronowitz2012} at $T=120$--140\,K to understand the effect of ice on the thermal performance of the JWST sunshield. They also match the data from \cite{bryson1974}, but only down to a temperature of 140\,K, where the sublimation flux begins to exceed the WMK models by factors 2--4. A similar trend is seen in the data from \cite{sack1993}, where for $T\leq140$\,K the measurements shortly after deposition\footnote{A clear statement is missing in \cite{sack1993}, but from their description we estimate about 15--90\,min between deposition and measurement, depending on chosen deposition temperature and thermal warm-up time.} showed sublimation rates temporarily increased by factors 2--5. This is explained by more volatile amorphous constituents that have not yet annealed into a more stable crystalline form upon heating the ice films to their desired temperature.

We note that the values measured by \cite{sack1993} systematically exceed the WMK models by respective factors of 1.4 and 2.1 at $T=180$\,K and $T=140$\,K (see Fig.~\ref{fig:sublimation_rate}). This is also seen in measurements done by ESA to scale \Euclid's decontamination mode (Szmolka \& Bras 2021, private communication; red dots in Fig.~\ref{fig:sublimation_rate}). A first explanation is that the measurements were done too soon after deposition, when the restrained amorphous ice or stacking-disordered ice still experience considerable annealing, in particular if the deposition rates were high \citep[Sects.~\ref{sec:ice_restrained} and \ref{sec:deposition_rates}, and][]{sack1993,pratte2006,smith2011,rosufinsen2022}. Indeed the measurements by \cite{woronowitz2012} showing lower sublimation fluxes were done over 40--60\,h, compared to 15\,min for \cite{sack1993}; no information about this is given in \cite{bryson1974}. A second explanation is that different coatings on the QCMs affected the ices' surface topography (Sect.~\ref{sec:surface_properties}) and thus altered the sublimation flux. The QCM used by \cite{sack1993} was coated with gold; the QCM\footnote{A CrystalTek Cryo QCM, \url{https://crystaltekcorp.com/products/cqcm}} used by ESA for our tests was also gold-plated; \cite{woronowitz2012} did not comment on possible coatings of their QCM. \cite{sack1993} accounted for this by including an effective surface-area factor in their fit.

\cite{sack1993} fitted their measured sublimation flux (in molecules\,m$^{-2}$\,s$^{-1}$) with a semi-empirical model,
\begin{equation}
    \label{sack_sublimation_rate}
    \Phi_{\rm SB93}(T) = a\,T^{3.5}\;{\rm exp}\,\left(\frac{-E_{\rm sub}}{k_{\rm B}\,T}\right)\;,
\end{equation}
where $a=1.82\times10^{25}$\,molecules\,m$^{-2}$\,s$^{-1}$\,K$^{-3.5}$ is a constant prefactor. The model is shown as the blue line in Fig.~\ref{fig:sublimation_rate} for $E_{\rm sub}=0.45$\,eV. We note that changing $E_{\rm sub}$ to 0.46\,eV makes this fit consistent with the WMK models within 20\% in the 120--160\,K range. Considerable discrepancies below 120\,K arise because $E_{\rm sub}$ is actually temperature-dependent: \citet{feistel2007} compute that the sublimation enthalpy $E_{\rm sub}$ decreases by 0.008\,eV from 140\,K to 90\,K, which has a pronounced effect on the sublimation curve. We conclude that Eq.~(\ref{sack_sublimation_rate}) is less suitable to accurately describe sublimation fluxes over a very large temperature range that extends below 120\,K, and that the WMK models are preferred.

\subsubsection{Estimates for amorphous ice\label{sec:sublimation_amorphous}}
Amorphous ice can absorb large quantities of gas thanks to its porosity \citep{talewar2019}, making it an important constituent in the colder parts of the Solar System \citep{guilbert2012}. Large amounts of gas can be released when the pores of amorphous ice collapse during crystallisation. This may enlarge the sublimation rate of amorphous ice by many orders of magnitude down to 50\,K \citep[e.g.][]{notesco2003,barnun2007,drobyshev2007,prialnik2022}, a phenomenon referred to as `molecular volcano' \citep{may2013} seen mostly in amorphous ice exceeding several micrometer thickness. Even small fractions of a few percent of absorbed trace gases can increase the sublimation rate of water substantially. This is not relevant for \Euclid, where ice films are expected to be thinner (Sect.~\ref{sec:contamination_forecast}) and decontamination will occur sooner (second paper). 

Sublimation measurements of pure amorphous ice at temperatures below 120\,K are difficult due to limited instrumental sensitivity. \cite{kouchi1987} show that the saturation vapour pressure in amorphous ice depends strongly on the deposition temperature and the rate of deposition, and estimate it to be 10--100 times higher than in crystalline ice \citep[see also][]{sack1993}. More recent work suggests that the sublimation flux of amorphous ice is enhanced by a factor of ten or less compared to crystalline ice, once annealing effects immediately after deposition have settled. \cite{fraser2001} compute that amorphous ice has a 4.7 times shorter half-life time compared to crystalline ice at 120\,K, increasing to 7.6 times at 90\,K. \cite{smith2011} measure the desorption rates at 137--150\,K for amorphous and crystalline ice. Using their estimates of $E_{\rm sub}$ and ignoring its temperature dependence \citep{feistel2007}, we extrapolate to lower temperatures and find that the sublimation rate of amorphous ice increased by factors 3.3 and 5.1 at $T=120$\, and 90\,K, respectively. \cite{nachbar2018b} find a factor 2--3 increase of the saturation vapour pressure at 130\,K for amorphous ice on flat gold and copper substrates, with an upward trend towards lower temperatures.

Hence the sublimation flux of amorphous ice gradually increases over that of crystalline ice for decreasing temperatures. Given the uncertainties just outlined, we estimate the sublimation flux for amorphous ice by shifting the WMK models -- that is  Eqs.~(\ref{eq:wagner}) and (\ref{eq:murphy}) -- by 3\,K to lower temperatures,\begin{equation}
j_{\rm s}^{\;\rm amorph}(T) = j_{\rm s}^{\;\rm crystal}(T+3\,K)\;.
\label{eq:subflux_amorph}
\end{equation}
This results in respective factors 8.4 and 3.4 enhancement of the sublimation flux at 90\,K and 120\,K, and is shown by the dashed pink line in Fig.~\ref{fig:sublimation_rate}. We assume that below $110$--$115$\,K any ice deposits in \Euclid are amorphous and will remain amorphous  (Fig.~\ref{fig:crystal_timescale_euclid}), applicable to M2, the external baffle, and the NISP detectors, all of which are at $T<110$\,K (see Table \ref{euclid_temperatures}, and Sect.~\ref{sec:lowdensity_am_ice}). 

We summarise that the sublimation flux is a very steep function of temperature 
(Fig.~\ref{fig:sublimation_rate}). Estimates for various PLM components are given in Table \ref{sublimation_rate_table} using Eqs.~(\ref{eq:murphy}) and (\ref{eq:subflux_amorph}) for operational and decontamination temperatures. The actual sublimation fluxes in \Euclid might deviate by a factor of a few, depending on the substrates and the in-flight temperatures.

%In the presence of crystals embedded in amorphous ice, some of the amorphous ice will migrate and attach to the crystals rather than evaporate, reducing the current (and future) sublimation flux. We can ignore this effect Such surface diffusion below 115\,K is very difficult though, and thus we ignore it.

\subsection{Vapour pressure in \Euclid cavities\label{sec:vapour_pressure}}
The last information we need for our water transport model is whether the pressure in the sublimate is negligible. Indeed, the molecules are in free molecular flow, that is they travel along straight lines between point of sublimation and point of adsorption without collision. This is shown as follows.

The probability $f_{\rm stick}$ of a water molecule to adhere to an ice surface upon impact  -- the `sticking coefficient' -- has been analysed by \cite{batista2005} and \cite{gibson2011}; see also \cite{suliga2020}.
Dependencies on kinetic energy, impact angle, surface topography, and temperature can be safely ignored in \Euclid conditions, resulting in high values of $f_{\rm stick}=0.98$--$1.00$. This is because the energy transfer from the impinging molecule to molecules of equal mass in the bulk ice is maximal, and because the kinetic energy quickly dissipates in the bulk ice \citep{brown1996a}. Thus the molecules are effectively 
removed from the gas phase upon surface contact in \Euclid. We adopt a conservative $f_{\rm stick}=0.97$, measured at 120\,K and $p=10^{-10}$\,mbar by \cite{brown1996a}. 
%\citep[see also][]{bryson1972,padowitz1989,sack1993,kossacki1999,labello2011}.

To estimate the gas pressure and mean free path length, we approximate \Euclid's telescope cavity with a cylinder (Table \ref{plm_model_dimensions} and Fig.~\ref{plm_model} in the Appendix). We also assume that the cavity wall is in thermal equilibrium with the gas phase -- which is incorrect (Sect.~\ref{sec:fluxcomputation}) -- but has no practical implications for our deduction of the mean free path length. The wall of the cylinder (\Euclid's external baffle) has a temperature of 100\,K and its bottom (PLM baseplate and M1) 120\,K. All surfaces are assumed to be iced. Using Eqs.~(\ref{eq:murphy}) and (\ref{eq:subflux_amorph}), the total sublimation flux into the cylinder is $n_{\rm sub}=3.24\times10^{13}$\,molecules\,s$^{-1}$, 99.9\% of which coming from the warmer bottom. The escape fraction, $f_{\rm esc}$, through the front telescope aperture on direct paths is 3.5\% (Appendix \ref{sec:escapefraction}). We adopt a typical distance of $s=1.0$\,m, travelled by a water molecule before its adsorption, with a mean velocity of $\langle v\rangle=374$\,m\,s$^{-1}$, that is the mean of the Maxwell-Boltzmann distribution at 120\,K. The number $N$ of molecules in the cylinder at any time is then 
\begin{equation}
    N = n_{\rm sub}\,\frac{s}{\langle v\rangle}\,(1-f_{\rm esc})\,\sum_{k=0}^\infty\,(1-f_{\rm stick})^k 
      = n_{\rm sub}\,\frac{s}{\langle v\rangle}\,\frac{1-f_{\rm esc}}{f_{\rm stick}},
\end{equation}
where the rapidly converging sum represents the molecules that do not stick after $k$ surface impacts. With these conditions we have $N=8.6\times10^{10}$ molecules in the cylinder at any time. In an ideal gas, the pressure is then $p=3.1\times10^{-13}$\,mbar and the mean free path length is $167\,000$\,km, using $0.28$\,nm for the diameter of the water molecule. Therefore, the gas in the telescope cavity is in free molecular flow; sublimed molecules travel in straight lines from their point of sublimation to their point of impact, where they stick. 

The realisation of free molecular flow implies that the sublimate is not in thermal equilibrium with the mechanical surfaces, and that its velocity distribution is dominated by the processes in the surface-gas interface (Sect.~\ref{sec:fluxcomputation}). Any effects from the resulting non-Maxwellian velocity distribution are negligible for the conclusion of free molecular flow.

\begin{figure}[t]
\centering
\includegraphics[angle=0,width=0.8\hsize]{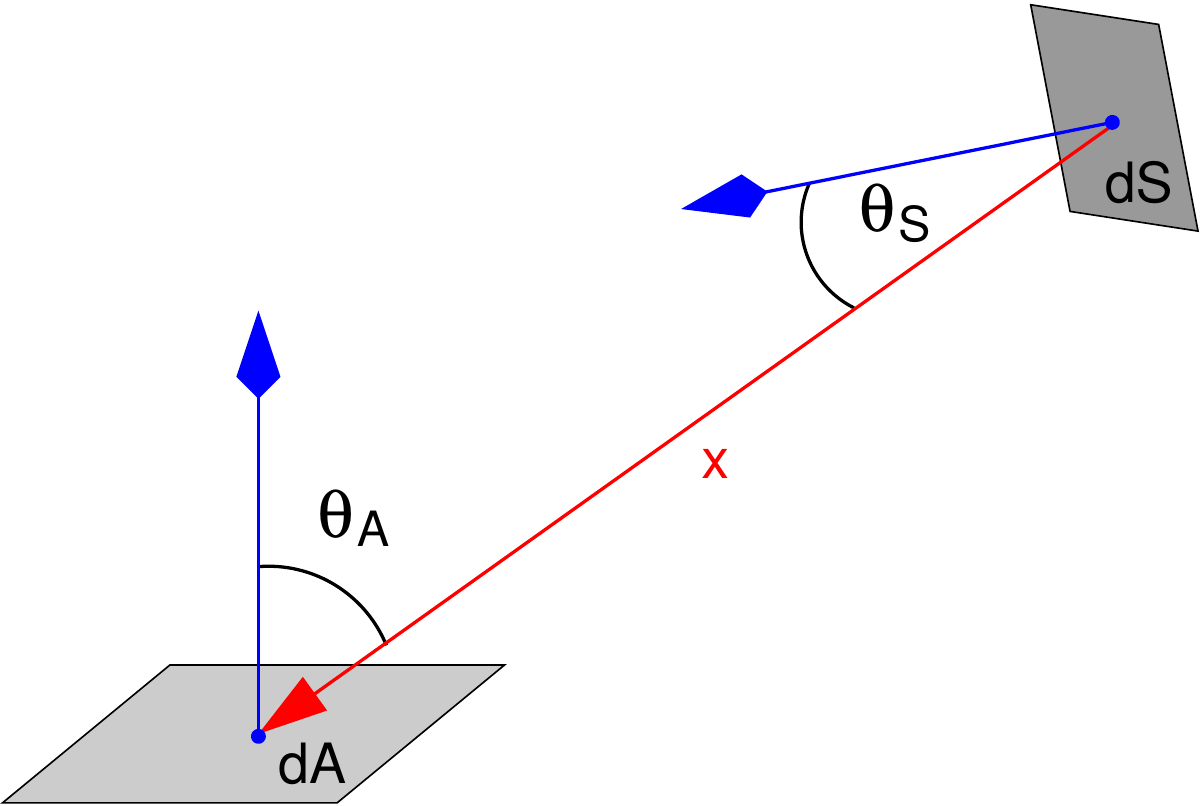}
\caption{Parametrisation of the sublimation geometry. The water flux is emitted by the surface element ${\rm d}S$ and received at the surface element ${\rm d}A$. The blue vectors represent the respective surface normal vectors. For details, see Sect.~\ref{sec:fluxcomputation}.}
\label{fig:sublimation_geometry}
\end{figure}

\subsection{Computation of incident water flux\label{sec:fluxcomputation}}
We consider the total flux $\Phi_{\rm tot}(T)$ of water molecules subliming from a surface element ${\rm d}S$ (Fig.~\ref{fig:sublimation_geometry}), in units of molecules\,m$^{-2}$\,s$^{-1}$, 
\begin{equation}
\Phi_{\rm tot}(T) = m\,j_{\rm s}(T)\,.
\end{equation}
Here, $m=2.99\times10^{-26}$\,kg is the mass of a water molecule, and $j_{\rm s}(T)$ is computed from Eqs.~(\ref{eq:hertzknudsen}) and (\ref{eq:murphy}) for crystalline ice; for amorphous ice, we use Eq.~(\ref{eq:subflux_amorph}). The dependence of the emitted flux on the angle $\theta_{\rm S}$ with respect to the surface normal is commonly described as ${\rm cos}\,\theta_{\rm S}$ \citep[Knudsen cosine law, see also][]{greenwood2002}. Lower-resolution experimental data initially supported this, as was shown by \cite{bryson1974} for H$_2$O and CO ice at \Euclid temperatures and by \citet{padowitz1989} for NO ice. This is questionable though, given the complex surface topography of ice (Sects.~\ref{sec:surface_properties} and \ref{sec:surface_RMS_estimates}); newer experiments suggest angular dependencies that are considerably more -- or less -- focused \citep[][and references therein]{todorov}. Closely related to the violation of the Knudsen cosine law is the fact that the velocity distribution of subliming particles can be sub- or super-Maxwellian; this is a consequence of the complex short- and long-range atomic forces at play in the desorption processes and in the surface-gas interface \citep{kann2016}. 

In the absence of experimental data providing more realistic angular and velocity distributions for the sublimates in \Euclid, we revert to the Knudsen cosine law and assume that the sublimate and the cavity are in thermal equilibrium. This and the free molecular flow established in Sect.~\ref{sec:vapour_pressure} allow us to treat the problem in analogy to the photon emission of a luminous surface\footnote{In principle, a code that computes the radiative heat exchange between surfaces can also compute contamination, by replacing the photon flux with the sublimation flux \citep[as e.g. in][]{brieda2022}.}.

Accordingly, the flux $f$ (in molecules\,s$^{-1}$) received by a surface element ${\rm d}A$ from the sublimating surface element ${\rm d}S$ is
\begin{equation}
    f(x,\theta,T) = {\rm d}S\,\Phi_0(T)\,x^{-2}\,{\rm cos}\,\theta_{\rm S}\,{\rm cos}\,\theta_{\rm A}\,{\rm d}A\,,
    \label{eq:sublimation_flux}
\end{equation}
where $\theta_{\rm S}$ and $\theta_{\rm A}$ are the respective angles to the surface normal vectors, $x$ is the distance between the two surface elements, and $\Phi_0(T)$ is the peak sublimation flux emitted at $\theta_{\rm S}=0$. We compute $\Phi_0(T)$ by determining the total sublimation flux emitted by the unit area into the hemisphere above,  
\begin{equation}
    \Phi_{\rm tot}(T) 
    = \int\displaylimits_0^{2\pi} \int\displaylimits_0^{\pi/2} \,\Phi_0(T) \, {\rm cos}\,\theta_{\rm S}\, {\rm sin}\,\theta_{\rm S}\, 
    {\rm d}\theta_{\rm S}\, {\rm d}\varphi
    = \pi\,\Phi_0(T)\;.
    \label{eq:phitots1}
\end{equation}
%\begin{alignat}{2}
%    \Phi_{\rm tot,S}(T) & = {\rm d}S\, \int\displaylimits_0^{2\pi} \int\displaylimits_0^{\pi/2} \,\Phi_0(T) \, x^{-2}\, {\rm cos}\,\theta_{\rm S}\, x^2\, {\rm sin}\,\theta_{\rm S}\, 
%    {\rm d}\theta_{\rm S}\, {\rm d}\varphi
%    \label{eq:phitots1}\\
%    & = \pi\,\Phi_0(T)\,{\rm d}S\;,
%\end{alignat}
%
Here, we assumed azimuthal symmetry in the angle $\varphi$. 
%This can be compared directly to Eq.~(\ref{eq:murphy}); solving for $\Phi_0(T)$, we obtain for crystalline ice
%\begin{equation}
%    \Phi_0(T) = \frac{1}{\pi}\,a\,T^{3.5}\;{\rm exp}\,\left(\frac{-E_{\rm sub}}{k_{\rm B}T}\right)\;.
%    \label{eq:flux_norm}
%\end{equation}
%
%One can then compute the contamination rate of one surface by another using Eq.~(\ref{eq:sublimation_flux}), using the sublimation-flux normalisation in Eq.~(\ref{eq:flux_norm}).

\subsection{Contamination forecasts\label{sec:contamination_forecast}}
%After commissioning and during routine phase operations, some volatiles may still be present in the PLM, either due to outgassing effects, or because they haven't yet manage to evacuate the PLM. These volatiles (mostly water ice) will deposit on optical surfaces, decreasing optical transmission and increasing scattered light. At some point, the performance of the spacecraft and our capability to accurately calibrate the data become sufficiently hampered, warranting a removal of the contaminants. GAIA, for example, required three decontamination procedures during its commissioning, and three more during the first three years of operation before reaching a mostly stable state.
The telescope cavity consists of mirrors M1, M2, and the external telescope baffle, and is directly exposed to open space at the telescope's front aperture. The instrument cavity is mounted on the PLM baseplate and is located below M1, containing folding optics and the instruments (see Fig.~\ref{fig:PLM}). The instrument cavity is connected to open space only through the bore hole in M1, limiting the capability of water escape.

\subsubsection{\label{contamination_summary_telescopecavity}Telescope cavity}
In Appendix \ref{apdx_contamination_M1M2} we introduce a cylindrical model of the telescope cavity to compute the contamination rates for M1 and M2 (see Table \ref{contamination_rate_table}), using the formalism developed earlier in this section. In this model, M1 can be contaminated by ice subliming from the interior wall of the external baffle, from M2, and from a `front ring' that reduces the telescope aperture. Likewise, M2 can be contaminated by sublimation from the baffle, from M1, and from a `back ring', that is the structural parts visible between M1 and the baffle wall. 

We compute the contamination for the nominal temperatures and the warm comparison case 
(Table~\ref{euclid_temperatures}). In flight, temperatures are expected to stay within a few kelvin of the nominal case.

The following are some of our findings for the nominal temperatures and the glacial scenario:
(i) 99.6\% of the water escaping through the front aperture is subliming from M1 and the back ring. 
(ii) 11\% (6\%) of the ice subliming from the baffle (M1) escape the telescope cavity on direct paths, the rest will redeposit.
(iii) M1 slowly decontaminates at $-0.33$\,nm\,month$^{-1}$. Despite being very cold, M2 will contaminate only slowly at $+0.13$\,nm\,month$^{-1}$.
(iv) The thickness variation of the ice on M1 and M2 is about 1\% or less and thus very uniform (Fig.~\ref{fig:mirror_surfaces}).

\begin{table}[t]
\caption{Total contamination rates for the nominal and warm case (Table~\ref{euclid_temperatures}).}
\smallskip
\label{contamination_rate_table}
\smallskip
\begin{tabular}{lrrr}
\hline
\rowcolor{gray!20}
Common to & \multicolumn{2}{c}{${\rm d}z/{\rm d}t\;(T_{\rm nominal})$} & ${\rm d}z/{\rm d}t\;(T_{\rm warm})$ \\
\rowcolor{gray!20}
VIS and NISP & \multicolumn{2}{c}{[nm\,month$^{-1}$]} & [nm\,month$^{-1}$]\\
\rowcolor{gray!20}
\hline
& WMK & Sack+ (1993) & WMK\\
\hline
M1  & $-0.33$ & $-1.1$ & $-4.0$ \\
M2  & $0.13$ & $+0.3$ & $+1.1$ \\
FoM1 & $+0.08$ & $-0.16$ & $-14$ \\
FoM2 & $+1.4$& $+3.4$ & $+0.50$ \\
M3 & $+1.4$ & $+3.4$ & $-25$ \\
Dichroic &$+1.4$  & $+3.4$ & $+0.50$ \\
\hline
\rowcolor{gray!20}
NISP path && & \\
\hline
NISP lenses &$-101$ & $-238$ & $-89$ \\
Detector &$+9.9$ & $+27$ & $+22$ \\
\hline
\rowcolor{gray!20}
VIS path && & \\
\hline
FoM3 &$+3.6$ & $+9.6$ & $+9.3$ \\
Detector &$-44\,000$ & $-79\,000$ & $-122\,000$ \\
\hline
\end{tabular}
\footnotesize
\tablefoot{Negative values mean that sublimation is more efficient than condensation. We note that in most related technical publications, the sublimation and condensation fluxes are given in units of g\,cm$^{-2}$\,s$^{-1}$. For us the optical effects are of interest, hence we parameterise contamination rates as a change of ice thickness $z$.}
\end{table}

\subsubsection{\label{contamination_summary_instrumentcavity}Instrument cavity}
In Appendix \ref{apdx_contamination_NISPVIS} we present a hemispherical model to compute contamination rates in the instrument cavity (Table \ref{contamination_rate_table}). In a hemispherical model, the flux of water is incoming from the $2\pi$\,sr solid angle above the point under consideration and is independent of the hemisphere's radius. For a simple estimate we can thus ignore the much more complex geometry of the instrument cavity (Figs.~\ref{fig:euclid_CAD_annotated} and \ref{fig:euclid_instcavity}), as long as the solid angle is filled with emitting surfaces at the same temperature.

For nominal operating temperatures and the glacial scenario we find:
(i) If the NISP optics are initially free of ice, then they will stay free of ice. A surface in the NISP optics will effectively sublime 101\,nm\,month$^{-1}$ since it is comparatively warm.
(ii) The NISP optics can decontaminate themselves during the time between launch and the arrival at Lagrange point L2, unless they get initially contaminated with more than 100\,nm per surface.
(iii) The NISP detectors will accumulate a substantial amount of 10\,nm\,month$^{-1}$, as they are considerably colder than their environment. 
(iv) Any contamination on FoM1 will remain unchanged.
(v) FoM2, M3, the dichroic, and in particular FoM3 will accumulate ice.
(vi) The VIS detector effectively decontaminates itself, as it is much warmer than its environment.
(vii) Actual contamination rates are highly sensitive to temperature changes as small as $1$--$2$\,K.

We note that the above statements about the NISP optics remaining free of ice only hold if the optics are exposed to the instrument cavity. In the as-built instrument this is not the case. The NISP optics, the filter wheel, and the grism wheel are encapsulated in a SiC box that has very small venting holes, only. NISP itself is wrapped in MLI (Fig.~\ref{fig:euclid_instcavity}, effectively forming a closed system with its own contamination dynamics.

\subsubsection{Industry forecast by Airbus Defence and Space (ADS)\label{airbus}}
ADS has performed a molecular contamination analysis of the PLM, using 3D geometric models and a distribution of various outgassing materials and molecular species. The expected water contamination ranges from 0.1\,nm\,month$^{-1}$ for the telescope cavity, to 1\,nm\,month$^{-1}$ for the instrument cavity. This is about the same order of magnitude as the water exchange from sublimation in our stationary glacial model (see Table \ref{contamination_rate_table}). We note that the ADS estimates are subject to the same uncertainties as outlined in Sects.~\ref{sec:diffusion} and \ref{sec:theoretical_sublim_rates}, and could be a factor of a few (or more) higher or lower. More quantitative estimates about the actual uncertainty cannot be made with the data at our hands.

%\begin{figure}[t]
%\begin{center}
%\includegraphics[angle=0,width=1.0\hsize]{sublimationrate_annotated.pdf}
%\end{center}
%\caption{sublimation flux of crystalline ice following Eq.~(\ref{eq:murphy}). Overlaid are the rates for the decontamination temperatures without Sun exposure of various components (see Tables \ref{euclid_temperatures} and \ref{sublimation_rate_table}). With Sun exposure as currently planned, the PLM baseplate and the baffle move up to about 205\,K. See also Table \ref{sublimation_rate_table}.}
%\label{fig:sublimation_rate_annotated}
%\end{figure}

\subsubsection{Ammonia contamination from thruster firings\label{sec:hydrazine}}
In this subsection we deviate shortly from water ice. \Euclid carries 137.5\,kg of pure hydrazine propellant (\ce{N2H4}), sufficient for a L2 halo-orbit insertion, a six-year mission, a potential 1--2 year mission extension, and an end-of-life insertion into a heliocentric graveyard orbit \citep{racca2016}. Halo-orbit correction manoeuvres are carried out every four weeks during a reserved 6\,h window \citep{scaramella2022}. Thruster firings will in general contaminate a spacecraft through expansion of the supersonic flow in a vacuum \citep{chen2000,dettleff2011,lee2017,yarygin2017}. Some of \Euclid's hydrazine thrusters are shown in Fig.~\ref{fig:euclid_thrusters_lowres}. Thales Alenia Space -- who built \Euclid's SVM --  have modelled \Euclid's thruster contamination and found it to be negligible, but no details could be communicated that would allow us to independently verify their conclusions. Therefore, here we make a simple worst-case estimate of the expected contamination, and confirm that it is negligible.

Hydrazine is a monopropellant -- that is it does not need an oxidiser -- with the following two reactions when pushed through the catalytic bed of a thruster \citep{price1968,makled2009},
\begin{eqnarray}
\ce{3 N2H4 -> 4 NH3 + N2}\;&\;\;{\rm and}\label{eq:nh3_1}\\
\ce{4 NH3 -> 2 N2 + 6 H2}\;.&\label{eq:nh3_2}
\end{eqnarray}
The first reaction is fast and exothermic and happens at the beginning of the catalyst bed, whereas the second reaction is slow and endothermic and occurs at the end of the catalyst bed. For thruster purposes the second reaction should be suppressed, that is as much \ce{NH3} as possible should be preserved to achieve a hot exhaust jet with high specific impulse \citep{Pakdehi2019}. The fraction of unspent \ce{NH3} is controlled by the thruster design\footnote{A comparatively cold but very gas-rich stream emerges if most \ce{NH3} is spent, in which case the catalytic chamber serves as a gas producer for various different technical purposes.}.

\begin{figure}[t]
  \begin{minipage}[t]{1.0\linewidth}
    \centering
    \includegraphics[width=\linewidth]{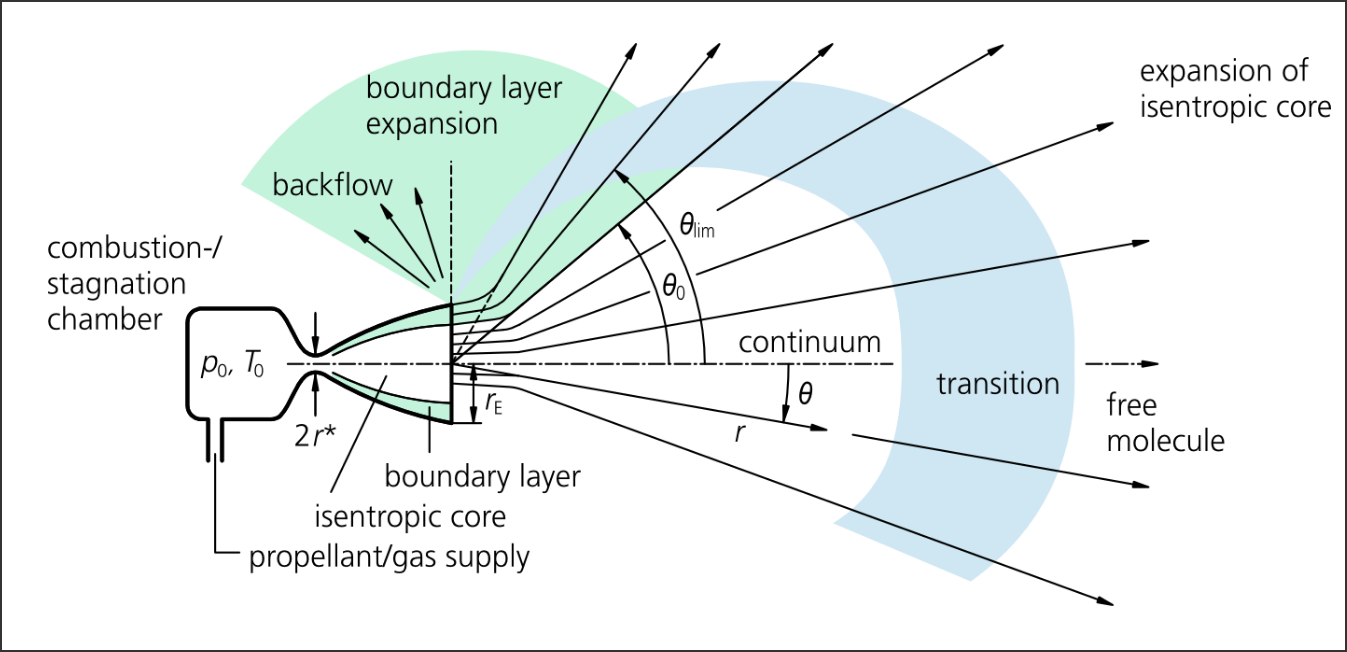}
    \caption{Schematic view of the flow regime of a typical small chemical or cold-gas thruster on a spacecraft, expanding into vacuum. The backflow might contaminate the spacecraft. Figure credit: \cite{dettleff2011}.}
    \label{fig:nozzle_backflow}
  \end{minipage}
  \begin{minipage}[t]{0.95\linewidth}
  \vspace{0.0cm}\phantom{x} 
  \end{minipage}
  
  \begin{minipage}[t]{1.0\linewidth}
    \centering
    \includegraphics[width=\linewidth]{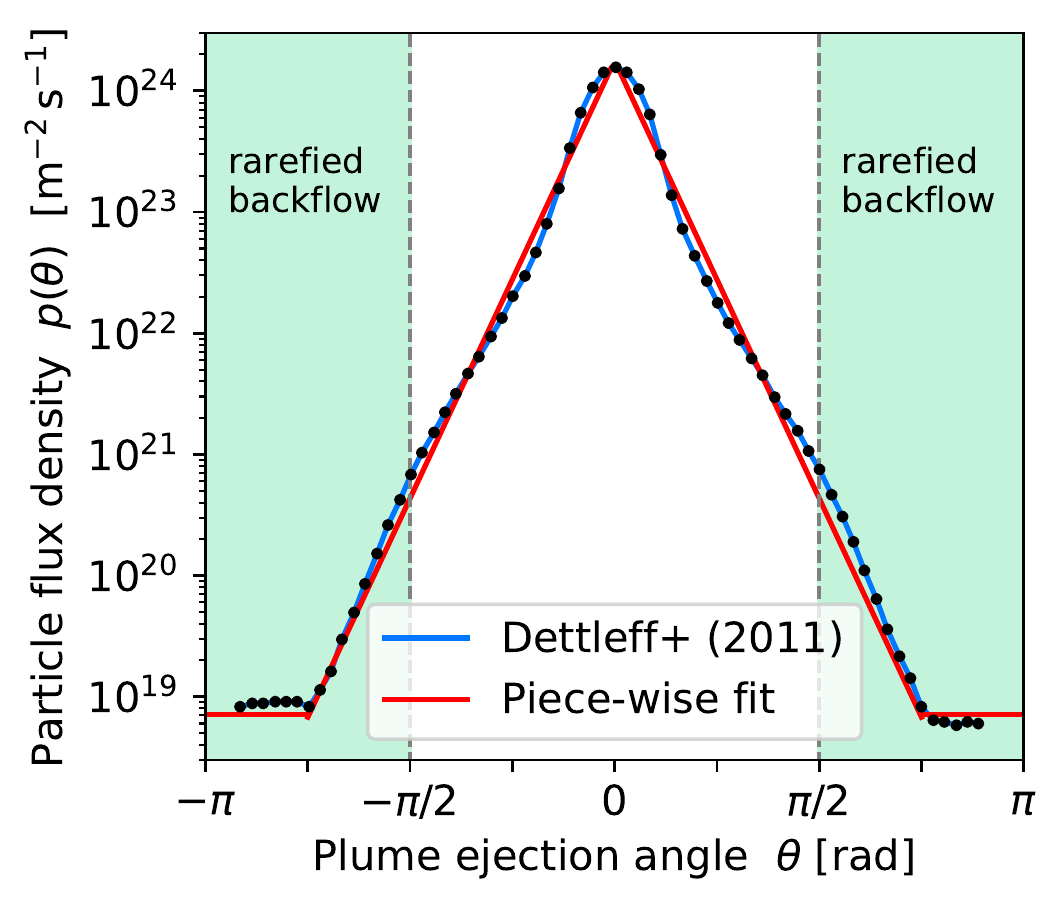}
    \caption{Measured particle flux density across the flow of a cold-gas thruster. Data taken from  \cite{dettleff2011}. The red line shows our piece-wise fit, symmetric around $\theta=0$, and extrapolated to the edges of the $[-\pi,+\pi]$ range.}
    \label{fig:jetprofile}
  \end{minipage}
\end{figure}

For the worst-case L2 halo-orbit correction manoeuvre we assume the following: 1\,kg of hydrazine is used to achieve a velocity change of $\Delta v=0.5$\,m\,s$^{-1}$, the latter being a worst-case assumption by \Euclid's flight-dynamics team; all \ce{NH3} is preserved, that is a maximum of 1.42\,kg of \ce{NH3} are produced; and the entire rarefied backflow (Fig.~\ref{fig:nozzle_backflow}) from the thruster's jet gets deposited uniformly on all \Euclid surfaces. 

To estimate the amount of \ce{NH3} that could contaminate \Euclid, we must determine the fraction of mass contained in the backflow. We digitised\footnote{Using {\tt WebPlotDigitizer} \citep{rohatgi2022}} the measurements in figure 28 of \cite{dettleff2011}, showing the particle flux density in the supersonic flow, and reproduce it in our Fig.~\ref{fig:jetprofile}. We then approximate the flow as
\[
    {\rm log_{10}}\,\left[\frac{p(\theta)}{1\,{\rm m^2\,s}}\right]= 
\begin{cases}
    -2.30\,|\theta| + 24.25\,, & {\rm for}\;|\theta|\leq 0.75\,\pi\;\;(\ang{135;;})\\
    18.85\,,              & {\rm for}\;0.75\,\pi<|\theta|<\pi\;.
\end{cases}
\]
Here, $\theta$ is the ejection angle with respect to the nozzle's axis and $p(\theta)$ is the particle flux per area measured in ${\rm m}^{-2}\,{\rm s}^{-1}$. Assuming radial symmetry around the nozzle's ejection axis, we integrate over the particle flux density and find that the backflow ($|\theta|>\pi/2$) contains 0.65\% of the total mass ejected.
This translates to 9.2\,g of \ce{NH3} in the backflow. We ignored any mass-segregation effects \citep{price1968}, that is \ce{NH3} and \ce{N2} are homogeneously distributed in the flow. 

Approximating \Euclid with a cylinder of 4.5\,m height and 3.1\,m width, it would have a surface area of 59\,m$^2$, of which 1.2\,m$^2$ are for the M1 mirror. Assuming \ce{NH3} is uniformly distributed over this surface, M1 would then accumulate 0.19\,g of NH$_3$. Solid NH$_3$ has a density of 0.9\,g\,cm$^{-3}$ at 100\,K \citep{satorre2013}, similar to the density of crystalline water ice; the \ce{NH3} layer would be 169\,nm thick. \cite{brown2007} show that the desorption rate of solid NH$_3$ in a vacuum at 100--120\,K is 6--8 orders of magnitude higher than that of water ice \citep[see also][]{zhang2009}. Hence this layer of NH$_3$ would sublime in about 4\,h at 110\,K (see also Fig.~\ref{fig:sublimation_rate}). That is consistent with \cite{dawes2007} and references therein, who report the occurrence of multilayer desorption of NH$_3$ at temperatures of 100\,K and above. In reality, not all of the backflow will deposit on \Euclid, and only a very small fraction will enter the telescope aperture that faces away from the thrusters' nozzle axes (see Fig.~\ref{fig:euclid_thrusters_lowres}). Any \ce{NH3} deposits from orbit maintenance will have sublimed before science operations resume.

\ce{N2} and \ce{H2} ices from hydrazine breakdown cannot form on \Euclid due to its comparatively high temperature. We have not considered unspent hydrazine, which may constitute 1\% of the mass in the exhaust \citep{chen2000}, but we note that hydrazine is dissociated by UV-photons with wavelengths shorter than 250\,nm \citep{vaghjiani1993}.

\subsection{\label{decontamination_procedure}Decontamination procedure}
The thermal decontamination scheme for \Euclid foresees temperatures of $140$--$270$\,K, using heaters and partial Sun exposure. The thermal cycle alone takes about 18 days. During the first three days, the spacecraft's decontamination heaters are turned on with full power demand. Because of \Euclid's compact design, it can generate only a limited amount of heating power from its on-board solar cells. Therefore, on the fourth day, the telescope's solar aspect angle\footnote{The solar aspect angle is the angle between \Euclid's viewing direction and the Sun. During routine operations it is kept between \ang{87;;}--\ang{120;;} to maximise thermal stability \citep{scaramella2022}.} will be reduced to \ang{45;;}. This allows the external telescope baffle to reach a temperature of up to 200\,K. Only a small part of the baffle is directly exposed to the Sun, but since it is made of aluminium, which conducts heat easily, the parts remaining in the shadow will also decontaminate. The demanded heater power is reduced during this time, acknowledging the reduced effective area of the solar panel. 

The telescope stays at full decontamination power for two days. The sublimation itself takes a few minutes only once maximum temperatures are reached: 0.23\,\micron\;and 3.6\,\micron\;of ice sublime per second at 200\,K and 220\,K, respectively. While the ice may evaporate rapidly, additional time is required to give sublimates a chance to find their way out of the cavities and leave the spacecraft. For example, according to our model only 6\% of the water molecules evaporating in the telescope cavity escape on direct paths, the rest will undergo numerous redeposition and sublimation cycles before eventually escaping. Furthermore, the high decontamination temperatures result in a decreased sticking coefficient \citep{kossacki1999,batista2005,gibson2011,brieda2022} and a massively increased evaporation rate. Consequently, the mean free path length might become comparable to -- or even smaller than -- the size of the spacecraft, in which case pressure effects would have to be taken into account for more accurate evacuation-time estimates.

Six days after beginning of the decontamination, the spacecraft is restored to a nominal solar aspect angle, and a controlled cool-down begins. The latter takes approximately twelve days, keeping optical elements warmer than their surroundings so that any water residuals can condense on colder surfaces. A series of recalibration steps is executed as soon as the instruments have reached operational temperatures. A possible optical realignment and further on-sky recalibrations can only occur once the telescope optics are stable again. The duration of a full decontamination cycle including all recalibrations is expected to last up to 25 days and potentially longer.

We expect essentially all superficial ice to evacuate from the telescope cavity during a full decontamination cycle. The situation for the instrument cavity is different, as the opening to the telescope cavity is comparatively small and the geometry of the instrument cavity complex. In particular in NISP, which is mostly enclosed in MLI, water has reduced escape capabilities and could eventually recontaminate the detector. Given a clear indication of ice on the NISP detectors -- for example modulated quantum efficiency, spectral absorption or structures in the flat fields, as elaborated in \cite{holmes2016} and in our second paper -- a partial decontamination only for NISP detectors could be considered. This would mean a smaller thermal disturbance to the telescope than a full decontamination, yet preliminary thermal considerations indicate that 3--4 days would still be required. 

Other than the NISP detectors, surgical decontamination of individual components is not possible with \Euclid, as the mounted optics and the PLM baseplate are fully constructed in SiC (see also Fig.~\ref{fig:euclid_CAD_annotated}). Any heat applied locally to an optical element would quickly propagate within the instrument cavity to other areas due to the high thermal conductivity of SiC, thus introducing a global thermal state change.
Fine temperature control of optical elements in \Euclid is not possible, as most heater controllers operate in an on-off fashion, providing full power when on.

To mitigate uncertainties in contamination, \Euclid will undergo an immediate post-launch thermal decontamination, being kept at 200--273\,K for four days. We expect that all water trapped on surfaces will desorb, and that a large fraction of it evacuate the telescope and instrument cavities.

%For completeness, we mention here that non-thermal decontamination schemes exist as well. For example, in low-Earth orbits (LEOs) abundant amounts of atomic oxygen (ATOX) exist, which is highly reactive and chemically degrades multiple components of spacecraft \citep{banks2003,samwel2014}, including MLI and optical elements. Protective coatings such as SiO$_2$ are required \citep{ferreira2021,zhang2021}. ATOX can be used to remove water ice from optics, by pointing a spacecraft in LEO in into the ram. Alternatively, onboard oxygen plasma supplies can be used (M. Portaluppi, private communication, 2023). 

\section{Results from thermal vacuum tests\label{tvac}}
\subsection{Vacuum tests of the PLM}
In 2021 the \Euclid PLM underwent extensive vacuum tests for 60 days at a pressure of $10^{-6}$\,mbar, simulating space conditions in a vacuum chamber at the Centre Spatial de Liège (CSL), Belgium. To cool down to its operational temperatures, the PLM must see a colder object, provided by a liquid helium shroud, which itself sits in a nitrogen shroud. Once the chamber was evacuated, everything was kept at ambient temperature for 4.5 days for initial outgassing of all components in the chamber. \Euclid was then cooled down and kept at operating temperatures for 30 days. Afterwards, a full decontamination was run (11 days), followed by another cool-down to operating temperatures (9 days) before the final-warm-up.

Witness samples for non-volatile organic contaminants were placed inside the PLM's instrument and telescope cavities. These contaminants are heavier than water and outgas at higher temperatures. No organic contamination could be found after the tests. This confirms the efficient bake-out of all components during construction, in a vacuum and at temperatures of $80^\circ$\,C to $120^\circ$\,C, much higher than \Euclid's decontamination temperatures. While heavier organic compounds might still be dissolved in some materials after bake-out, they are not expected to outgas in flight at cryogenic temperatures, nor during the thermal decontamination that reaches at most room temperatures. We are thus confident that \Euclid will not be contaminated by organic species; water ice remains the only concern. 

No signatures of contamination -- from water ice or else -- were detected in the test data taken by the cold PLM instruments, for example NISP flat fields and images of an artificial star. However, the test was short compared to \Euclid's in-orbit life; slowly growing ice films could simply not have had enough time to become thick enough for detection during the test. Moreover, the in-flight calibration observations in zero gravity and with low background, with the optics at its full performance, will be much more powerful in detecting contamination, as we show in our second paper.
 
During the vacuum tests the same type of witness samples were placed inside the shrouds. A residual gas analyser faced the helium shroud MLI from nearby, as it must sample the contamination plume from a close distance. No emission from the helium shroud was detected, and the witness samples remained clean. While these measurements did not probe the PLM, they show that the vacuum tests were nominal from a contamination perspective.

\subsection{Vacuum tests of the fully assembled spacecraft}
In 2022 the fully integrated spacecraft was tested in space conditions for another month at Thales Alenia Space in France (TAS-F). At this point in time, with the PLM being handed over by ADS to TAS-F, the entrance aperture of the PLM was sealed to avoid particulate contamination of interior surfaces. No measurement probes and witness samples were allowed anymore inside the PLM. A QCM placed in the chamber showed no excess contamination during the test in comparison to a reference blank run without the flight hardware. For more details about the vacuum tests see for example \cite{poidomani2020}.

\section{Conclusions and outlook\label{sec:conclusions}}
This paper is the first in a series of two about water-ice contamination processes in spacecraft, and \Euclid specifically. To the best of our knowledge, this is the first presentation of the subject from a first-principles perspective. We review the outgassing and contamination records of a dozen different spacecraft and instruments, and we conclude that contamination is a highly dynamic and very long-lived process. The dominant reservoir of water in spacecraft such as \Euclid is the MLI used for thermal insulation. In worst-case conditions, it will take years for the MLI to fully dry up. Consequently, we expect molecular contamination to be active throughout \Euclid's six-year mission duration (Sect.~\ref{sec:spacecraftlessons}), with a forecast of low water contamination overall, albeit with considerable uncertainty.

To better understand the contamination process of optical surfaces themselves, and ultimately the performance impact on the data (evaluated in our second paper), we have reviewed the current knowledge of the creation of thin ice films on different substrates. We find that the structure and topography of the ice films is highly dependent on the substrate material. Most of the coating materials used for \Euclid's optical surfaces are not disclosed to us by the manufacturers, hence we cannot make accurate forecasts about their optical properties -- such as scattering losses -- when iced. Even if the coatings were known, including the exact crystalline or amorphous atomic structure exposed at their surfaces, current theories are not able to reliably predict the growth and structure of deposited ice films (Sect.~\ref{sec:ice}).

Quantitative estimates of the in-flight outgassing and contamination rates remain rather uncertain. At \Euclid's typical temperatures, even small changes of a few kelvin accelerate the diffusion speed of water in the MLI, and the subsequent sublimation flux, by a factor of a few. There is also a strong dependence on the MLI's chemical composition and molecular structure, and uncertainties when extrapolating the outgassing rates measured at room temperature to \Euclid temperatures. Small deviations of \Euclid's in-flight temperatures from their pre-launch expectations can therefore have a considerable impact on the actual contamination rates (Sects.~\ref{sec:diffusion} and \ref{sec:theoretical_sublim_rates}).

The matter is complicated further since crystalline ice layers on top of outgassing substrates may act as diffusion barriers (Sects.~\ref{sec:diffusion}). Thus, at low temperatures, existing thin ice films on \Euclid's non-optical surfaces might actually be beneficial. However, \Euclid is only 10--20\,K below the point of 140--150\,K where sublimation and diffusion accelerate rapidly in an exponential fashion. Forecasts of the absolute amount of contamination are therefore hard; they also require full 3D modelling of the emitting and contaminating surfaces, as was done for example for JWST in \cite{brieda2022}, which is well beyond the scope of this paper. 

To estimate the contamination dynamics from a sublimation perspective alone, we assumed that all surfaces in \Euclid are already iced and that these ice layers act as effective diffusion barriers, such that diffusion can be neglected. We could then compute the water exchange rates between surfaces using a semi-empirical model of the sublimation flux at cryogenic temperatures, without the uncertainties inherent to direct material outgassing. We find typical water-contamination rates of up to $10$\,nm\,month$^{-1}$ for the various optical surfaces (Table~\ref{contamination_rate_table}). The coldest surfaces in \Euclid are at greatest risk of contamination because they have the lowest sublimation fluxes. The NISP detectors at 95\,K will act as a cold trap for water vapour \citep[see also][]{holmes2016}, which does not have many escape paths from the NISP's MLI enclosure (Sects.~\ref{sec:fluxcomputation} and \ref{sec:contamination_forecast}). Fortunately, the NISP detectors have separate heaters, and a partial decontamination can be considered if necessary, compared to a full decontamination of the entire spacecraft.
%The effect of ice contamination on the quantum efficiency of the NISP detectors is studied in \cite{holmes2016}, and we account for their results in our second paper.

Our contamination rates estimated from sublimation alone are comparable to the water contamination rates estimated by ADS. The latter ones are computed directly for diffusion outgassing and range from  0.1\,nm\,month$^{-1}$ in the telescope cavity to 1.0\,nm\,month$^{-1}$ in the instrument cavity. These estimates are uncertain by a factor of a few or more, as argued in Sects.~\ref{sec:diffusion} and \ref{sec:theoretical_sublim_rates}. Organic contamination is not expected for \Euclid, owing to extensive bake-out at high temperatures, which was confirmed during the on-ground thermal vacuum tests (Sect.~\ref{tvac}). The Gaia and XMM-\textit{Newton} OM experiences, though, caution us that considerable in-flight contamination is not necessarily anticipated by on-ground tests, and suitable calibration and decontamination plans must be in place for \Euclid's operational phase.

%The SpaceX standard launch scenario foresees venting of the Falcon 9's fairing with gaseous nitrogen for the last 1.5 hours prior to launch. This will decrease the initial amount of water condensed on \Eulid's surfaces, but carries an increased risk of electrostatic discharge. 

In the second paper, we examine the optical effects of water ice on \Euclid's spectrophotometric data. We look at absorption, interference, scattering, polarisation, apodisation, and phase shifts, and investigate \Euclid's sensitivity to these effects. Our estimates are based on theoretical calculations as well as dedicated optical experiments of contaminated mirror coating samples. Overall, we find \Euclid in a great position to detect even very small amounts of a few to a few tens of nanometer water ice in its optical path, using -- among others -- regular observations of a stellar self-calibration field. This sensitivity, however, also implies that already small amounts of ice must be tracked and accounted for in the data analysis. Decontamination must occur when our calibration requirements cannot be met anymore by the corrected data.

For future missions where contamination could be relevant and which cannot be decontaminated easily, the installation of QCMs with suitable viewing angles near critical surfaces would be beneficial. QCMs are capable of detecting even fractional monolayers of water ice and other contaminants, thus providing accurate real-time knowledge of actual contamination levels. This was demonstrated successfully by the MSX experiment \citep{uy1998,uy2003,wood2003}. 
%
%\begin{figure}[t]
%\centering
%\includegraphics[angle=0,width=1.0\hsize]{decontamination_spacecraft.png}
%\caption{Direct Sun exposure (reducing the solar aspect angle to $45^\circ$) will decontaminate the PLM external baffle and the M2 baffle.}
%\label{decontamination_spacecraft}
%\end{figure}
%
%\begin{table}[t]
%\caption{\small{Surface micro-roughness for various optical elements in Euclid, measured on the substrate (SiC) %over areas of approximately 1 mm$^2$ before coating with reflective layers. The coatings are not expected to %change the rms significantly. The $^*$ indicates worst case values.}}
%\small
%\smallskip
%\label{surface_roughness_data}
%\smallskip
%\begin{tabular}{|l|l|l|l|l|l|l|}
%\hline
%\rowcolor{gray!30}
% & M1 & M2 & M3 & FoM1$^*$ & FoM2$^*$  & FoM3$^*$  \\
%\hline
%$\sigma_{\rm SiC}$ & 1.1\,nm & 0.9\,nm & 0.925\,nm & 2.0\,nm & 2.4\,nm & 1.8\,nm \\
%\hline
%\end{tabular}
%\end{table}

\begin{acknowledgements}
The authors at MPIA acknowledge direct funding by the German DLR under grant numbers 50\,QE\,2003 and 50\,QE\,2303, and the support of our librarian Simone Kronewetter for providing the full texts of numerous non-astronomical references.\\
Most figures in this paper were prepared with {\tt Matplotlib} \citep{hunter20007}.\\
\AckEC
\end{acknowledgements}

\bibliographystyle{aa}

\begin{thebibliography}{201}
\expandafter\ifx\csname natexlab\endcsname\relax\def\natexlab#1{#1}\fi

\bibitem[{Abeel {et~al.}(2022)Abeel, Wooldridge, Calcabrini, Ward, \&
  Schmeitzky}]{abeel2022}
Abeel, A.~C., Wooldridge, E.~M., Calcabrini, M., Ward, J.~O., \& Schmeitzky, O.
  2022, in Space Systems Contamination: Prediction, Control, and Performance
  2022, ed. C.~E. Soares, E.~M. Wooldridge, \& B.~A. Matheson, Vol. 12224,
  International Society for Optics and Photonics (SPIE), 122240D

\bibitem[{{Altobelli} {et~al.}(2016){Altobelli}, {Postberg}, {Fiege},
  {Trieloff}, {Kimura}, {Sterken}, {Hsu}, {Hillier}, {Khawaja},
  {Moragas-Klostermeyer}, {Blum}, {Burton}, {Srama}, {Kempf}, \&
  {Gruen}}]{altobelli2016}
{Altobelli}, N., {Postberg}, F., {Fiege}, K., {et~al.} 2016, Science, 352, 312

\bibitem[{Andreas(2007)}]{andreas2007}
Andreas, E.~L. 2007, Icarus, 186, 24

\bibitem[{{Baggett} \& {Gonzaga}(1998)}]{baggett1998}
{Baggett}, S. \& {Gonzaga}, S. 1998, WFPC2 Long-Term Photometric Stability,
  Space Telescope WFPC2 Instrument Science Report

\bibitem[{{Baggett} {et~al.}(2001){Baggett}, {Gonzaga}, {Biretta}, {Casertano},
  {Heyer}, {Koekemoer}, {Mack}, {McMaster}, {Riess}, {Schultz}, \&
  {Wiggs}}]{baggett2001}
{Baggett}, S., {Gonzaga}, S., {Biretta}, J., {et~al.} 2001, WFPC2 Cycle 8
  Closure Report, Space Telescope WFPC2 Instrument Science Report

\bibitem[{{Baggett} {et~al.}(1996){Baggett}, {Sparks}, {Ritchie}, \&
  {MacKenty}}]{baggett1996}
{Baggett}, S., {Sparks}, W., {Ritchie}, C., \& {MacKenty}, J. 1996,
  Contamination Correction in SYNPHOT for WFPC-2 and WF/PC-1, Space Telescope
  WFPC2 Instrument Science Report

\bibitem[{Banks {et~al.}(2003)Banks, Miller, de~Groh, \& Demko}]{banks2003}
Banks, B., Miller, S., de~Groh, K., \& Demko, R. 2003, in Scattered atomic
  oxygen effects on spacecraft materials, NASA Technical Memorandum,
  NASA/TM-2003-212484, available at
  \url{https://ntrs.nasa.gov/citations/20030062195}

\bibitem[{Bar-Nun {et~al.}(2007)Bar-Nun, Notesco, \& Owen}]{barnun2007}
Bar-Nun, A., Notesco, G., \& Owen, T. 2007, Icarus, 190, 655, deep Impact
  Mission to Comet 9P/Tempel 1, Part 2

\bibitem[{Bassler(2010)}]{bassler2010}
Bassler, N. 2010, Radiation and environmental biophysics, 49, 373

\bibitem[{Batista {et~al.}(2005)Batista, Ayotte, Bili\ifmmode~\acute{c}\else
  \'{c}\fi{}, Kay, \& J\'onsson}]{batista2005}
Batista, E.~R., Ayotte, P., Bili\ifmmode~\acute{c}\else \'{c}\fi{}, A., Kay,
  B.~D., \& J\'onsson, H. 2005, Phys. Rev. Lett., 95, 223201

\bibitem[{Belosludov {et~al.}(2008)Belosludov, Subbotin, Mizuseki, Rodger,
  Kawazoe, \& Belosludov}]{belosludov2008}
Belosludov, R.~V., Subbotin, O.~S., Mizuseki, H., {et~al.} 2008, The Journal of
  Chemical Physics, 129, 114507

\bibitem[{Bertie {et~al.}(1969)Bertie, Labbé, \& Whalley}]{bertie1969}
Bertie, J.~E., Labbé, H.~J., \& Whalley, E. 1969, The Journal of Chemical
  Physics, 50, 4501

\bibitem[{{Bhaskaran} {et~al.}(2004){Bhaskaran}, {Mastrodemos}, {Riedel}, \&
  {Synnott}}]{bhaskaran2004}
{Bhaskaran}, S., {Mastrodemos}, N., {Riedel}, J.~E., \& {Synnott}, S.~P. 2004,
  in ESA Special Publication, Vol. 548, 18th International Symposium on Space
  Flight Dynamics, 455

\bibitem[{Bieler {et~al.}(2016)Bieler, Altwegg, Balsiger, Berthelier, Calmonte,
  Combi, Keyser, Fiethe, Fuselier, Gasc, Gombosi, Hansen, Hässig, Korth, Roy,
  Mall, Rème, Rubin, Sémon, Tenishev, Tzou, Waite, \& Wurz}]{bieler2016}
Bieler, A., Altwegg, K., Balsiger, H., {et~al.} 2016, in Systems Contamination:
  Prediction, Control, and Performance 2016, ed. J.~Egges, C.~E. Soares, \&
  E.~M. Wooldridge, Vol. 9952, International Society for Optics and Photonics
  (SPIE), 120--129

\bibitem[{Bj{\"o}rneholm {et~al.}(2016)Bj{\"o}rneholm, Hansen, Hodgson, Liu,
  Limmer, Michaelides, Pedevilla, Rossmeisl, Shen, Tocci, Tyrode, Walz, Werner,
  \& Bluhm}]{bjoerneholm2016}
Bj{\"o}rneholm, O., Hansen, M.~H., Hodgson, A., {et~al.} 2016, Chemical
  Reviews, 116, 7698

\bibitem[{{Bohlin} \& {Deustua}(2019)}]{bohlin2019}
{Bohlin}, R.~C. \& {Deustua}, S.~E. 2019, \aj, 157, 229

\bibitem[{Bougoin \& Lavenac(2011)}]{bougoin2011}
Bougoin, M. \& Lavenac, J. 2011, in Optical Manufacturing and Testing IX, ed.
  J.~H. Burge, O.~W. Fähnle, \& R.~Williamson, Vol. 8126, International
  Society for Optics and Photonics (SPIE), 248--257

\bibitem[{{Bougoin} {et~al.}(2019){Bougoin}, {Mallet}, {Lavenac},
  {Gerbert-Gaillard}, {Ballhause}, \& {Chaumeil}}]{bougoin2019}
{Bougoin}, M., {Mallet}, F., {Lavenac}, J., {et~al.} 2019, in Society of
  Photo-Optical Instrumentation Engineers (SPIE) Conference Series, Vol. 11180,
  International Conference on Space Optics -- ICSO 2018, 111801P

\bibitem[{{Breeveld} {et~al.}(2010){Breeveld}, {Curran}, {Hoversten}, {Koch},
  {Landsman}, {Marshall}, {Page}, {Poole}, {Roming}, {Smith}, {Still},
  {Yershov}, {Blustin}, {Brown}, {Gronwall}, {Holland}, {Kuin}, {McGowan},
  {Rosen}, {Boyd}, {Broos}, {Carter}, {Chester}, {Hancock}, {Huckle}, {Immler},
  {Ivanushkina}, {Kennedy}, {Mason}, {Morgan}, {Oates}, {de Pasquale},
  {Schady}, {Siegel}, \& {vanden Berk}}]{breeveld2010}
{Breeveld}, A.~A., {Curran}, P.~A., {Hoversten}, E.~A., {et~al.} 2010, \mnras,
  406, 1687

\bibitem[{Brieda \& Laugharn(2020)}]{brieda2020}
Brieda, L. \& Laugharn, M. 2020, in Systems Contamination: Prediction, Control,
  and Performance 2020, ed. C.~E. Soares, E.~M. Wooldridge, \& B.~A. Matheson,
  Vol. 11489, International Society for Optics and Photonics (SPIE), 114890H

\bibitem[{Brieda {et~al.}(2022)Brieda, Laugharn, Woronowicz, Henderson-Nelson,
  May, \& Wooldridge}]{brieda2022}
Brieda, L., Laugharn, M., Woronowicz, M., {et~al.} 2022, in Space Systems
  Contamination: Prediction, Control, and Performance 2022, ed. C.~E. Soares,
  E.~M. Wooldridge, \& B.~A. Matheson, Vol. 12224, International Society for
  Optics and Photonics (SPIE), 1222403

\bibitem[{Brown \& George(1996)}]{brown1996}
Brown, D.~E. \& George, S.~M. 1996, The Journal of Physical Chemistry, 100,
  15460

\bibitem[{Brown {et~al.}(1996)Brown, George, Huang, Wong, Rider, Smith, \&
  Kay}]{brown1996a}
Brown, D.~E., George, S.~M., Huang, C., {et~al.} 1996, The Journal of Physical
  Chemistry, 100, 4988

\bibitem[{{Brown} \& {Bolina}(2007)}]{brown2007}
{Brown}, W.~A. \& {Bolina}, A.~S. 2007, \mnras, 374, 1006

\bibitem[{Bryson {et~al.}(1974)Bryson, Cazcarra, \& Levenson}]{bryson1974}
Bryson, C.~E., Cazcarra, V., \& Levenson, L.~L. 1974, Journal of Chemical \&
  Engineering Data, 19, 107

\bibitem[{{Burnett} {et~al.}(2005){Burnett}, {McNamara}, {Jurewicz}, \&
  {Woolum}}]{burnett2005}
{Burnett}, D.~S., {McNamara}, K.~M., {Jurewicz}, A.~J.~G., \& {Woolum}, D.~S.
  2005, in 36th Annual Lunar and Planetary Science Conference, ed.
  S.~{Mackwell} \& E.~{Stansbery}, Lunar and Planetary Science Conference, 2405

\bibitem[{{Calaway} {et~al.}(2006){Calaway}, {Stansbery}, \&
  {McNamara}}]{calaway2006}
{Calaway}, M.~J., {Stansbery}, E.~K., \& {McNamara}, K.~M. 2006, in 37th Annual
  Lunar and Planetary Science Conference, ed. S.~{Mackwell} \& E.~{Stansbery},
  Lunar and Planetary Science Conference, 1420

\bibitem[{{Cao}(2021)}]{cao2021}
{Cao}, H.~S. 2021, Journal of Physics D Applied Physics, 54, 203002

\bibitem[{Cepeda-Rizo {et~al.}(2021)Cepeda-Rizo, Gayle, \& Ravich}]{cepeda2021}
Cepeda-Rizo, J., Gayle, J., \& Ravich, J. 2021, Thermal and Structural
  Electronic Packaging Analysis for Space and Extreme Environments (CRC Press)

\bibitem[{{Chan} {et~al.}(2020){Chan}, {Stroud}, {Martins}, \&
  {Yabuta}}]{chan2020}
{Chan}, Q. H.~S., {Stroud}, R., {Martins}, Z., \& {Yabuta}, H. 2020, Space
  Science Reviews, 216, 56

\bibitem[{Chen {et~al.}(2016)Chen, Ding, Li, \& Wang}]{chen2016}
Chen, J., Ding, N., Li, Z., \& Wang, W. 2016, Progress in Aerospace Sciences,
  83, 37

\bibitem[{Chen {et~al.}(2000)Chen, Thomson, \& Woronowicz}]{chen2000}
Chen, P., Thomson, S., \& Woronowicz, M. 2000, Applying contamination modeling
  to spacecraft propulsion system designs and operations (American Institute of
  Aeronautics and Astronautics), AIAA 2000, 461

\bibitem[{{Chiggiato}(2020)}]{chiggiato2020}
{Chiggiato}, P. 2020, arXiv:2006.07124

\bibitem[{{Clampin}(1992)}]{clampin1992}
{Clampin}, M. 1992, System Level Contamination Issues for WFPC2 and COSTAR,
  Space Telescope WFPC2 Instrument Science Report

\bibitem[{Crouzet {et~al.}(2020)Crouzet, Shortt, Beaufort, Blommaert,
  Duinkerken, ter Haar, Kohley, Lemmel, van~der Luijt, Oosthoek, Smit,
  Tourneur-Silvain, \& Visser}]{crouzet2020}
Crouzet, P.-E., Shortt, B., Beaufort, T., {et~al.} 2020, in X-Ray, Optical, and
  Infrared Detectors for Astronomy IX, ed. A.~D. Holland \& J.~Beletic, Vol.
  11454, International Society for Optics and Photonics (SPIE), 43--60

\bibitem[{Cuppen {et~al.}(2022)Cuppen, Noble, Coussan, Redlich, \&
  Ioppolo}]{cuppen2022}
Cuppen, H.~M., Noble, J.~A., Coussan, S., Redlich, B., \& Ioppolo, S. 2022, The
  Journal of Physical Chemistry A, 126, 8859, pMID: 36383692

\bibitem[{{Dale} \& {Marshall}(1991)}]{dale1991}
{Dale}, C.~J. \& {Marshall}, P.~W. 1991, in Society of Photo-Optical
  Instrumentation Engineers (SPIE) Conference Series, Vol. 1447, Charge-Coupled
  Devices and Solid State Optical Sensors II, ed. M.~M. {Blouke}, 70--86

\bibitem[{{Dartois} {et~al.}(2015){Dartois}, {Aug{\'e}}, {Boduch}, {Brunetto},
  {Chabot}, {Domaracka}, {Ding}, {Kamalou}, {Lv}, {Rothard}, {da Silveira}, \&
  {Thomas}}]{dartois2015}
{Dartois}, E., {Aug{\'e}}, B., {Boduch}, P., {et~al.} 2015, \aap, 576, A125

\bibitem[{{Dartois} {et~al.}(2013){Dartois}, {Ding}, {de Barros}, {Boduch},
  {Brunetto}, {Chabot}, {Domaracka}, {Godard}, {Lv}, {Mej{\'\i}a Guam{\'a}n},
  {Pino}, {Rothard}, {da Silveira}, \& {Thomas}}]{dartois2013}
{Dartois}, E., {Ding}, J.~J., {de Barros}, A.~L.~F., {et~al.} 2013, \aap, 557,
  A97

\bibitem[{Dawes {et~al.}(2007)Dawes, Mukerji, Davis, Holtom, Webb, Sivaraman,
  Hoffmann, Shaw, \& Mason}]{dawes2007}
Dawes, A., Mukerji, R.~J., Davis, M.~P., {et~al.} 2007, The Journal of Chemical
  Physics, 126, 244711

\bibitem[{Decoster {et~al.}(2013)Decoster, Clerbaux, Baudrez, Dewitte, Ipe,
  Nevens, Blazquez, \& Cornelis}]{decoster2013}
Decoster, I., Clerbaux, N., Baudrez, E., {et~al.} 2013, Journal of Atmospheric
  and Oceanic Technology, 30, 496

\bibitem[{del Rosso {et~al.}(2020)del Rosso, Celli, Grazzi, Catti, Hansen,
  Fortes, \& Ulivi}]{delrosso2020}
del Rosso, L., Celli, M., Grazzi, F., {et~al.} 2020, Nat. Mater., 19, 663

\bibitem[{Dettleff \& Grabe(2011)}]{dettleff2011}
Dettleff, G. \& Grabe, M. 2011, in Models and Computational Methods for
  Rarefied Flows, RTO-EN-AVT-194, ed. N.~Science \& a.~a.~u.
  Technology~Organization (NATO Science and Technology Organization)

\bibitem[{{Dirri}(2016)}]{dirri2016}
{Dirri}, F. 2016, in 41st COSPAR Scientific Assembly, Vol.~41, G0.3--2--16

\bibitem[{{Dohn{\'a}lek} {et~al.}(2000){Dohn{\'a}lek}, {Kimmel}, {Ciolli},
  {Stevenson}, {Smith}, \& {Kay}}]{dohnalek2000}
{Dohn{\'a}lek}, Z., {Kimmel}, G.~A., {Ciolli}, R.~L., {et~al.} 2000, \jcp, 112,
  5932

\bibitem[{Drobyshev {et~al.}(2007)Drobyshev, Aldiyarov, Zhumagaliuly, Kurnosov,
  \& Tokmoldin}]{drobyshev2007}
Drobyshev, A., Aldiyarov, A., Zhumagaliuly, D., Kurnosov, V., \& Tokmoldin, N.
  2007, Low Temperature Physics, 33, 472

\bibitem[{{Engelhart} {et~al.}(2017){Engelhart}, {Cooper}, {Cowardin},
  {Maxwell}, {Plis}, {Ferguson}, {Barton}, {Schiefer}, \&
  {Hoffmann}}]{engelhart2017}
{Engelhart}, D.~P., {Cooper}, R., {Cowardin}, H., {et~al.} 2017, in Advanced
  Maui Optical and Space Surveillance (AMOS) Technologies Conference, ed.
  S.~{Ryan}, 21

\bibitem[{Epstein \& Ruth(1993)}]{epstein1993}
Epstein, G. \& Ruth, S. 1993, Honeycomb sandwich structures: Vented versus
  unvented designs for space systems, Tech. rep., Aerospace Corp, El Segundo,
  CA, USA, available at \url{https://apps.dtic.mil/sti/citations/ADA276713}

\bibitem[{{Euclid Collaboration: Scaramella} {et~al.}(2022){Euclid
  Collaboration: Scaramella}, {Amiaux}, {Mellier}, {Burigana}, {Carvalho},
  {Cuillandre}, {Da Silva}, {Derosa}, {Dinis}, {Maiorano}, {Maris}, {Tereno},
  {Laureijs}, {Boenke}, {Buenadicha}, {Dupac}, {Gaspar Venancio},
  {G{\'o}mez-{\'A}lvarez}, {Hoar}, {Lorenzo Alvarez}, {Racca},
  {Saavedra-Criado}, {Schwartz}, {Vavrek}, {Schirmer}, {Aussel}, {Azzollini},
  {Cardone}, {Cropper}, {Ealet}, {Garilli}, {Gillard}, {Granett}, {Guzzo},
  {Hoekstra}, {Jahnke}, {Kitching}, {Maciaszek}, {Meneghetti}, {Miller},
  {Nakajima}, {Niemi}, {Pasian}, {Percival}, {Pottinger}, {Sauvage},
  {Scodeggio}, {Wachter}, {Zacchei}, {Aghanim}, {Amara}, {Auphan}, {Auricchio},
  {Awan}, {Balestra}, {Bender}, {Bodendorf}, {Bonino}, {Branchini},
  {Brau-Nogue}, {Brescia}, {Candini}, {Capobianco}, {Carbone}, {Carlberg},
  {Carretero}, {Casas}, {Castander}, {Castellano}, {Cavuoti}, {Cimatti},
  {Cledassou}, {Congedo}, {Conselice}, {Conversi}, {Copin}, {Corcione},
  {Costille}, {Courbin}, {Degaudenzi}, {Douspis}, {Dubath}, {Duncan}, {Dusini},
  {Farrens}, {Ferriol}, {Fosalba}, {Fourmanoit}, {Frailis}, {Franceschi},
  {Franzetti}, {Fumana}, {Gillis}, {Giocoli}, {Grazian}, {Grupp}, {Haugan},
  {Holmes}, {Hormuth}, {Hudelot}, {Kermiche}, {Kiessling}, {Kilbinger},
  {Kohley}, {Kubik}, {K{\"u}mmel}, {Kunz}, {Kurki-Suonio}, {Lahav}, {Ligori},
  {Lilje}, {Lloro}, {Mansutti}, {Marggraf}, {Markovic}, {Marulli}, {Massey},
  {Maurogordato}, {Melchior}, {Merlin}, {Meylan}, {Mohr}, {Moresco}, {Morin},
  {Moscardini}, {Munari}, {Nichol}, {Padilla}, {Paltani}, {Peacock},
  {Pedersen}, {Pettorino}, {Pires}, {Poncet}, {Popa}, {Pozzetti}, {Raison},
  {Rebolo}, {Rhodes}, {Rix}, {Roncarelli}, {Rossetti}, {Saglia}, {Schneider},
  {Schrabback}, {Secroun}, {Seidel}, {Serrano}, {Sirignano}, {Sirri},
  {Skottfelt}, {Stanco}, {Starck}, {Tallada-Cresp{\'\i}}, {Tavagnacco},
  {Taylor}, {Teplitz}, {Toledo-Moreo}, {Torradeflot}, {Trifoglio}, {Valentijn},
  {Valenziano}, {Verdoes Kleijn}, {Wang}, {Welikala}, {Weller}, {Wetzstein},
  {Zamorani}, {Zoubian}, {Andreon}, {Baldi}, {Bardelli}, {Boucaud}, {Camera},
  {Di Ferdinando}, {Fabbian}, {Farinelli}, {Galeotta}, {Graci{\'a}-Carpio},
  {Maino}, {Medinaceli}, {Mei}, {Neissner}, {Polenta}, {Renzi}, {Romelli},
  {Rosset}, {Sureau}, {Tenti}, {Vassallo}, {Zucca}, {Baccigalupi},
  {Balaguera-Antol{\'\i}nez}, {Battaglia}, {Biviano}, {Borgani}, {Bozzo},
  {Cabanac}, {Cappi}, {Casas}, {Castignani}, {Colodro-Conde}, {Coupon},
  {Courtois}, {Cuby}, {de la Torre}, {Desai}, {Dole}, {Fabricius}, {Farina},
  {Ferreira}, {Finelli}, {Flose-Reimberg}, {Fotopoulou}, {Ganga}, {Gozaliasl},
  {Hook}, {Keihanen}, {Kirkpatrick}, {Liebing}, {Lindholm}, {Mainetti},
  {Martinelli}, {Martinet}, {Maturi}, {McCracken}, {Metcalf}, {Morgante},
  {Nightingale}, {Nucita}, {Patrizii}, {Potter}, {Riccio}, {S{\'a}nchez},
  {Sapone}, {Schewtschenko}, {Schultheis}, {Scottez}, {Teyssier}, {Tutusaus},
  {Valiviita}, {Viel}, {Vriend}, \& {Whittaker}}]{scaramella2022}
{Euclid Collaboration: Scaramella}, R., {Amiaux}, J., {Mellier}, Y., {et~al.}
  2022, \aap, 662, A112

\bibitem[{{Euclid Collaboration: Schirmer} {et~al.}(2022){Euclid Collaboration:
  Schirmer}, {Jahnke}, {Seidel}, {Aussel}, {Bodendorf}, {Grupp}, {Hormuth},
  {Wachter}, {Appleton}, {Barbier}, {Brinchmann}, {Carrasco}, {Castander},
  {Coupon}, {De Paolis}, {Franco}, {Ganga}, {Hudelot}, {Jullo}, {Lan{\c{c}}on},
  {Nucita}, {Paltani}, {Smadja}, {Strafella}, {Venancio}, {Weiler}, {Amara},
  {Auphan}, {Auricchio}, {Balestra}, {Bender}, {Bonino}, {Branchini},
  {Brescia}, {Capobianco}, {Carbone}, {Carretero}, {Casas}, {Castellano},
  {Cavuoti}, {Cimatti}, {Cledassou}, {Congedo}, {Conselice}, {Conversi},
  {Copin}, {Corcione}, {Costille}, {Courbin}, {Da Silva}, {Degaudenzi},
  {Douspis}, {Dubath}, {Dupac}, {Dusini}, {Ealet}, {Farrens}, {Ferriol},
  {Fosalba}, {Frailis}, {Franceschi}, {Franzetti}, {Fumana}, {Garilli},
  {Gillard}, {Gillis}, {Giocoli}, {Grazian}, {Guzzo}, {Haugan}, {Hoekstra},
  {Holmes}, {Hornstrup}, {K{\"u}mmel}, {Kermiche}, {Kiessling}, {Kilbinger},
  {Kitching}, {Kohley}, {Kunz}, {Kurki-Suonio}, {Laureijs}, {Ligori}, {Lilje},
  {Lloro}, {Maciaszek}, {Maiorano}, {Mansutti}, {Marggraf}, {Markovic},
  {Marulli}, {Massey}, {Maurogordato}, {Mellier}, {Meneghetti}, {Merlin},
  {Meylan}, {Moresco}, {Moscardini}, {Munari}, {Nakajima}, {Nichol}, {Niemi},
  {Padilla}, {Pasian}, {Pedersen}, {Percival}, {Pettorino}, {Pires}, {Poncet},
  {Popa}, {Pozzetti}, {Prieto}, {Raison}, {Rhodes}, {Rix}, {Roncarelli},
  {Rossetti}, {Saglia}, {Sartoris}, {Scaramella}, {Schneider}, {Secroun},
  {Serrano}, {Sirignano}, {Sirri}, {Stanco}, {Tallada-Cresp{\'\i}}, {Taylor},
  {Teplitz}, {Tereno}, {Toledo-Moreo}, {Torradeflot}, {Trifoglio}, {Valentijn},
  {Valenziano}, {Wang}, {Weller}, {Zamorani}, {Zoubian}, {Andreon}, {Bardelli},
  {Boucaud}, {Camera}, {Farinelli}, {Graci{\'a}-Carpio}, {Maino}, {Medinaceli},
  {Mei}, {Morisset}, {Polenta}, {Renzi}, {Romelli}, {Tenti}, {Vassallo},
  {Zacchei}, {Zucca}, {Baccigalupi}, {Balaguera-Antol{\'\i}nez}, {Biviano},
  {Blanchard}, {Borgani}, {Bozzo}, {Burigana}, {Cabanac}, {Cappi}, {Carvalho},
  {Casas}, {Castignani}, {Colodro-Conde}, {Cooray}, {Courtois}, {Crocce},
  {Cuby}, {Davini}, {de la Torre}, {Di Ferdinando}, {Escartin}, {Farina},
  {Ferreira}, {Finelli}, {Fotopoulou}, {Galeotta}, {Garcia-Bellido},
  {Gaztanaga}, {George}, {Gozaliasl}, {Hook}, {Ili{\'c}}, {Kansal},
  {Kashlinsky}, {Keihanen}, {Kirkpatrick}, {Lindholm}, {Mainetti}, {Maoli},
  {Martinelli}, {Martinet}, {Maturi}, {Mauri}, {McCracken}, {Metcalf},
  {Monaco}, {Morgante}, {Nightingale}, {Patrizii}, {Peel}, {Popa}, {Porciani},
  {Potter}, {Reimberg}, {Riccio}, {S{\'a}nchez}, {Sapone}, {Scottez},
  {Sefusatti}, {Teyssier}, {Tutusaus}, {Valieri}, {Valiviita}, {Viel}, \&
  {Hildebrandt}}]{schirmer2022}
{Euclid Collaboration: Schirmer}, M., {Jahnke}, K., {Seidel}, G., {et~al.}
  2022, \aap, 662, A92

\bibitem[{Evans(2000)}]{evans2000}
Evans, S.~W. 2000, in Natural environment near the sun/earth-moon L2 libration
  point, Prepared for the Next Generation Space Telescope Program, Marshall
  Space Flight Center (MSFC), Alabama, ed. S.W. Evans, available at
  \url{https://docslib.org/doc/9787540/natural-environment-near-the-sun-earth-moon-l2-libration-point}

\bibitem[{Famá {et~al.}(2010)Famá, Loeffler, Raut, \& Baragiola}]{fama2010}
Famá, M., Loeffler, M., Raut, U., \& Baragiola, R. 2010, Icarus, 207, 314

\bibitem[{{Feistel} \& {Wagner}(2007)}]{feistel2007}
{Feistel}, R. \& {Wagner}, W. 2007, Geochimica et Cosmochimica Acta, 71, 36

\bibitem[{{Fraser} {et~al.}(2001){Fraser}, {Collings}, {McCoustra}, \&
  {Williams}}]{fraser2001}
{Fraser}, H.~J., {Collings}, M.~P., {McCoustra}, M. R.~S., \& {Williams}, D.~A.
  2001, \mnras, 327, 1165

\bibitem[{{Gaia Collaboration: Prusti} {et~al.}(2016){Gaia Collaboration:
  Prusti}, {de Bruijne}, {Brown}, {Vallenari}, {Babusiaux}, {Bailer-Jones},
  {Bastian}, {Biermann}, {Evans}, \& et~al.}]{gaia2016}
{Gaia Collaboration: Prusti}, T., {de Bruijne}, J.~H.~J., {Brown}, A.~G.~A.,
  {et~al.} 2016, \aap, 595, A1

\bibitem[{{Garoli} {et~al.}(2020){Garoli}, {Rodriguez De Marcos}, {Larruquert},
  {Corso}, {Proietti Zaccaria}, \& {Pelizzo}}]{garoli2010}
{Garoli}, D., {Rodriguez De Marcos}, L.~V., {Larruquert}, J.~I., {et~al.} 2020,
  arXiv:2010.00045

\bibitem[{Gasser {et~al.}(2021)Gasser, Thoeny, Fortes, \&
  Loerting}]{gasser2021}
Gasser, T.~M., Thoeny, A.~V., Fortes, A.~D., \& Loerting, T. 2021, Nature
  Communications, 12, 1128

\bibitem[{Ghesquière {et~al.}(2015)Ghesquière, Mineva, Talbi, Theulé, Noble,
  \& Chiavassa}]{ghesquiere2015}
Ghesquière, P., Mineva, T., Talbi, D., {et~al.} 2015, Phys. Chem. Chem. Phys.,
  17, 11455

\bibitem[{Gibson {et~al.}(2011)Gibson, Killelea, Yuan, Becker, \&
  Sibener}]{gibson2011}
Gibson, K.~D., Killelea, D.~R., Yuan, H., Becker, J.~S., \& Sibener, S.~J.
  2011, The Journal of Chemical Physics, 134, 034703

\bibitem[{Goldberg {et~al.}(2011)Goldberg, Ohring, Butler, Cao, Datla,
  Doelling, Gärtner, Hewison, Iacovazzi, Kim, Kurino, Lafeuille, Minnis,
  Renaut, Schmetz, Tobin, Wang, Weng, Wu, Yu, Zhang, \& Zhu}]{goldberg2011}
Goldberg, M., Ohring, G., Butler, J., {et~al.} 2011, Bulletin of the American
  Meteorological Society, 92, 467

\bibitem[{{Gonzaga} {et~al.}(2010){Gonzaga}, {Biretta}, {Baggett}, {McMaster},
  {Shaw}, {Voit}, {Leitherer}, \& {Baum}}]{gonzaga2010}
{Gonzaga}, S., {Biretta}, J., {Baggett}, S., {et~al.} 2010, HST WFPC2 Data
  Handbook, v5.0, ed., Space Telescope WFC Instrument Science Report

\bibitem[{{Green}(2001)}]{green2001}
{Green}, D.~B. 2001, Satellite Contamination and Materials Outgassing
  Knowledgebase - An Interactive Database Reference, NASA STI/Recon Technical
  Report N

\bibitem[{Greenwood(2002)}]{greenwood2002}
Greenwood, J. 2002, Vacuum, 67, 217

\bibitem[{Gr\"un {et~al.}(1985)Gr\"un, Zook, Fechtig, \& Giese}]{gruen1985}
Gr\"un, E., Zook, H., Fechtig, H., \& Giese, R. 1985, Icarus, 62, 244

\bibitem[{{Gudipati} \& {Castillo-Rogez}(2013)}]{gudipati2013}
{Gudipati}, M.~S. \& {Castillo-Rogez}, J. 2013, The Science of Solar System
  Ices, Vol. 356 (Springer-Verlag, New York)

\bibitem[{Guilbert-Lepoutre(2012)}]{guilbert2012}
Guilbert-Lepoutre, A. 2012, The Astronomical Journal, 144, 97

\bibitem[{Haemmerle \& Gerhard(2006)}]{haemmerle2006}
Haemmerle, V. \& Gerhard, J. 2006, in SpaceOps 2006 Conference, SpaceOps
  Conferences, 5834

\bibitem[{Haq \& Hodgson(2007)}]{haq2007}
Haq, S. \& Hodgson, A. 2007, The Journal of Physical Chemistry C, 111, 5946

\bibitem[{Hasegawa {et~al.}(2019)Hasegawa, Akutsu, Kimura, Saito, Suzuki,
  Tomaru, Ueda, \& Miyoki}]{hasegawa2019}
Hasegawa, K., Akutsu, T., Kimura, N., {et~al.} 2019, Phys. Rev. D, 99, 022003

\bibitem[{{He} {et~al.}(2019){He}, {Clements}, {Emtiaz}, {Toriello}, {Garrod},
  \& {Vidali}}]{he2019}
{He}, J., {Clements}, A.~R., {Emtiaz}, S., {et~al.} 2019, \apj, 878, 94

\bibitem[{He {et~al.}(2022)He, Diamant, Wang, Yu, Rocha, Rachid, \&
  Linnartz}]{he2022}
He, J., Diamant, S. J.~M., Wang, S., {et~al.} 2022, The Astrophysical Journal,
  925, 179

\bibitem[{He {et~al.}(2009)He, Tilocca, Dulub, Selloni, \& Diebold}]{he2009}
He, Y., Tilocca, A., Dulub, O., Selloni, A., \& Diebold, U. 2009, Nat. Mater.,
  8, 585

\bibitem[{Hessinger \& Pohl(1996)}]{hessinger1996a}
Hessinger, J. \& Pohl, R. 1996, Journal of Non-Crystalline Solids, 208, 151

\bibitem[{Hessinger {et~al.}(1996)Hessinger, White, \& Pohl}]{hessinger1996}
Hessinger, J., White, B., \& Pohl, R. 1996, Planetary and Space Science, 44,
  937, solar System Ices

\bibitem[{Hodgson \& Haq(2009)}]{hodgson2009}
Hodgson, A. \& Haq, S. 2009, Surface Science Reports, 64, 381

\bibitem[{Holmes {et~al.}(2016)Holmes, McKenney, Barbier, Cho, Cillis, Clemens,
  Dawson, Delo, Ealet, Feizi, Ferraro, Foltz, Goodsall, Hickey, Hwang,
  Israelsson, Jhabvala, Kahle, Kan, Kan, Lotkin, Maciaszek, McClure, Miko,
  Nguyen, Pravdo, Prieto, Powers, Seiffert, Strada, Tucker, Turck, Waczynski,
  Wang, Weber, \& Williams}]{holmes2016}
Holmes, W., McKenney, C., Barbier, R., {et~al.} 2016, in Space Telescopes and
  Instrumentation 2016: Optical, Infrared, and Millimeter Wave, ed. H.~A.
  MacEwen, G.~G. Fazio, M.~Lystrup, N.~Batalha, N.~Siegler, \& E.~C. Tong, Vol.
  9904, International Society for Optics and Photonics (SPIE), 99042R

\bibitem[{Holmes-Siedle {et~al.}(1991)Holmes-Siedle, Holland, Johlander, \&
  Adams}]{holmes1991}
Holmes-Siedle, A., Holland, A., Johlander, B., \& Adams, L. 1991, in RADECS 91
  First European Conference on Radiation and its Effects on Devices and Systems
  (IEEE), 338--342

\bibitem[{{Holtzman} {et~al.}(1995){Holtzman}, {Burrows}, {Casertano},
  {Hester}, {Trauger}, {Watson}, \& {Worthey}}]{holtzmann1995}
{Holtzman}, J.~A., {Burrows}, C.~J., {Casertano}, S., {et~al.} 1995, \pasp,
  107, 1065

\bibitem[{Hondoh(2015)}]{hondoh2015}
Hondoh, T. 2015, Philosophical Magazine, 95, 3590

\bibitem[{Hui {et~al.}(2022)Hui, Buditama, Liu, Alaan, \& Dunscombe}]{hui2022}
Hui, A.~O., Buditama, A.~N., Liu, D.-L., Alaan, D.~R., \& Dunscombe, S.~L.
  2022, in Space Systems Contamination: Prediction, Control, and Performance
  2022, ed. C.~E. Soares, E.~M. Wooldridge, \& B.~A. Matheson, Vol. 12224,
  International Society for Optics and Photonics (SPIE), 122240S

\bibitem[{Hunter(2007)}]{hunter20007}
Hunter, J.~D. 2007, Computing in Science \& Engineering, 9, 90

\bibitem[{{Jenniskens} \& {Blake}(1994)}]{jenniskens1994}
{Jenniskens}, P. \& {Blake}, D.~F. 1994, Science, 265, 753

\bibitem[{Kann \& Skinner(2016)}]{kann2016}
Kann, Z.~R. \& Skinner, J.~L. 2016, The Journal of Chemical Physics, 144,
  154701

\bibitem[{Kazmerski(2012)}]{kazmerski2012}
Kazmerski, L. 2012, Polycrystalline and amorphous thin films and devices
  (Academic Press)

\bibitem[{{Kimoto}(2017)}]{kimoto2017}
{Kimoto}, Y. 2017, in Society of Photo-Optical Instrumentation Engineers (SPIE)
  Conference Series, Vol. 10563, 105630N

\bibitem[{{Kinser} {et~al.}(1991){Kinser}, {Weller}, {Mendenhall},
  {Wiedlocher}, {Nichols}, {Tucker}, \& {Whitaker}}]{kinser1991}
{Kinser}, D.~L., {Weller}, R.~A., {Mendenhall}, M.~H., {et~al.} 1991, in NASA
  Conference Publication, Vol. 3134, LDEF, 69 months in space : first
  post-retrieval symposium, ed. A.~S. {Levine}, 1377

\bibitem[{Kirsch {et~al.}(2005)Kirsch, Abbey, Altieri, Baskill, Dennerl, van
  Dooren, Fauste, Freyberg, Gabriel, Haberl, Hartmann, Hartner, Meidinger,
  Metcalfe, Olabarri, Pollock, Read, Rives, Sembay, Smith, Stuhlinger, \&
  Talavera}]{kirsch2005}
Kirsch, M.~G., Abbey, A., Altieri, B., {et~al.} 2005, in UV, X-Ray, and
  Gamma-Ray Space Instrumentation for Astronomy XIV, ed. O.~H.~W. Siegmund,
  Vol. 5898, International Society for Optics and Photonics (SPIE), 58980S

\bibitem[{K\"onig(1943)}]{koenig1943}
K\"onig, H. 1943, Zeitschrift für Kristallographie - Crystalline Materials,
  105, 279

\bibitem[{{Korsch}(1977)}]{korsch1977}
{Korsch}, D. 1977, \ao, 16, 2074

\bibitem[{{Kossacki} {et~al.}(1999){Kossacki}, {Markiewicz}, {Skorov}, \&
  {K{\"o}mle}}]{kossacki1999}
{Kossacki}, K.~J., {Markiewicz}, W.~J., {Skorov}, Y., \& {K{\"o}mle}, N.~I.
  1999, \planss, 47, 1521

\bibitem[{{Kouchi}(1987)}]{kouchi1987}
{Kouchi}, A. 1987, \nat, 330, 550

\bibitem[{{Kouchi} \& {Kuroda}(1990)}]{kouchi1990}
{Kouchi}, A. \& {Kuroda}, T. 1990, \nat, 344, 134

\bibitem[{{Kouchi} {et~al.}(1994){Kouchi}, {Yamamoto}, {Kozasa}, {Kuroda}, \&
  {Greenberg}}]{kouchi1994}
{Kouchi}, A., {Yamamoto}, T., {Kozasa}, T., {Kuroda}, T., \& {Greenberg}, J.~M.
  1994, \aap, 290, 1009

\bibitem[{Krijger {et~al.}(2014)Krijger, Snel, van Harten, Rietjens, \&
  Aben}]{krijger2014}
Krijger, J.~M., Snel, R., van Harten, G., Rietjens, J. H.~H., \& Aben, I. 2014,
  Atmospheric Measurement Techniques, 7, 3387

\bibitem[{Kuhs {et~al.}(2012)Kuhs, Sippel, Falenty, \& Hansen}]{kuhs2012}
Kuhs, W.~F., Sippel, C., Falenty, A., \& Hansen, T.~C. 2012, Proceedings of the
  National Academy of Sciences, 109, 21259

\bibitem[{{Kuin} {et~al.}(2015){Kuin}, {Landsman}, {Breeveld}, {Page},
  {Lamoureux}, {James}, {Mehdipour}, {Still}, {Yershov}, {Brown}, {Carter},
  {Mason}, {Kennedy}, {Marshall}, {Roming}, {Siegel}, {Oates}, {Smith}, \& {De
  Pasquale}}]{kuin2015}
{Kuin}, N.~P.~M., {Landsman}, W., {Breeveld}, A.~A., {et~al.} 2015, \mnras,
  449, 2514

\bibitem[{{La Spisa} {et~al.}(2001){La Spisa}, {Waldheim}, {Lintemoot},
  {Thomas}, {Naff}, \& {Robinson}}]{laspisa2001}
{La Spisa}, S., {Waldheim}, M., {Lintemoot}, J., {et~al.} 2001, \jgr, 106,
  33351

\bibitem[{Labello(2011)}]{labello2011}
Labello, J.~M. 2011, Water Ice Films in Cryogenic Vacuum Chambers, Ph.D.
  Dissertation, The University of Tennessee

\bibitem[{{Lallo}(2012)}]{lallo2012}
{Lallo}, M.~D. 2012, Optical Engineering, 51, 011011

\bibitem[{{Laureijs} {et~al.}(2011){Laureijs}, {Amiaux}, {Arduini},
  {Augu{\`e}res}, {Brinchmann}, {Cole}, {Cropper}, {Dabin}, {Duvet}, {Ealet},
  \& et~al.}]{laureijs2011}
{Laureijs}, R., {Amiaux}, J., {Arduini}, S., {et~al.} 2011, arXiv:1110.3193

\bibitem[{Lee(2017)}]{lee2017}
Lee, K.~H. 2017, PLOS ONE, 12, 1

\bibitem[{Leger \& Bricker(1972)}]{leger1972}
Leger, L.~J. \& Bricker, R.~W. 1972, Apollo Experience Report, Apollo Window
  Contamination, Tech. Rep. TN-D-6721, NASA, available at
  \url{https://ntrs.nasa.gov/api/citations/19720012257/downloads/19720012257.pdf}

\bibitem[{Liebing {et~al.}(2018)Liebing, Krijger, Snel, Bramstedt, Noël,
  Bovensmann, \& Burrows}]{liebing2018}
Liebing, P., Krijger, J., Snel, R., {et~al.} 2018, Atmospheric Measurement
  Techniques, 11, 265

\bibitem[{Lightsey {et~al.}(2012)Lightsey, Atkinson, Clampin, \&
  Feinberg}]{lightsey2012}
Lightsey, P.~A., Atkinson, C.~B., Clampin, M.~C., \& Feinberg, L.~D. 2012,
  Optical Engineering, 51, 1

\bibitem[{{Limmer} \& {Chandler}(2014)}]{limmer2014}
{Limmer}, D.~T. \& {Chandler}, D. 2014, Proceedings of the National Academy of
  Science, 111, 9413

\bibitem[{Lin {et~al.}(2018)Lin, Corem, Godsi, Alexandrowicz, Darling, \&
  Hodgson}]{lin2018}
Lin, C., Corem, G., Godsi, O., {et~al.} 2018, Journal of the American Chemical
  Society, 140, 15804

\bibitem[{{Luey} {et~al.}(2018){Luey}, {Olson}, \& {Coleman}}]{luey2018}
{Luey}, K.~T., {Olson}, K.~R., \& {Coleman}, D.~J. 2018, Journal of
  Astronomical Telescopes, Instruments, and Systems, 4, 036001

\bibitem[{{Maciaszek} {et~al.}(2016){Maciaszek}, {Ealet}, {Jahnke}, {Prieto},
  {Barbier}, {Mellier}, {Beaumont}, {Bon}, {Bonnefoi}, {Carle}, {Caillat},
  {Costille}, {Dormoy}, {Ducret}, {Fabron}, {Febvre}, {Foulon}, {Garcia},
  {Gimenez}, {Grassi}, {Laurent}, {Le Mignant}, {Martin}, {Rossin}, {Pamplona},
  {Sanchez}, {Vives}, {Cl{\'e}mens}, {Gillard}, {Niclas}, {Secroun}, {Serra},
  {Kubik}, {Ferriol}, {Amiaux}, {Barri{\`e}re}, {Berthe}, {Rosset},
  {Macias-Perez}, {Auricchio}, {De Rosa}, {Franceschi}, {Guizzo}, {Morgante},
  {Sortino}, {Trifoglio}, {Valenziano}, {Patrizii}, {Chiarusi}, {Fornari},
  {Giacomini}, {Margiotta}, {Mauri}, {Pasqualini}, {Sirri}, {Spurio}, {Tenti},
  {Travaglini}, {Dusini}, {Dal Corso}, {Laudisio}, {Sirignano}, {Stanco},
  {Ventura}, {Borsato}, {Bonoli}, {Bortoletto}, {Balestra}, {D'Alessandro},
  {Medinaceli}, {Farinelli}, {Corcione}, {Ligori}, {Grupp}, {Wimmer},
  {Hormuth}, {Seidel}, {Wachter}, {Padilla}, {Lamensans}, {Casas}, {Lloro},
  {Toledo-Moreo}, {Gomez}, {Colodro-Conde}, {Liz{\'a}n}, {Diaz}, {Lilje},
  {Toulouse-Aastrup}, {Andersen}, {S{\o}rensen}, {Jakobsen}, {Hornstrup},
  {Jessen}, {Thizy}, {Holmes}, {Israelsson}, {Seiffert}, {Waczynski},
  {Laureijs}, {Racca}, {Salvignol}, {Boenke}, \& {Strada}}]{maciaszek2016}
{Maciaszek}, T., {Ealet}, A., {Jahnke}, K., {et~al.} 2016, in Society of
  Photo-Optical Instrumentation Engineers (SPIE) Conference Series, Vol. 9904,
  Space Telescopes and Instrumentation 2016: Optical, Infrared, and Millimeter
  Wave, ed. H.~A. {MacEwen}, G.~G. {Fazio}, M.~{Lystrup}, N.~{Batalha},
  N.~{Siegler}, \& E.~C. {Tong}, 99040T

\bibitem[{{MacKenty} {et~al.}(1993){MacKenty}, {Schneider}, {Baggett},
  {Mitchell}, {Ritchie}, \& {Sparks}}]{mackenty1993}
{MacKenty}, J.~W., {Schneider}, G., {Baggett}, S.~M., {et~al.} 1993, in Society
  of Photo-Optical Instrumentation Engineers (SPIE) Conference Series, Vol.
  1945, Space Astronomical Telescopes and Instruments II, ed. P.~Y. {Bely} \&
  J.~B. {Breckinridge}, 340--345

\bibitem[{Maier(2018)}]{maier2018}
Maier, S. 2018, in Encyclopedia of Interfacial Chemistry, ed. K.~Wandelt
  (Oxford: Elsevier), 304--313

\bibitem[{Makled \& Belal(2009)}]{makled2009}
Makled, A. \& Belal, H. 2009, in 13th International Conference on Aerospace
  Sciences \& Aviation Technology, Cairo, ASAT--13, 26--28

\bibitem[{Malkin {et~al.}(2015)Malkin, Murray, Salzmann, Molinero, Pickering,
  \& Whale}]{malkin2015}
Malkin, T.~L., Murray, B.~J., Salzmann, C.~G., {et~al.} 2015, Phys. Chem. Chem.
  Phys., 17, 60

\bibitem[{{Mason} {et~al.}(2001){Mason}, {Breeveld}, {Much}, {Carter},
  {Cordova}, {Cropper}, {Fordham}, {Huckle}, {Ho}, {Kawakami}, {Kennea},
  {Kennedy}, {Mittaz}, {Pandel}, {Priedhorsky}, {Sasseen}, {Shirey}, {Smith},
  \& {Vreux}}]{mason2001}
{Mason}, K.~O., {Breeveld}, A., {Much}, R., {et~al.} 2001, \aap, 365, L36

\bibitem[{{Massey} {et~al.}(2014){Massey}, {Schrabback}, {Cordes}, {Marggraf},
  {Israel}, {Miller}, {Hall}, {Cropper}, {Prod'homme}, \& {Niemi}}]{massey2014}
{Massey}, R., {Schrabback}, T., {Cordes}, O., {et~al.} 2014, \mnras, 439, 887

\bibitem[{{Mastrapa} \& {Brown}(2006)}]{mastrapa2006}
{Mastrapa}, R. M.~E. \& {Brown}, R.~H. 2006, \icarus, 183, 207

\bibitem[{{Mastrapa} {et~al.}(2013){Mastrapa}, {Grundy}, \&
  {Gudipati}}]{mastrapa2013}
{Mastrapa}, R. M.~E., {Grundy}, W.~M., \& {Gudipati}, M.~S. 2013, Amorphous and
  Crystalline H$_{2}$O-Ice, Vol. 356 (Springer-Verlag, New York), 371

\bibitem[{Matthews {et~al.}(2005)Matthews, Priestley, Spence, Cooper, \&
  Walikainen}]{matthews2005}
Matthews, G., Priestley, K., Spence, P., Cooper, D., \& Walikainen, D. 2005, in
  Earth Observing Systems X, ed. J.~J. Butler, Vol. 5882, International Society
  for Optics and Photonics (SPIE), 588212

\bibitem[{May {et~al.}(2013)May, Smith, \& Kay}]{may2013}
May, R., Smith, R., \& Kay, B. 2013, The Journal of Chemical Physics, 138,
  104501

\bibitem[{{McElwain} {et~al.}(2023){McElwain}, {Feinberg}, {Perrin}, {Clampin},
  {Mountain}, {Lallo}, {Lajoie}, {Kimble}, {Bowers}, {Stark}, {Acton},
  {Aiello}, {Atkinson}, {Barinek}, {Barto}, {Basinger}, {Beck}, {Bergkoetter},
  {Bluth}, {Boucarut}, {Brady}, {Brooks}, {Brown}, {Byard}, {Carey},
  {Carrasquilla}, {Celeste}, {Chae}, {Chaney}, {Chayer}, {Chonis}, {Cohen},
  {Cole}, {Comeau}, {Coon}, {Coppock}, {Coyle}, {Davis}, {Dean}, {Dziak},
  {Eisenhower}, {Flagey}, {Franck}, {Gallagher}, {Gilman}, {Glassman},
  {Golnik}, {Green}, {Grieco}, {Haase}, {Hadjimichael}, {Hagopian}, {Hahn},
  {Hartig}, {Havey}, {Hayden}, {Hellekson}, {Hicks}, {Holfeltz}, {Howard},
  {Huguet}, {Jahne}, {Johnson}, {Johnston}, {Jurling}, {Kegley}, {Kennard},
  {Keski-Kuha}, {Knight}, {Kulp}, {Levi}, {Levine}, {Lightsey}, {Luetgens},
  {Mather}, {Matthews}, {McKay}, {Mehalick}, {Mel{\'e}ndez}, {Messer},
  {Mosier}, {Murphy}, {Nelan}, {Niedner}, {No{\"e}l}, {Ohara}, {Ohl}, {Olczak},
  {Osborne}, {Park}, {Patton}, {Perrygo}, {Pueyo}, {Quesnel}, {Ranck},
  {Redding}, {Regan}, {Reynolds}, {Rifelli}, {Rigby}, {Sabatke}, {Saif},
  {Scorse}, {Seo}, {Shi}, {Sigrist}, {Smith}, {Smith}, {Smith}, {Sohn},
  {Spina}, {Stahl}, {Telfer}, {Terlecki}, {Texter}, {Van Buren}, {Van Campen},
  {Vila}, {Voyton}, {Waldman}, {Walker}, {Weiser}, {Wells}, {West}, {Whitman},
  {Wick}, {Wolf}, {Young}, \& {Zielinski}}]{mcelwain2023}
{McElwain}, M.~W., {Feinberg}, L.~D., {Perrin}, M.~D., {et~al.} 2023,
  arXiv:2301.01779

\bibitem[{Mitlin \& Leung(2002)}]{mitlin2002}
Mitlin, S. \& Leung, K.~T. 2002, The Journal of Physical Chemistry B, 106, 6234

\bibitem[{{Mora} {et~al.}(2016){Mora}, {Biermann}, {Bombrun}, {Boyadjian},
  {Chassat}, {Corberand}, {Davidson}, {Doyle}, {Escolar}, {Gielesen},
  {Guilpain}, {Hernandez}, {Kirschner}, {Klioner}, {Koeck}, {Laine},
  {Lindegren}, {Serpell}, {Tatry}, \& {Thoral}}]{mora2016}
{Mora}, A., {Biermann}, M., {Bombrun}, A., {et~al.} 2016, in Society of
  Photo-Optical Instrumentation Engineers (SPIE) Conference Series, Vol. 9904,
  Space Telescopes and Instrumentation 2016: Optical, Infrared, and Millimeter
  Wave, ed. H.~A. {MacEwen}, G.~G. {Fazio}, M.~{Lystrup}, N.~{Batalha},
  N.~{Siegler}, \& E.~C. {Tong}, 99042D

\bibitem[{Murphy \& Koop(2005)}]{murphy2005}
Murphy, D.~M. \& Koop, T. 2005, Quarterly Journal of the Royal Meteorological
  Society, 131, 1539

\bibitem[{Nachbar {et~al.}(2018{\natexlab{a}})Nachbar, Duft, \&
  Leisner}]{nachbar2018a}
Nachbar, M., Duft, D., \& Leisner, T. 2018{\natexlab{a}}, Atmospheric Chemistry
  and Physics, 18, 3419

\bibitem[{Nachbar {et~al.}(2018{\natexlab{b}})Nachbar, Duft, \&
  Leisner}]{nachbar2018b}
Nachbar, M., Duft, D., \& Leisner, T. 2018{\natexlab{b}}, The Journal of
  Physical Chemistry B, 122, 10044

\bibitem[{Notesco {et~al.}(2003)Notesco, Bar-Nun, \& Owen}]{notesco2003}
Notesco, G., Bar-Nun, A., \& Owen, T. 2003, Icarus, 162, 183

\bibitem[{Padowitz \& Sibener(1989)}]{padowitz1989}
Padowitz, D.~F. \& Sibener, S.~J. 1989, Surface Science, 217, 233

\bibitem[{Pakdehi {et~al.}(2019)Pakdehi, Shirvani, \& Zolfaghari}]{Pakdehi2019}
Pakdehi, S., Shirvani, F., \& Zolfaghari, R. 2019, Archives of Thermodynamics,
  vol. 40, 151

\bibitem[{Palusinski {et~al.}(2009)Palusinski, Walters, Matson, Fuqua, Jenkins,
  Barrie, Meshishnek, Messenger, Geis, Jackson, \& Lorentzen}]{palusinski2009}
Palusinski, I.~A., Walters, R.~J., Matson, L.~E., {et~al.} 2009, AIP Conference
  Proceedings, 1087, 249

\bibitem[{{Patel} {et~al.}(2019){Patel}, {Dean}, {Salinas}, {Shiraishi}, \&
  {Newlin}}]{patel2019}
{Patel}, N., {Dean}, Z., {Salinas}, Y., {Shiraishi}, L., \& {Newlin}, L. 2019,
  Life Sciences and Space Research, 23, 22

\bibitem[{Persad \& Ward(2016)}]{persad2016}
Persad, A.~H. \& Ward, C.~A. 2016, Chemical Reviews, 116, 7727

\bibitem[{Plis {et~al.}(2019)Plis, Engelhart, Cooper, Johnston, Ferguson, \&
  Hoffmann}]{plis2019}
Plis, E.~A., Engelhart, D.~P., Cooper, R., {et~al.} 2019, Applied Sciences, 9

\bibitem[{{Plucinsky} {et~al.}(2012){Plucinsky}, {Beardmore}, {DePasquale},
  {Dewey}, {Foster}, {Haberl}, {Miller}, {Pollock}, {Posson-Brown}, {Sembay},
  \& {Smith}}]{plucinsky2012}
{Plucinsky}, P.~P., {Beardmore}, A.~P., {DePasquale}, J.~M., {et~al.} 2012, in
  Society of Photo-Optical Instrumentation Engineers (SPIE) Conference Series,
  Vol. 8443, Space Telescopes and Instrumentation 2012: Ultraviolet to Gamma
  Ray, ed. T.~{Takahashi}, S.~S. {Murray}, \& J.-W.~A. {den Herder}, 844312

\bibitem[{Plucinsky {et~al.}(2020)Plucinsky, Bogdan, \&
  Marshall}]{plucinsky2020}
Plucinsky, P.~P., Bogdan, A., \& Marshall, H.~L. 2020, in Space Telescopes and
  Instrumentation 2020: Ultraviolet to Gamma Ray, ed. J.-W.~A. den Herder,
  S.~Nikzad, \& K.~Nakazawa, Vol. 11444, International Society for Optics and
  Photonics (SPIE), 1333--1345

\bibitem[{{Plucinsky} {et~al.}(2018){Plucinsky}, {Bogdan}, {Marshall}, \&
  {Tice}}]{plucinsky2018}
{Plucinsky}, P.~P., {Bogdan}, A., {Marshall}, H.~L., \& {Tice}, N.~W. 2018, in
  Society of Photo-Optical Instrumentation Engineers (SPIE) Conference Series,
  Vol. 10699, Space Telescopes and Instrumentation 2018: Ultraviolet to Gamma
  Ray, ed. J.-W.~A. {den Herder}, S.~{Nikzad}, \& K.~{Nakazawa}, 106996B

\bibitem[{Poidomani {et~al.}(2020)Poidomani, Gottero, Ferrero, Filleul, \&
  Stramaccioni}]{poidomani2020}
Poidomani, G., Gottero, M., Ferrero, A., Filleul, D., \& Stramaccioni, D. 2020,
  in The Thermal Testing of Euclid STM, International Conference on
  Environmental Systems, ICES-2020-491

\bibitem[{{Poole} {et~al.}(2008){Poole}, {Breeveld}, {Page}, {Landsman},
  {Holland}, {Roming}, {Kuin}, {Brown}, {Gronwall}, {Hunsberger}, {Koch},
  {Mason}, {Schady}, {vanden Berk}, {Blustin}, {Boyd}, {Broos}, {Carter},
  {Chester}, {Cucchiara}, {Hancock}, {Huckle}, {Immler}, {Ivanushkina},
  {Kennedy}, {Marshall}, {Morgan}, {Pandey}, {de Pasquale}, {Smith}, \&
  {Still}}]{poole2008}
{Poole}, T.~S., {Breeveld}, A.~A., {Page}, M.~J., {et~al.} 2008, \mnras, 383,
  627

\bibitem[{Postberg {et~al.}(2009)Postberg, Kempf, Rost, Stephan, Srama,
  Trieloff, Mocker, \& Goerlich}]{postberg2009}
Postberg, F., Kempf, S., Rost, D., {et~al.} 2009, Planetary and Space Science,
  57, 1359

\bibitem[{Pratte {et~al.}(2006)Pratte, van~den Bergh, \& Rossi}]{pratte2006}
Pratte, P., van~den Bergh, H., \& Rossi, M. 2006, The journal of physical
  chemistry. A, 110, 3042

\bibitem[{{Prialnik} \& {Jewitt}(2022)}]{prialnik2022}
{Prialnik}, D. \& {Jewitt}, D. 2022, arXiv:2209.05907

\bibitem[{Price \& Evans(1968)}]{price1968}
Price, T.~W. \& Evans, D.~D. 1968, in Jet Propulsion Laboratory Technical
  Report 32-1227, available at
  \url{https://ntrs.nasa.gov/api/citations/19680006875/downloads/19680006875.pdf}

\bibitem[{Quast {et~al.}(2019)Quast, Giering, Govaerts, Rüthrich, \&
  Roebeling}]{quast2019}
Quast, R., Giering, R., Govaerts, Y., Rüthrich, F., \& Roebeling, R. 2019,
  Remote Sensing, 11, 480

\bibitem[{{Racca} {et~al.}(2016){Racca}, {Laureijs}, {Stagnaro}, {Salvignol},
  {Lorenzo Alvarez}, {Saavedra Criado}, {Gaspar Venancio}, {Short}, {Strada},
  {B{\"o}nke}, {Colombo}, {Calvi}, {Maiorano}, {Piersanti}, {Prezelus},
  {Rosato}, {Pinel}, {Rozemeijer}, {Lesna}, {Musi}, {Sias}, {Anselmi},
  {Cazaubiel}, {Vaillon}, {Mellier}, {Amiaux}, {Berth{\'e}}, {Sauvage},
  {Azzollini}, {Cropper}, {Pottinger}, {Jahnke}, {Ealet}, {Maciaszek},
  {Pasian}, {Zacchei}, {Scaramella}, {Hoar}, {Kohley}, {Vavrek}, {Rudolph}, \&
  {Schmidt}}]{racca2016}
{Racca}, G.~D., {Laureijs}, R., {Stagnaro}, L., {et~al.} 2016, in Society of
  Photo-Optical Instrumentation Engineers (SPIE) Conference Series, Vol. 9904,
  Space Telescopes and Instrumentation 2016: Optical, Infrared, and Millimeter
  Wave, ed. H.~A. {MacEwen}, G.~G. {Fazio}, M.~{Lystrup}, N.~{Batalha},
  N.~{Siegler}, \& E.~C. {Tong}, 99040O

\bibitem[{{Raut} {et~al.}(2008){Raut}, {Fam{\'a}}, {Loeffler}, \&
  {Baragiola}}]{raut2008}
{Raut}, U., {Fam{\'a}}, M., {Loeffler}, M.~J., \& {Baragiola}, R.~A. 2008,
  \apj, 687, 1070

\bibitem[{{Riello} {et~al.}(2021){Riello}, {De Angeli}, {Evans}, {Montegriffo},
  {Carrasco}, {Busso}, {Palaversa}, {Burgess}, {Diener}, {Davidson}, {Rowell},
  {Fabricius}, {Jordi}, {Bellazzini}, {Pancino}, {Harrison}, {Cacciari}, {van
  Leeuwen}, {Hambly}, {Hodgkin}, {Osborne}, {Altavilla}, {Barstow}, {Brown},
  {Castellani}, {Cowell}, {De Luise}, {Gilmore}, {Giuffrida}, {Hidalgo},
  {Holland}, {Marinoni}, {Pagani}, {Piersimoni}, {Pulone}, {Ragaini}, {Rainer},
  {Richards}, {Sanna}, {Walton}, {Weiler}, \& {Yoldas}}]{riello2021}
{Riello}, M., {De Angeli}, F., {Evans}, D.~W., {et~al.} 2021, \aap, 649, A3

\bibitem[{{Rodmann} {et~al.}(2019){Rodmann}, {Miller}, {Bunte}, \&
  {Millinger}}]{rodmann2019}
{Rodmann}, J., {Miller}, A., {Bunte}, K.~D., \& {Millinger}, M. 2019, in First
  International Orbital Debris Conference, Vol. 2109, 6070

\bibitem[{Rohatgi(2022)}]{rohatgi2022}
Rohatgi, A. 2022, Webplotdigitizer: Version 4.6

\bibitem[{{Roming} {et~al.}(2005){Roming}, {Kennedy}, {Mason}, {Nousek}, {Ahr},
  {Bingham}, {Broos}, {Carter}, {Hancock}, {Huckle}, {Hunsberger}, {Kawakami},
  {Killough}, {Koch}, {McLelland}, {Smith}, {Smith}, {Soto}, {Boyd},
  {Breeveld}, {Holland}, {Ivanushkina}, {Pryzby}, {Still}, \&
  {Stock}}]{roming2005}
{Roming}, P. W.~A., {Kennedy}, T.~E., {Mason}, K.~O., {et~al.} 2005, \ssr, 120,
  95

\bibitem[{Rosu-Finsen {et~al.}(2022)Rosu-Finsen, Chikani, \&
  Salzmann}]{rosufinsen2022}
Rosu-Finsen, A., Chikani, B., \& Salzmann, C.~G. 2022, Monthly Notices of the
  Royal Astronomical Society, 517, 1919

\bibitem[{Rosu-Finsen {et~al.}(2023)Rosu-Finsen, Davies, Amon, Wu, Sella,
  Michaelides, \& Salzmann}]{rosufinsen2023}
Rosu-Finsen, A., Davies, M.~B., Amon, A., {et~al.} 2023, Science, 379, 474

\bibitem[{Rothard {et~al.}(2017)Rothard, Domaracka, Boduch, Palumbo,
  Strazzulla, da~Silveira, \& Dartois}]{rothard2017}
Rothard, H., Domaracka, A., Boduch, P., {et~al.} 2017, Journal of Physics B:
  Atomic, Molecular and Optical Physics, 50, 062001

\bibitem[{Roussel {et~al.}(2016)Roussel, Vanhove, Tondu, Faye, \&
  Guigue}]{roussel2016}
Roussel, J.-F., Vanhove, E., Tondu, T., Faye, D., \& Guigue, P. 2016, Journal
  of Spacecraft and Rockets, 53, 1159

\bibitem[{{Sack} \& {Baragiola}(1993)}]{sack1993}
{Sack}, N.~J. \& {Baragiola}, R.~A. 1993, \prb, 48, 9973

\bibitem[{Salzmann(2019)}]{salzmann2019}
Salzmann, C.~G. 2019, The Journal of Chemical Physics, 150, 060901

\bibitem[{Samwel(2014)}]{samwel2014}
Samwel, S. 2014, Space Research Journal, 7, 1

\bibitem[{Sandford {et~al.}(2020)Sandford, Bierhaus, Antreasian, Leonard,
  Materese, May, Songer, Dworkin, Lauretta, \& Rizk}]{sandford2020}
Sandford, S.~A., Bierhaus, E.~B., Antreasian, P., {et~al.} 2020, Acta
  Astronautica, 166, 391

\bibitem[{Satorre {et~al.}(2013)Satorre, Leliwa-Kopystynski, Santonja, \&
  Luna}]{satorre2013}
Satorre, M., Leliwa-Kopystynski, J., Santonja, C., \& Luna, R. 2013, Icarus,
  225, 703

\bibitem[{{Schartel} {et~al.}(2022){Schartel}, {Gonz{\'a}lez-Riestra},
  {Kretschmar}, {Kirsch}, {Rodr{\'\i}guez-Pascual}, {Rosen}, {Santos-Lle{\'o}},
  {Smith}, {Stuhlinger}, \& {Verdugo-Rodrigo}}]{schartel2022}
{Schartel}, N., {Gonz{\'a}lez-Riestra}, R., {Kretschmar}, P., {et~al.} 2022,
  arXiv:2212.10995

\bibitem[{Scherillo {et~al.}(2014)Scherillo, Petretta, Galizia, La~Manna,
  Musto, \& Mensitieri}]{scherillo2014}
Scherillo, G., Petretta, M., Galizia, M., {et~al.} 2014, Frontiers in
  Chemistry, 2

\bibitem[{{Schl{\"a}ppi} {et~al.}(2010){Schl{\"a}ppi}, {Altwegg}, {Balsiger},
  {H{\"a}Ssig}, {J{\"a}Ckel}, {Wurz}, {Fiethe}, {Rubin}, {Fuselier},
  {Berthelier}, {de Keyser}, {R{\`e}Me}, \& {Mall}}]{schlaeppi2010}
{Schl{\"a}ppi}, B., {Altwegg}, K., {Balsiger}, H., {et~al.} 2010, Journal of
  Geophysical Research (Space Physics), 115, A12313

\bibitem[{Schriver-Mazzuoli {et~al.}(2000)Schriver-Mazzuoli, Schriver, \&
  Hallou}]{schrivermazzuoli2000}
Schriver-Mazzuoli, L., Schriver, A., \& Hallou, A. 2000, Journal of Molecular
  Structure, 554, 289

\bibitem[{Scialdone(1993)}]{scialdone1993}
Scialdone, J.~J. 1993, Journal of Spacecraft and Rockets, 30, 208

\bibitem[{Shakeel {et~al.}(2018)Shakeel, Wei, \& Pomeroy}]{shakeel2018}
Shakeel, H., Wei, H., \& Pomeroy, J. 2018, The Journal of chemical
  thermodynamics, 118, 127

\bibitem[{{Shanahan} {et~al.}(2017){Shanahan}, {Gosmeyer}, \&
  {Baggett}}]{shanahan2017}
{Shanahan}, C.~E., {Gosmeyer}, C.~M., \& {Baggett}, S. 2017, {2017 Update on
  the WFC3/UVIS Stability and Contamination Monitor}, Space Telescope WFC
  Instrument Science Report

\bibitem[{Sharma {et~al.}(2018)Sharma, Kroonblawd, Sun, \&
  Glascoe}]{sharma2018}
Sharma, H.~N., Kroonblawd, M.~P., Sun, Y., \& Glascoe, E.~A. 2018, Scientific
  Reports, 8, 16889

\bibitem[{Sheikh {et~al.}(2008)Sheikh, Connell, \& Dummer}]{sheik2008}
Sheikh, D.~A., Connell, S.~J., \& Dummer, R.~S. 2008, in Space Telescopes and
  Instrumentation 2008: Optical, Infrared, and Millimeter, ed. J.~M.~O. Jr.,
  M.~W.~M. de~Graauw, \& H.~A. MacEwen, Vol. 7010, International Society for
  Optics and Photonics (SPIE), 1277--1281

\bibitem[{{Simonetto} {et~al.}(2020){Simonetto}, {Marmonti}, \&
  {Potenza}}]{simonetto2020}
{Simonetto}, F., {Marmonti}, M., \& {Potenza}, M. A.~C. 2020, Journal of
  Astronomical Telescopes, Instruments, and Systems, 6, 038004

\bibitem[{Smith {et~al.}(2012)Smith, Mutlow, Delderfield, Watkins, \&
  Mason}]{smith2012b}
Smith, D., Mutlow, C., Delderfield, J., Watkins, B., \& Mason, G. 2012, Remote
  Sensing of Environment, 116, 4, advanced Along Track Scanning Radiometer
  (AATSR) Special Issue

\bibitem[{Smith(2012)}]{smith2012a}
Smith, D.~L. 2012, ENVISAT AATSR Instrument Performance - End of Mission
  Report, Science and Technology Facilities Council, available at
  \url{https://atsrsensors.org/pdf/ENVISAT\%20AATSR\%20Instrument\%20Performance\%20-\%20End\%20of\%20Mission\%20Report\%20-\%20Issue\%201.0.pdf\#page=32}

\bibitem[{Smith {et~al.}(2011)Smith, Matthiesen, Knox, \& Kay}]{smith2011}
Smith, R.~S., Matthiesen, J., Knox, J., \& Kay, B.~D. 2011, The Journal of
  Physical Chemistry A, 115, 5908, pMID: 21218834

\bibitem[{Spallino {et~al.}(2021)Spallino, Angelucci, Pasqualetti, Battes, Day,
  Grohmann, Majorana, Ricci, \& Cimino}]{spallino2021}
Spallino, L., Angelucci, M., Pasqualetti, A., {et~al.} 2021, Phys. Rev. D, 104,
  062001

\bibitem[{Steinlechner \& Martin(2019)}]{steinlechner2019}
Steinlechner, J. \& Martin, I.~W. 2019, Phys. Rev. Research, 1, 013008

\bibitem[{Stramaccioni {et~al.}(2000)Stramaccioni, Faust, \&
  Hinger}]{stramaccioni2000}
Stramaccioni, D., Faust, T., \& Hinger, J. 2000, SAE Transactions, 109, 499

\bibitem[{Suliga {et~al.}(2020)Suliga, Ergincan, \& Rampini}]{suliga2020}
Suliga, A., Ergincan, O., \& Rampini, R. 2020, in Systems Contamination:
  Prediction, Control, and Performance 2020, ed. C.~E. Soares, E.~M.
  Wooldridge, \& B.~A. Matheson, Vol. 11489, International Society for Optics
  and Photonics (SPIE), 114890E

\bibitem[{Sun {et~al.}(2020)Sun, Abshire, Lauenstein, Babu, Beck, Sullivan, \&
  Hubbs}]{sun2020}
Sun, X., Abshire, J., Lauenstein, J.-M., {et~al.} 2020, IEEE Transactions on
  Nuclear Science, 68, 1

\bibitem[{Talewar {et~al.}(2019)Talewar, Halukeerthi, Riedlaicher, Shephard,
  Clout, Rosu-Finsen, Williams, Langhoff, Johannsmann, \&
  Salzmann}]{talewar2019}
Talewar, S.~K., Halukeerthi, S.~O., Riedlaicher, R., {et~al.} 2019, The Journal
  of Chemical Physics, 151, 134505

\bibitem[{Tamijani {et~al.}(2020)Tamijani, Bjorklund, Augustine, Catalano, \&
  Mason}]{tamijani2020}
Tamijani, A.~A., Bjorklund, J.~L., Augustine, L.~J., Catalano, J.~G., \& Mason,
  S.~E. 2020, Langmuir, 36, 13166

\bibitem[{Tanioka {et~al.}(2020)Tanioka, Hasegawa, \& Aso}]{tanioka2020}
Tanioka, S., Hasegawa, K., \& Aso, Y. 2020, Phys. Rev. D, 102, 022009

\bibitem[{{Th{\"u}rmer} \& {Bartelt}(2008)}]{thuermer2008}
{Th{\"u}rmer}, K. \& {Bartelt}, N.~C. 2008, \prb, 77, 195425

\bibitem[{Th{\"u}rmer \& {Nie}(2013)}]{thuermer2013}
Th{\"u}rmer, K. \& {Nie}, S. 2013, Proceedings of the National Academy of
  Sciences, 110, 11757

\bibitem[{Th{\"u}rmer {et~al.}(2014)Th{\"u}rmer, Nie, Feibelman, \&
  Bartelt}]{thuermer2014}
Th{\"u}rmer, K., Nie, S., Feibelman, P.~J., \& Bartelt, N.~C. 2014, The Journal
  of Chemical Physics, 141, 18C520

\bibitem[{Todorov \& Bloch(2017)}]{todorov}
Todorov, P. \& Bloch, D. 2017, The Journal of Chemical Physics, 147, 194202

\bibitem[{Trost(2015)}]{trost2015}
Trost, M. 2015, Light scattering and roughness properties of optical components
  for 13.5 nm; PhD thesis, Friedrich-Schiller-Universit\"at Jena

\bibitem[{{Tveekrem} {et~al.}(1996){Tveekrem}, {Leviton}, {Fleetwood}, \&
  {Feinberg}}]{tveekrem1996}
{Tveekrem}, J.~L., {Leviton}, D.~B., {Fleetwood}, C.~M., \& {Feinberg}, L.~D.
  1996, in Society of Photo-Optical Instrumentation Engineers (SPIE) Conference
  Series, Vol. 2864, Optical System Contamination V, and Stray Light and System
  Optimization, ed. A.~P.~M. {Glassford}, R.~P. {Breault}, \& S.~M. {Pompea},
  246--257

\bibitem[{{Uy} {et~al.}(1998){Uy}, {Benson}, {Erlandson}, {Silver}, {Lesho},
  {Galica}, {Green}, {Boies}, {Wood}, \& {Hall}}]{uy1998}
{Uy}, O.~M., {Benson}, R.~C., {Erlandson}, R.~E., {et~al.} 1998, in Society of
  Photo-Optical Instrumentation Engineers (SPIE) Conference Series, Vol. 3427,
  Optical Systems Contamination and Degradation, ed. P.~T. {Chen}, W.~E.
  {McClintock}, \& G.~J. {Rottman}, 28--43

\bibitem[{{Uy} {et~al.}(2003){Uy}, {Green}, {Wood}, {Galica}, {Boies}, {Lesho},
  {Cain}, \& {Hall}}]{uy2003}
{Uy}, O.~M., {Green}, B.~D., {Wood}, B.~E., {et~al.} 2003, in ESA Special
  Publication, Vol. 540, Materials in a Space Environment, ed. K.~{Fletcher},
  197--202

\bibitem[{Vaghjiani(1993)}]{vaghjiani1993}
Vaghjiani, G.~L. 1993, The Journal of Chemical Physics, 98, 2123

\bibitem[{Venancio {et~al.}(2014)Venancio, Laureijs, Lorenzo, Salvignol, Short,
  Strada, Vavrek, Vaillon, Gennaro, Amiaux, \& Prieto}]{venancio2014}
Venancio, L. M.~G., Laureijs, R., Lorenzo, J., {et~al.} 2014, in Space
  Telescopes and Instrumentation 2014: Optical, Infrared, and Millimeter Wave,
  ed. J.~M.~O. Jr., M.~Clampin, G.~G. Fazio, \& H.~A. MacEwen, Vol. 9143,
  International Society for Optics and Photonics (SPIE), 91430I

\bibitem[{Wagner {et~al.}(2011)Wagner, Riethmann, Feistel, \&
  Harvey}]{wagner2011}
Wagner, W., Riethmann, T., Feistel, R., \& Harvey, A. 2011, Journal of Physical
  and Chemical Reference Data, 40, 3103

\bibitem[{Wang {et~al.}(2013)Wang, Wu, Weng, \& Goldberg}]{wang2013}
Wang, L., Wu, X., Weng, F., \& Goldberg, M. 2013, IEEE Transactions on
  Geoscience and Remote Sensing, 51, 1224

\bibitem[{Watanabe \& Kouchi(2008)}]{watanabe2008}
Watanabe, N. \& Kouchi, A. 2008, Progress in Surface Science, 83, 439

\bibitem[{Wilkes \& Zwiener(1999)}]{wilkes1999}
Wilkes, D.~R. \& Zwiener, J.~M. 1999, in Rough Surface Scattering and
  Contamination, ed. Z.-H. Gu, A.~A. Maradudin, P.~T.~C. Chen, Z.-H. Gu, \&
  A.~A. Maradudin, Vol. 3784, International Society for Optics and Photonics
  (SPIE), 72--83

\bibitem[{{Willson} {et~al.}(2018){Willson}, {Ricco}, {Gold}, {McKay},
  {Bonaccorsi}, \& {Adams}}]{willson2018}
{Willson}, D., {Ricco}, A., {Gold}, R.~E., {et~al.} 2018, in 42nd COSPAR
  Scientific Assembly, Vol.~42, B5.3--64--18

\bibitem[{{Wood} {et~al.}(2003){Wood}, {Uy}, {Green}, {Bertrand}, {Lesho}, \&
  {Hall}}]{wood2003}
{Wood}, B.~E., {Uy}, O.~M., {Green}, B.~D., {et~al.} 2003, in ESA Special
  Publication, Vol. 540, Materials in a Space Environment, ed. K.~{Fletcher},
  503--508

\bibitem[{Woronowicz \& Meadows(2012)}]{woronowitz2012}
Woronowicz, M. \& Meadows, G. 2012, in Optical System Contamination: Effects,
  Measurements, and Control 2012, ed. S.~A. Straka, N.~Carosso, \& J.~Egges,
  Vol. 8492, International Society for Optics and Photonics (SPIE), 849209

\bibitem[{Wright {et~al.}(2015)Wright, Wright, Goodson, Rieke, Aitink-Kroes,
  Amiaux, Aricha-Yanguas, Azzollini, Banks, Barrado-Navascues,
  Belenguer-Davila, Bloemmart, Bouchet, Brandl, Colina, Örs Detre,
  Diaz-Catala, Eccleston, Friedman, Garc{\'{\i}}a-Mar{\'{\i}}n, Güdel, Glasse,
  Glauser, Greene, Groezinger, Grundy, Hastings, Henning, Hofferbert, Hunter,
  Jessen, Justtanont, Karnik, Khorrami, Krause, Labiano, Lagage, Langer, Lemke,
  Lim, Lorenzo-Alvarez, Mazy, McGowan, Meixner, Morris, Morrison, Müller,
  rgaard Nielson, Olofsson, O'Sullivan, Pel, Penanen, Petach, Pye, Ray,
  Renotte, Renouf, Ressler, Samara-Ratna, Scheithauer, Schneider, Shaughnessy,
  Stevenson, Sukhatme, Swinyard, Sykes, Thatcher, Tikkanen, van Dishoeck,
  Waelkens, Walker, Wells, \& Zhender}]{wright2015}
Wright, G.~S., Wright, D., Goodson, G.~B., {et~al.} 2015, Publications of the
  Astronomical Society of the Pacific, 127, 595

\bibitem[{Yang {et~al.}(1985)Yang, Koros, Hopfenberg, \& Stannett}]{yang1985}
Yang, D.~K., Koros, W.~J., Hopfenberg, H.~B., \& Stannett, V.~T. 1985, Journal
  of Applied Polymer Science, 30, 1035

\bibitem[{Yang {et~al.}(1986)Yang, Koros, Hopfenberg, \& Stannett}]{yang1986}
Yang, D.~K., Koros, W.~J., Hopfenberg, H.~B., \& Stannett, V.~T. 1986, Journal
  of Applied Polymer Science, 31, 1619

\bibitem[{Yarygin {et~al.}(2017)Yarygin, Gerasimov, Krylov, Prikhodko,
  Skorovarov, \& Yarygin}]{yarygin2017}
Yarygin, V.~N., Gerasimov, Y.~I., Krylov, A.~N., {et~al.} 2017, Journal of
  Physics: Conference Series, 925, 012003

\bibitem[{Zhang \& Paige(2009)}]{zhang2009}
Zhang, J.~A. \& Paige, D.~A. 2009, Geophysical Research Letters, 36, L16203

\bibitem[{Zhao {et~al.}(2009)Zhao, Shen, Xing, \& Ma}]{zhao2009}
Zhao, X., Shen, Z., Xing, Y., \& Ma, S. 2009, Hangkong Xuebao/Acta Aeronautica
  et Astronautica Sinica, 30, 159

\bibitem[{Zitouni \& von Germersheim(2020)}]{zitouni2020}
Zitouni, B. \& von Germersheim, L. 2020, in Systems Contamination: Prediction,
  Control, and Performance 2020, ed. C.~E. Soares, E.~M. Wooldridge, \& B.~A.
  Matheson, Vol. 11489, International Society for Optics and Photonics (SPIE),
  114890F

\end{thebibliography}

\begin{appendix}
\onecolumn

\section{Non-exhaustive list of publications about contamination in spacecraft\label{sec:apdx-library}}

This list offers different entry points into contamination of spacecraft, from an empirical point-of-view and about evenly balanced between astrophysical space observatories and Earth-observation satellites (EOS). However, this balance is not representative, as the number of commercial and scientific EOS far exceeds that of astrophysical space observatories. There is a great wealth of literature available about the effects of contamination in EOS that can be explored following the references in the publications below.

\cite{schartel2022} and \cite{kirsch2005}, {\bf XMM-\textit{Newton}}: Hydrocarbon and other contaminants affect the EPIC and RGS X-ray instruments. A substantial throughput loss of 16\% to 56\% from $V$-band to the UV manifested during commissioning, with further degradation during the mission, from non-volatile organic contaminants. 

\cite{riello2021} and \cite{mora2016}, {\bf Gaia}: An initial heavy contamination with impact on throughput (likely scatter loss) and PSF had to be cleared by six thermal decontaminations over two years. The decontamination affected the optical alignment, that is SiC structures do not exactly revert to their original state. The contamination moved around the spacecraft, affecting different optical parts at different times.

\cite{wang2013}, {\bf GOES-12}: The Geostationary Operational Environmental Satellite experienced considerable water-ice contamination that resulted in throughput modulation from interference and absorption in its mid-infrared channels. Four thermal decontaminations were required. The ice thickness could have exceeded 2\,\micron, but a quantitative analysis is considered difficult.

\cite{hasegawa2019}, \cite{steinlechner2019}, \cite{tanioka2020}, and \cite{spallino2021}, {\bf KAGRA}: The Kamioka Gravitational Wave Detector uses cryogenic mirrors in a vacuum for enhanced sensitivity, and thus experiences the same contamination problems from outgassing as spacecraft.

\cite{quast2019} and \cite{decoster2013}, {\bf Meteosat}: The visible channel onboard the Meteosat First Generation (MFG) Meteosat-7 lost 20-30\% of its spectral transmission in the 450--900\,nm band between 1997 and 2017. The contamination is not specified, but was likely non-volatile and affected mostly the visible channel and not the infrared channels. An exponential fit described the overall degradation, and an additional linear factor the dependence on wavelength, with larger flux losses at shorter wavelengths. The same correction law is applicable to several other instruments onboard other first and second generation Meteosat spacecraft in geostationary orbits. Some of the data show typical interference effects from growing contamination layers, but they are either not recognised or not considered further. Degradation from atomic oxygen in low Earth orbits require nonlinear corrections in wavelength.

\cite{smith2012a}, \cite{smith2012b}, \cite{krijger2014}, and \cite{liebing2018}, {\bf ENVISAT AATSR and SCIAMACHY}: Heavy water contamination on a relay lens in AATSR caused periodic throughput oscillations with an amplitude of up to 1\% from interference fringes in the 560, 660, and 870\,nm channels. These were superimposed on a generic exponential throughput decay. AATSR was decontaminated about 50 times during 4000 days of operation, every time the ice reached a thickness of 10\,micron. The contamination eventually stabilised at around 70\,nm\,day$^{-1}$. In SCIAMACHY, molecular contamination on scan mirrors and an optical bench required optical modelling of the contaminant layer, with thicknesses varying between 1 up to 40\,nm. In 2009, a decontamination procedure resulted in increased contamination until 2011.

\cite{plucinsky2012}, {\bf Chandra}: X-ray telescopes can identify their own contaminants in spectra. Outgassing active for decades, and increased over time. Temperature changes due to deteriorating thermal insulation probably activated different outgassing sources.

\cite{goldberg2011}, {\bf GSICS}: The Global Space-Based Intercalibration System was established by the World Meteorological Organization (WMO) and the Coordination Group for Meteorological Satellites (CGMS). It allows cross-calibration of the international fleet of meteorological satellites, based on stable reference targets on Earth (deserts, ice shields) as well as the Sun, the Moon, and stars. In this way, individual contamination issues can be overcome for more accurate weather observations and long-term climate monitoring.

\cite{schlaeppi2010}, {\bf Rosetta}: Outgassing is highly susceptible to attitude changes. Over 100 different chemical substances were discovered in Rosetta's outgassing cloud. The pressure around the spacecraft was measurably increased compared to interplanetary levels. The contaminants travelled around the spacecraft and contaminated it on sides opposite to the outgassing sources (backscattering). The acceleration due to outgassing exceeded the Solar radiation pressure by an order of magnitude after attitude changes that exposed previously unilluminated surfaces. Instrument switch-ons had a notable impact on outgassing constituents.

\cite{haemmerle2006} and \cite{postberg2009}, {\bf Cassini NAC and CDA}: Direct evidence of scatter loss was evident in NAC images due to ice contamination (halos around stars). A fast decontamination (rapid heat-up) triggered more contamination. An organic layer on the CDA probably originated from UV-photolysed hydrocarbons elsewhere in the spacecraft. 

\cite{burnett2005} and \cite{calaway2006}, {\bf Genesis:} The Solar wind sample container showed pervasive dark stains, either from vacuum pyrolysis or UV photo-polymerisation.

\cite{matthews2005} and \cite{kinser1991}, {\bf CERES, GOME, LDEF}: The Cloud’s and the Earth’s Radiant Energy System (CERES) consists of four instruments measuring in three wide spectral passbands. The short-wave instruments cover the 0.3--5\micron~band and experienced considerable throughput degradation between 200 and 450\,nm. Similar throughput losses at short wavelengths were seen by the Global Ozone Monitoring Experiment (GOME). This is explained by atomic oxygen present in low Earth orbits that reacts with outgassing materials that then get baked onto optical surfaces by UV radiation. This was already recognised by the Long Duration Exposure Facility (LDEF) that -- among many other experiments -- exposed a range of materials and optical elements in 1984 before retrieving them again in 1990.

\cite{poole2008}, \cite{breeveld2010}, and \cite{kuin2015}, {\bf OM onboard SWIFT}: The optical monitor on SWIFT is a clone of XMM-\textit{Newton}. Owing to improved contamination control schemes it has shown little to no contamination.

\cite{bhaskaran2004}, {\bf STARDUST}: The optical navigation camera experienced recurrent contamination. Regular decontamination was needed including pointing the radiators into the Sun for sufficient heat generation. Long exposure times to mitigate throughput losses resulted sometimes in trailed images.

\cite{wilkes1999}, {\bf OPM on Mir space station}: The Optical Properties Monitor determined the reflectance, UV spectroscopy, and total integrated scatter loss of various materials exposed for eight months outside the Mir station. Degraded reflectance and increased scattering from particulate contamination were found.

\cite{uy1998}, \cite{uy2003}, and \cite{wood2003}, {\bf MSX}: The Midcourse Space Experiment found high a susceptibility of the outgassing rate to attitude changes. The contamination reservoirs did not deplete even after repeated Sun exposures, and contamination rates even increased over time.

\cite{clampin1992}, \cite{mackenty1993}, \cite{holtzmann1995}, \cite{tveekrem1996}, \cite{baggett1996}, \cite{baggett1998}, \cite{baggett2001}, \cite{gonzaga2010}, \cite{shanahan2017}, and \cite{bohlin2019}, {\bf HST WFPC1, WFPC2, WFC3}: Heavy non-volatile contamination impacted WFPC1, resulting in large throughput losses in the UV. WFPC2 was mostly contaminated by water, and had to be decontaminated every 28 days and then 49 days until decommissioning. The WFC3/IR channel sees very small throughput losses likely from photopolymerisation of non-volatile contaminants. 

\cite{leger1972}, {\bf Mercury, Gemini, Apollo}: The command capsule windows were contaminated by engine plumes, expulsed waste water, and outgassing sealants affected docking manoeuvres and other activities where clear visibility was required.

\section{Pertinent renderings and photographs\label{sec:apdx_photos}}
\begin{figure}[ht]
\centering
\includegraphics[angle=0,width=1.0\hsize]{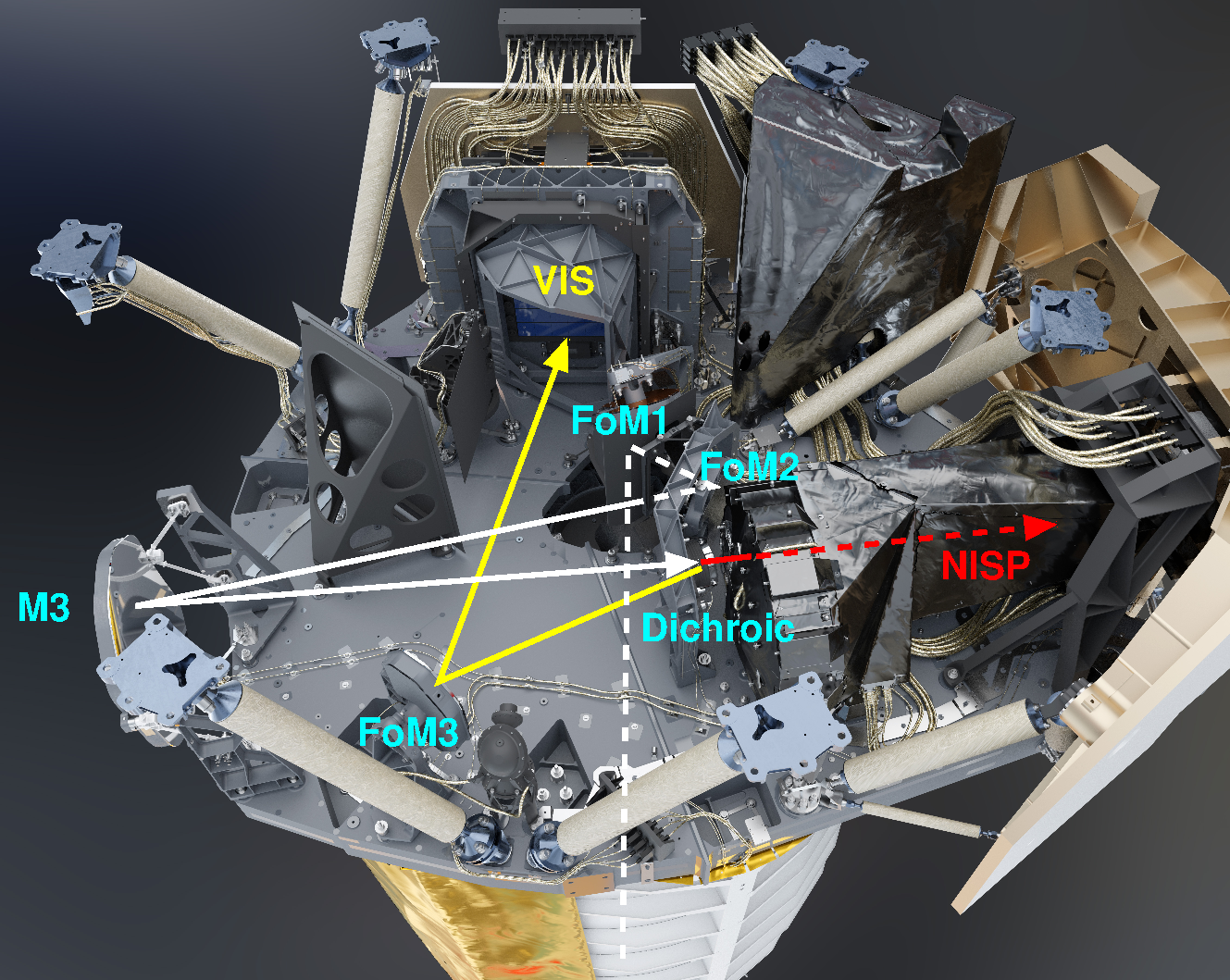}
\caption{3D CAD (computer-aided design) rendering of the instrument cavity. In this orientation the telescope is below the assembly and observing towards the bottom. For clarity, we have added the principal light path and optical components to the rendering. The similarity with the actual flight hardware shown in Fig.~\ref{fig:euclid_instcavity} is striking. It is evident from this figure that accurate models of water-ice contamination in the instrument cavity would need to account for a complex geometry of optical surfaces and water emitters. The NISP instrument begins immediately after the dichroic element and is completely covered in its own MLI; water in NISP will be mostly trapped and just redistributed during thermal decontamination. The large white structure to the right of NISP is its outward-facing radiator. It can be clearly seen in the photographs shown in Figs.~\ref{fig:euclid_mating_plm} and \ref{fig:euclid_thrusters_lowres}. A high-resolution image is available at \url{https://sci.esa.int/web/euclid/-/61034-euclid-payload-module}. Copyright of the rendering: Airbus Defence and Space - Toulouse.}
\label{fig:euclid_CAD_annotated}
\end{figure}

\begin{figure}[ht]
\centering
\includegraphics[angle=0,width=1.0\hsize]{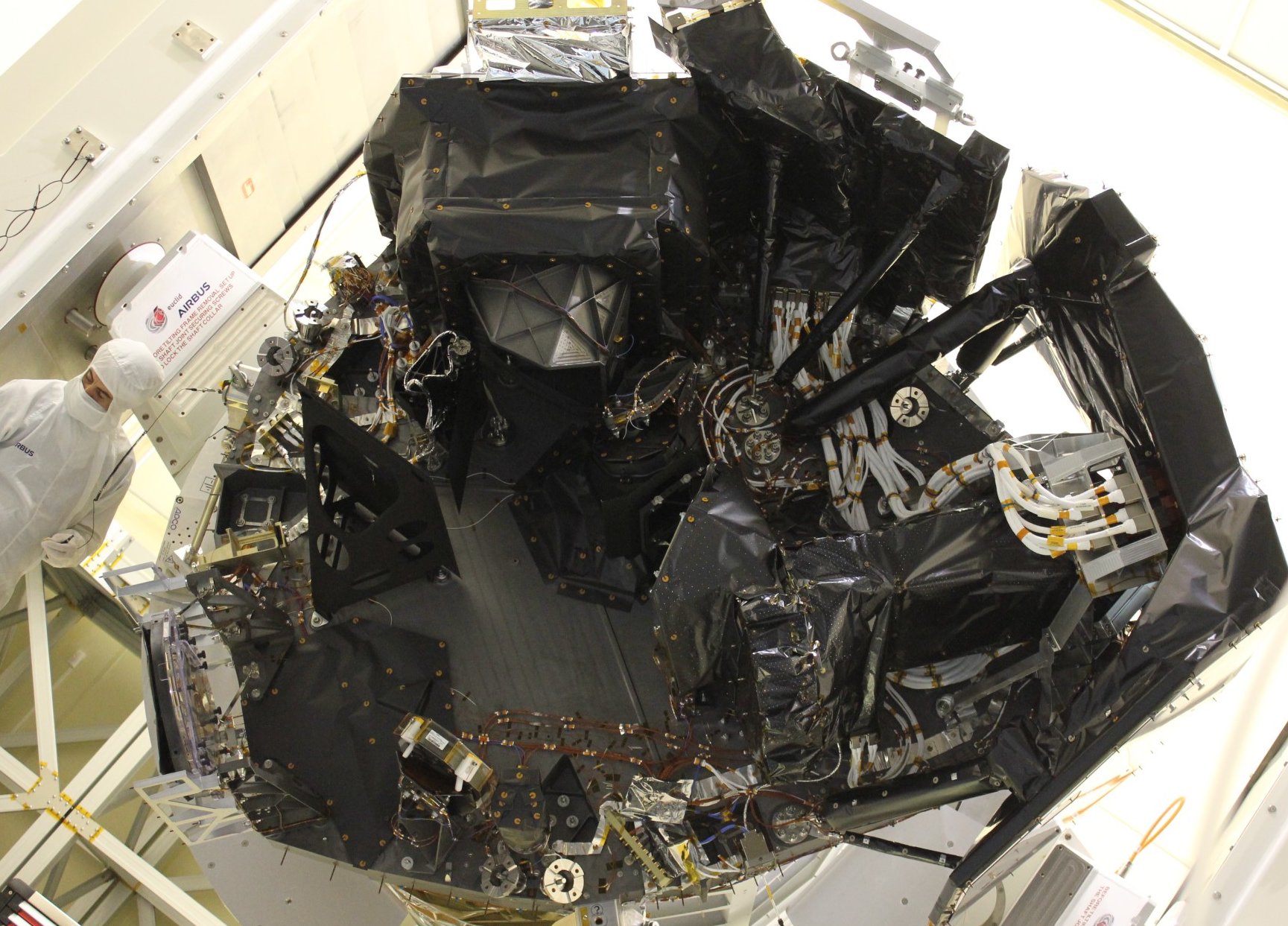}
\caption{Photograph of the flight-hardware instrument cavity (to be compared with Fig.~\ref{fig:euclid_CAD_annotated}). Evident are the large amounts of MLI, specifically the carbon-charged black Kapton that shield NISP (lower right) and numerous structural components. The orange-brown anti-static tapes are made from uncoated Kapton and are omnipresent. Kapton and MLI form primary water reservoirs. A high-resolution image is available at \url{https://sci.esa.int/web/euclid/-/the-euclid-payload-module}. Copyright: Airbus Defence and Space - Toulouse.} 
\label{fig:euclid_instcavity}
\end{figure}

\begin{figure}[ht]
\centering
\includegraphics[angle=0,width=1.0\hsize]{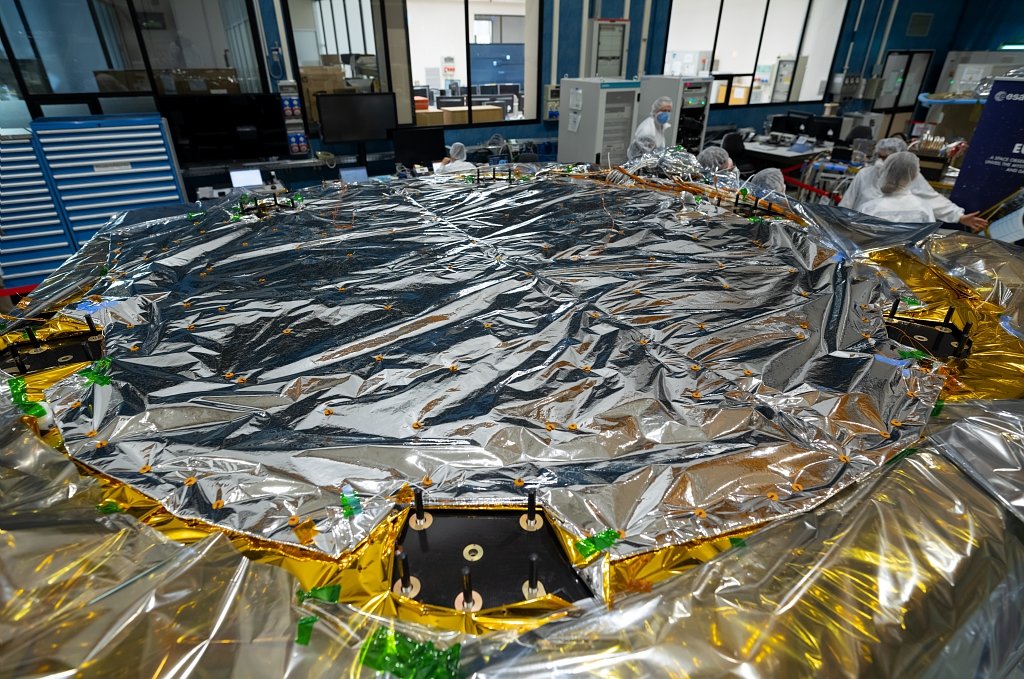}
\caption{Upper side of \Euclid's SVM that interfaces the PLM. The SVM is shielded with an aluminised SLI that is well visible in this photograph. The SLI prevents contaminants from escaping the SVM and entering the PLM with the scientific instruments and the telescope. The SVM was built by Thales Alenia Space. Credit: ESA - S. Corvaja.} 
\label{fig:euclid_svm2}
\end{figure}

\begin{figure}[ht]
\centering
\includegraphics[angle=0,width=1.0\hsize]{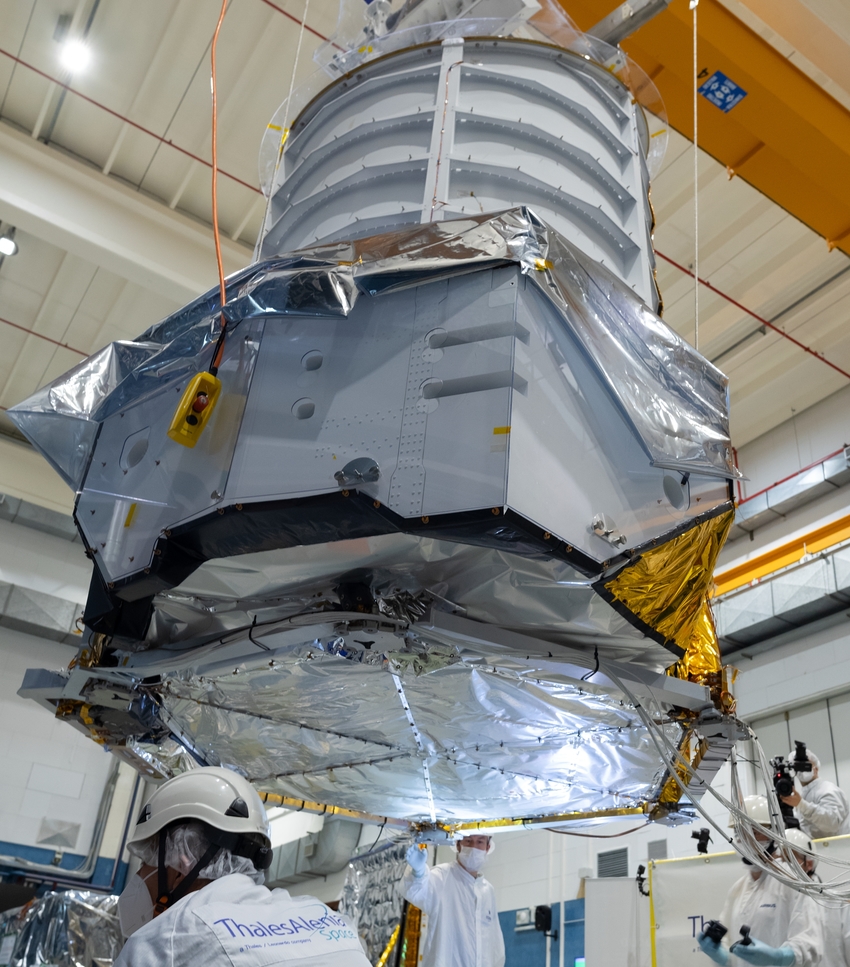}
\caption{Bottom of the PLM, shortly before the mating with the SVM shown in Fig.~\ref{fig:euclid_svm2}. The bottom is covered with MLI to provide thermal insulation and to prevent molecular contamination entering from the SVM. The large white polygonal structure girding the lower parts is the NISP radiator. A high-resolution image is available after account creation from the ESA Photo Library For Professionals at \url{https://www.esa-photolibrary.com/ESA/media/65700}. The PLM was built by Airbus Defence and Space. Credit: ESA - S. Corvaja.} 
\label{fig:euclid_mating_plm}
\end{figure}

\begin{figure}[ht]
\centering
\includegraphics[angle=0,width=1.0\hsize]{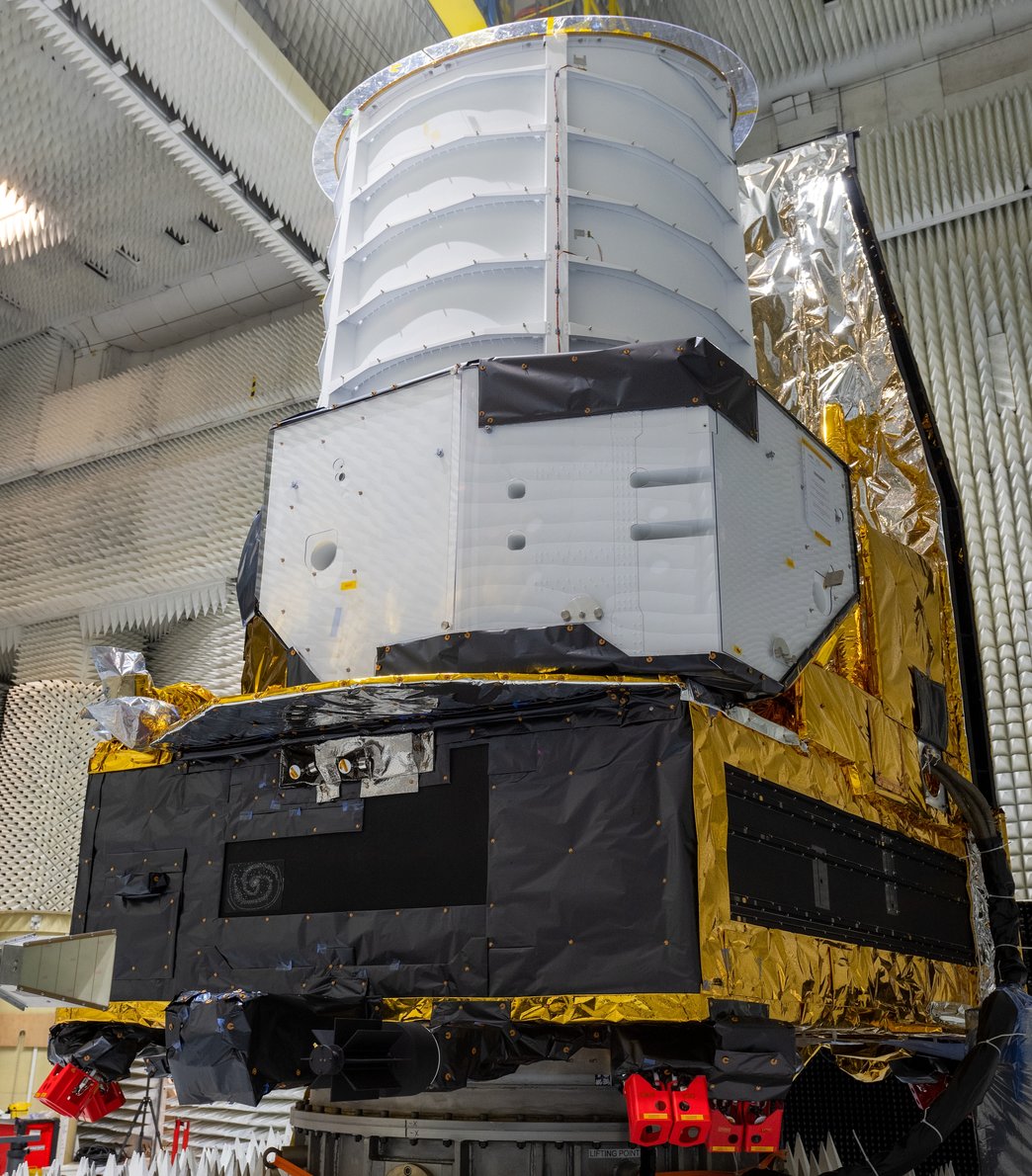}
\caption{Fully assembled spacecraft on February 2023 in the anechoic chamber of Thales Alenia Space in France, after completing final electromagnetic compatibility (EMC) tests. The side shown here will always face away from the Sun. The hydrazine thrusters on the SVM still have their protective red covers on, and point away from the PLM to minimise contamination in flight (see Sect.~\ref{sec:hydrazine}). The plaque with the miniaturised fingerprint galaxy created by Euclid Consortium members can be seen at the lower left of the SVM (\url{https://www.esa.int/ESA_Multimedia/Videos/2022/07/The_Fingertip_Galaxy_Reflecting_Euclid_in_art}). A high-resolution image is available at \url{https://www.esa-photolibrary.com/ESA/media/73223}. Credit: ESA - M. P\'edoussaut.} 
\label{fig:euclid_thrusters_lowres}
\end{figure}

\clearpage

\section{Tables}

\begin{table*}[ht]
\caption{Pure sublimation fluxes for various \Euclid components.}
\smallskip
\label{sublimation_rate_table}
\smallskip
\begin{tabular}{llll}
\hline
\rowcolor{gray!20}
Common path & ${\rm d}z/{\rm d}t\;\left(T_{\rm nominal}\right)$ & ${\rm d}z/{\rm d}t\;\left(T_{\rm warm}\right)$ & ${\rm d}z/{\rm d}t\;\left(T_{\rm decont.}\right)$ \\
\hline
M1   & $-$0.3\,nm\,month$^{-1}$ & $-$4.0\,nm\,month$^{-1}$ & $-$3.6\,\micron\,s$^{-1}$ \\
M2   & $-0.0$\,nm\,year$^{-1}$ & $-$1.0\,nm\,year$^{-1}$ & $-$2498\,\micron\,s$^{-1}$  \\
FoM1 & $-$4.0\,nm\,month$^{-1}$ & $-$27\,nm\,month$^{-1}$ & $-$3.6\,\micron\,s$^{-1}$ \\
FoM2 & $-$2.7\,nm\,month$^{-1}$ & $-$13\,nm\,month$^{-1}$ & $-$4.1\,\micron\,s$^{-1}$\\
M3   & $-$2.7\,nm\,month$^{-1}$ & $-$39\,nm\,month$^{-1}$ & $-$3.6\,\micron\,s$^{-1}$ \\
Dichroic & $-$2.7\,nm\,month$^{-1}$ & $-$13\,nm\,month$^{-1}$ & $-$4.1\,\micron\,s$^{-1}$ \\
\hline
\rowcolor{gray!20}
NISP path & & & \\
\hline
Corrector lens &  $-$55\,nm\,month$^{-1}$ & $-$79\,nm\,month$^{-1}$ & $-$2.8\,\micron\,s$^{-1}$ \\
Filter / Grism &  $-$156\,nm\,month$^{-1}$  & $-$157\,nm\,month$^{-1}$ & $-$0.56\,\micron\,s$^{-1}$ \\
Camera lenses & $-$111\,nm\,month$^{-1}$ & $-$111\,nm\,month$^{-1}$ & $-$0.42\,\micron\,s$^{-1}$ \\
Detector & $-$0.0\,nm\,year$^{-1}$ & $-$0.0\,nm\,year$^{-1}$ & $-$0.23\,\micron\,s$^{-1}$ \\
\hline
\rowcolor{gray!20}
VIS path & & & \\
\hline
FoM3     &  $-$0.5\,nm\,month$^{-1}$  & $-$4.0\,nm\,month$^{-1}$ & $-$3.6\,\micron\,s$^{-1}$ \\
Detector &  $-$1.5\,\micron\,day$^{-1}$ & $-$4.1\,\micron\,day$^{-1}$ & $-$579\,\micron\,s$^{-1}$ \\
\hline
\rowcolor{gray!20}
Structural elements & & & \\
\hline
Baffle global & $-$0.0\,nm\,year$^{-1}$ & $-0.3$\,nm\,year$^{-1}$ & $-$0.49\,\micron\,s$^{-1}$\\
PLM baseplate &  $-$1.2\,nm\,month$^{-1}$ & $-$8.8\,nm\,month$^{-1}$ & $-$0.64\,\micron\,s$^{-1}$ \\
\hline
\end{tabular}
\footnotesize
\tablefoot{These sublimation fluxes are pure, that is they neglect simultaneous condensation. They  were computed for the temperatures in Table \ref{euclid_temperatures} using Eqs.~(\ref{eq:murphy}) and (\ref{eq:subflux_amorph}), and are expressed as a change of ice film thickness. Total effective rates (sublimation and condensation) are given in Table \ref{contamination_rate_table}. The sublimation flux alone is visualised in Fig.~\ref{fig:sublimation_rate}.}
\end{table*}

\section{\label{apdx_contamination_M1M2}Contamination estimates for the telescope cavity (mirrors M1 and M2)}
\subsection{Telescope model and assumptions}
\begin{figure*}[ht]
\begin{center}
\includegraphics[angle=0,width=0.7\hsize]{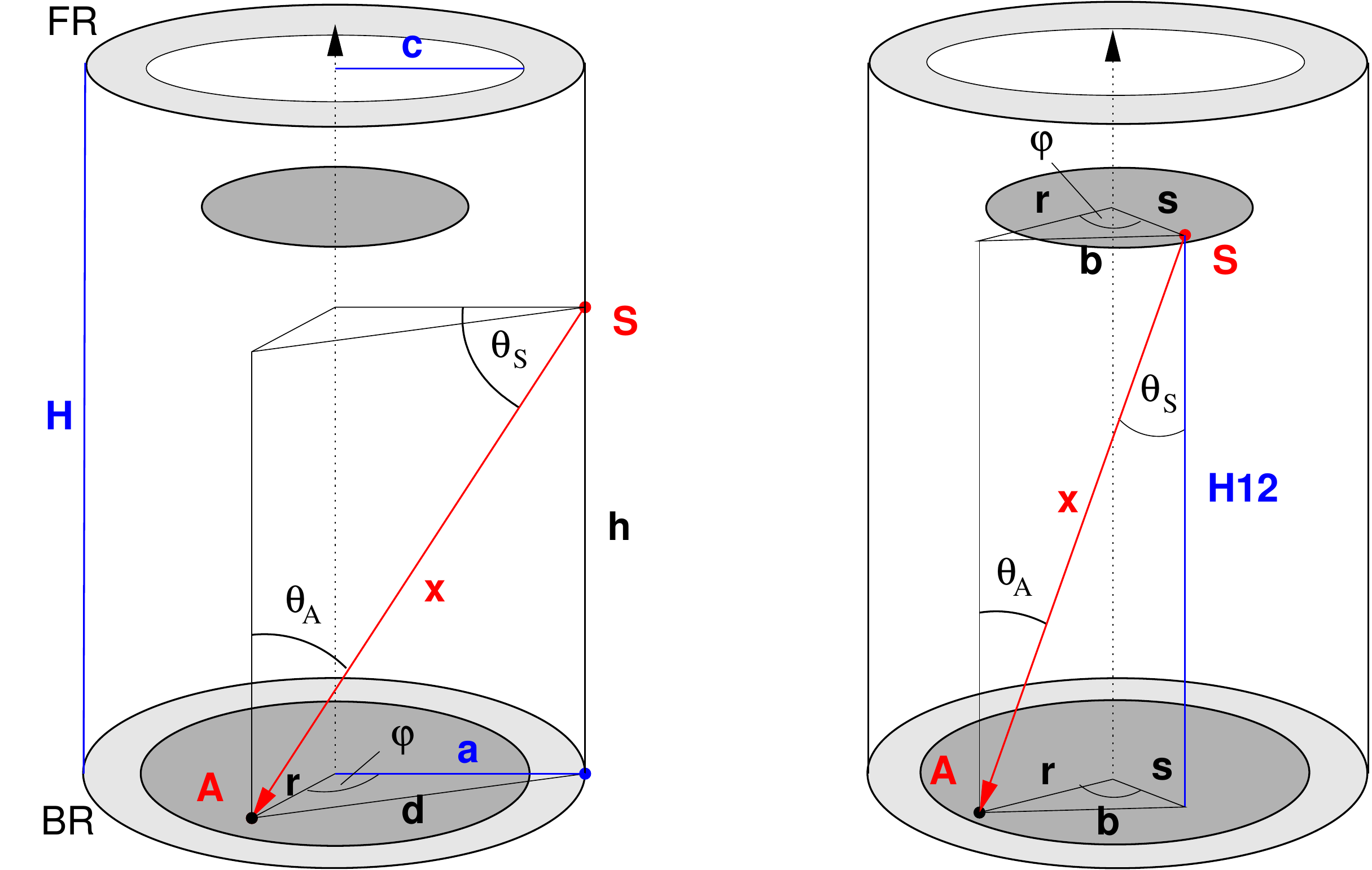}
\end{center}
\caption{Geometric model of the telescope to compute contamination on the optical surfaces. \textit{Left panel}: Contamination from a point $Q$ on the baffle wall onto a point $P$ on M1 (bottom grey disk). \textit{Right panel}: Contamination from a point $Q$ on M2 (top grey disk) onto $P$. The same setup can be used to compute contamination of M1 by sublimation from the `front ring' (FR), contamination of M2 from M1 and the `back ring' BR, and the water flux loss through the front aperture, simply by adjusting the respective integral bounds. The blue lines indicate fixed parameters listed in Table \ref{plm_model_dimensions}.}
\label{plm_model}
\end{figure*}

In this section we compute the sublimation-condensation rates on M1 and M2 for a glacial scenario (Sect.~\ref{sec:diffusion}), in which all surfaces including M1 and M2 are covered in ice. The ice is assumed to be thick, such that it does not deplete by sublimation. The geometric model of the cavity is shown in Fig.~\ref{plm_model}, and its dimensions are given in Table \ref{plm_model_dimensions}. We make the following simplifying assumptions: the surfaces of M1 and M2 are flat, the surface of M1 is located at the bottom of the external baffle, the baffle has a cylindrical interior wall, the temperature of the front ring (FR; reduced telescope aperture) is the same as the that of the baffle, and the temperature of the back ring (BR; structural parts between M1 and the baffle) is the same as that of the PLM baseplate. Further internal baffles, the hole in M1, the M2 telescope spider, and any other structures are ignored.

\begin{table}
\caption{Parameters of the geometric telescope model in Fig.~\ref{plm_model}.}
\label{plm_model_dimensions}
\smallskip
\begin{tabular}{lll}
\hline
Parameter & Value [m] & Comment \\
\hline
$H$ & 2.07 & Height of the external baffle. M1 is assumed to be flush with the bottom.\\
$H_{12}$ & 1.756 & Distance between M1 and M2 \\
$R_{\rm M1}$ & 0.625 & Radius of M1 \\
$R_{\rm M2}$ & 0.175 & Radius of M2 \\
$c$ & 0.6 & Radius of the front telescope aperture \\
$a$ & 0.855 & Radius of the baffle \\
\hline
\end{tabular}
\end{table}

\subsection{Flux from the baffle onto M1 and M2}
We consider a point $S$ on the baffle wall, at a height $h$ above the surface of M1 (left panel in Fig.~\ref{plm_model}). Point $S$ sublimes a molecule at angle $\theta_{\rm S}$ with respect to the surface normal, hitting a point $A$ on M1 at angle $\theta_{\rm A}$, with the following geometrical relations:
\begin{equation}
    {\rm cos}\,\theta_{\rm S} = \frac{a}{x}\;,\;\;\;
    {\rm cos}\,\theta_{\rm A} = \frac{h}{x}\;,\;\;\; 
    x=\sqrt{h^2+d^2}\;,\;\;\;{\rm and}\;\;\;
    d^2 = r^2 + a^2 - 2\, r\, a\; {\rm cos}\,\varphi\;.
\end{equation}
Integrating over the wall of the baffle, the flux of molecules (in molecules\,m$^{-2}$\,s$^{-1}$) received at a radius $r$ on M1 is
\begin{equation}
    \Phi_{\rm{M1;B}}(T, r)
    = \Phi_0(T)\;\int\displaylimits_0^{2\pi} \int\displaylimits_0^H\, x^{-2}\,
    {\rm cos}\,\theta_{\rm S}\; {\rm cos}\,\theta_{\rm A}\;
    a\;{\rm d}h\,{\rm d}\varphi
    = \Phi_0(T)\;\int\displaylimits_0^{2\pi} \int\displaylimits_0^H\, \frac{a^2\,h}{\left(h^2+d^2\right)^{2}}\;{\rm d}h\,{\rm d}\varphi\;.
\end{equation}
The flux from the baffle onto M2, $\Phi_{\rm{M2,B}}(T, r)$, is computed likewise, 
\begin{equation}
    \Phi_{\rm{M2;B}}(T, r) = \Phi_0(T)\;\int\displaylimits_0^{2\pi} \int\displaylimits_0^{H_{12}}\, 
    \frac{a^2\,h}{\left(h^2+d^2\right)^{2}}\;{\rm d}h\,{\rm d}\varphi\;.
\end{equation}

\subsection{Flux from M2 and the front ring onto M1}
The sublimation point $S$ is at a height $H_{12}$ above M1 (right panel in Fig.~\ref{plm_model}), with the geometrical relations
\begin{equation}
    {\rm cos}\,\theta_{\rm S} = \frac{H_{12}}{x}\;,\;\;\;
    \theta_{\rm A} = \theta_{\rm S}\;,\;\;\;
    x = \sqrt{H_{12}^2+b^2}\;,\;\;\;{\rm and}\;\;\;
    b^2 = r^2 + s^2 - 2\, r\, s\, {\rm cos}\,\varphi\;.
\end{equation}
Integrating over the surface of M2, the flux of molecules received at a radius $r$ on M1 is
\begin{equation}
    \Phi_{\rm{M1;M2}}(T, r) 
    = \Phi_0(T)\;\int\displaylimits_0^{2\pi} \int\displaylimits_0^{R_{\rm M2}}\,x^{-2}\,{\rm cos}^2\,\theta_{\rm S}\;
    s\;{\rm d}s\,{\rm d}\varphi
    = \Phi_0(T)\;\int\displaylimits_0^{2\pi} \int\displaylimits_0^{R_{\rm M2}}\,
    \frac{s\,H_{12}^2}{\left(H_{12}^2+b^2\right)^2}\;{\rm d}s\,{\rm d}\varphi\;.
\end{equation}
The flux from the front ring onto M1 is computed likewise,
\begin{equation}
    \Phi_{\rm{M1;FR}}(T, r) = \Phi_0(T)\;\int\displaylimits_0^{2\pi} \int\displaylimits_c^a\,
    \frac{s\,H^2}{\left(H^2+b^2\right)^2}\;{\rm d}s\,{\rm d}\varphi\;.
\end{equation}

\subsection{Flux from M1 and the back ring onto M2}
The mirror case, contamination of M2 by M1 and the back ring, is done the same way:
\begin{equation}
    \Phi_{\rm{M2;M1}}(T, r) = \Phi_0(T)\;\int\displaylimits_0^{2\pi} \int\displaylimits_0^{R_{\rm M1}}\,
    \frac{s\,H_{12}^2}{\left(H_{12}^2+b^2\right)^2}\;{\rm d}s\,{\rm d}\varphi\;.
\end{equation}
\begin{equation}
    \Phi_{\rm{M2;BR}}(T, r) = \Phi_0(T)\;\int\displaylimits_0^{2\pi} \int\displaylimits_{R_{\rm M1}}^a\,
    \frac{s\,H_{12}^2}{\left(H_{12}^2+b^2\right)^2}\;{\rm d}s\,{\rm d}\varphi\;.
\end{equation}

\subsection{\label{apdx_totalfluxM1M2}Total contamination flux on M1 and M2}
The total net sublimation-condensation rates for M1 and M2, in molecules\,m$^{-2}$\,s$^{-1}$, are given by the sum of the individual contributors, minus the sublimation fluxes from M1 and M2 themselves, 
\begin{equation}
    \Phi_{\rm M1}(r) = \Phi_{\rm{M1;B}}(T_{\rm B},r) + \Phi_{\rm{M1;M2}}(T_{\rm M2}, r) + \Phi_{\rm{M1;FR}}(T_{\rm FR}, r) - \Phi_{\rm tot}(T_{\rm M1})
\end{equation}
and
\begin{equation}
    \Phi_{\rm M2}(r) = \Phi_{\rm{M2;B}}(T_{\rm B},r) + \Phi_{\rm{M2;M1}}(T_{\rm M1}, r) + \Phi_{\rm{M2;BR}}(T_{\rm BR}, r) - \Phi_{\rm tot}(T_{\rm M2})\;.
\end{equation}
We convert the molecular fluxes into a growth rate ${\rm d}z/{\rm d}t$ for the thickness $z$ of the ice films. For this we need the mass of water molecules, $m=2.99\times10^{-26}$\,kg, and the density of ice \Ih, $\rho=917$\,kg\,m$^{-3}$. The change in ice film thickness on M1 and M2 is then
\begin{equation}
    \frac{{\rm d}z(r)}{{\rm d}t} = \frac{m}{\rho}\;\Phi_{\rm M1; M2}(r)\;,    
\end{equation}
which we scale to units of nm\,month$^{-1}$ (where a month is taken as 30 days).
%The dependence on radius is very small, of the order of $1$\% (Fig.~\ref{fig:mirror_surfaces}).
Mean contamination rates are given in Table \ref{contamination_rate_table}. The actual density of the ice may be smaller than $\rho=917$\,kg\,m$^{-3}$, in particular if amorphous components are still embedded in the crystalline ice, or if the bulk ice is amorphous throughout.

\subsection{Amount of water escaping through front aperture\label{sec:escapefraction}}
Using the same formalism, we compute the total flux of water molecules escaping through the front aperture. We take into account the reduced aperture size and the obstruction by M2. 
From the flux that sublimes from the baffle towards the front aperture, one must subtracted the flux that is deposited onto M2, and then integrate over the aperture annulus:

\begin{equation}
    \Phi_{\rm AP;B} = 
    %2\pi\, \Phi_0(T) \int\displaylimits_0^c \int\displaylimits_0^{2\pi} %\int\displaylimits_0^H d^2\,(h^2+d^2)^{-2}\,r\,a\,{\rm d}h\,{\rm d}\varphi\,{\rm d}r
    %- 2\pi \int\displaylimits_0^{\rm R_{M2}}\Phi_{\rm{M2;B}}(T, r)\,r\,{\rm d}r
    2\pi\, \Phi_0(T) \int\displaylimits_0^c
    \int\displaylimits_0^{2\pi} \int\displaylimits_0^H\, \frac{a^2\,h}{\left(h^2+d^2\right)^{2}}\;
    r\;{\rm d}h\,{\rm d}\varphi\,{\rm d}r
    - 2\pi \int\displaylimits_0^{R_{\rm M2}}\Phi_{\rm{M2;B}}(T, r)\,r\,{\rm d}r\;.
\end{equation}
Similarly, the total escaping fluxes from M1 and BR are
\begin{equation}
    \Phi_{\rm AP;M1} = 2\pi\, \Phi_0(T) 
    \int\displaylimits_0^c
    \int\displaylimits_0^{2\pi} 
    \int\displaylimits_0^{R_{\rm M1}}\,
    \frac{s\,H^2}{\left(H^2+b^2\right)^2}\;
    r\;{\rm d}s\,{\rm d}\varphi\,{\rm d}r\;
    - 2\pi \int\displaylimits_0^{R_{\rm M2}} \Phi_{\rm{M2;M1}}(T, r)\,r\,{\rm d}r\;,
\end{equation}
\begin{equation}
    \Phi_{\rm AP;BR} = 2\pi\, \Phi_0(T) 
    \int\displaylimits_0^c
    \int\displaylimits_0^{2\pi} 
    \int\displaylimits_{R_{\rm M1}}^a\,
    \frac{s\,H^2}{\left(H^2+b^2\right)^2}\;
    r\;{\rm d}s\,{\rm d}\varphi\,{\rm d}r\;
    -2\pi \int\displaylimits_0^{R_{\rm M2}} \Phi_{\rm{M2;BR}}(T, r)\,r\,{\rm d}r\;.
\end{equation}
The total flux of water molecules escaping through the front aperture is
\begin{equation}
    \Phi_{\rm AP}^{\rm tot} = \Phi_{\rm AP;B} + \Phi_{\rm AP;M1} + \Phi_{\rm AP;BR}\;.
\end{equation}

Under nominal operational temperatures (Table \ref{euclid_temperatures}), $1.1\times10^{12}$\,molecules\,s$^{-1}$ are escaping into space through the telescope aperture, corresponding to 89\,\textmu g\,month$^{-1}$. The contributions from the baffle, M1 and the back ring are 1\%, 26\% and 73\%, respectively.
For the warm case, $1.0\times10^{13}$\,molecules\,s$^{-1}$ or 788\,\textmu g\,month$^{-1}$ escape, 3\%, 36\% and 61\% coming from the baffle, M1 and the back ring, respectively. With all systems at their decontamination temperature, 898, 257 and 35\,mg\,s$^{-1}$
%without Sun exposure, 0.00145, 324 and 0.900\,mg\,s$^{-1}$ 
escape from the baffle, M1 and the back ring.

\begin{figure}[ht]
\begin{center}
\includegraphics[angle=0,width=1.0\hsize]{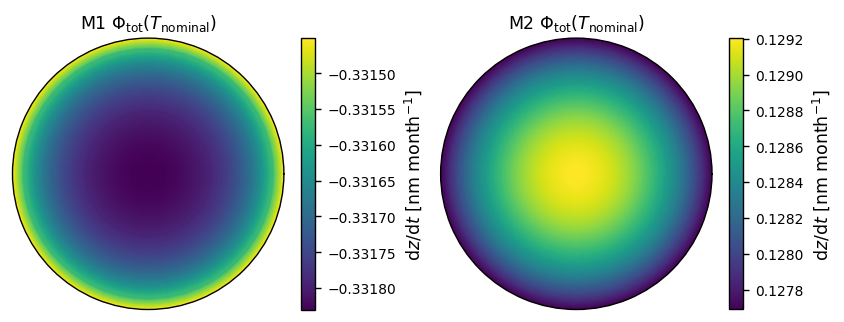}
\end{center}
\caption{Radial dependence of the contamination rate (change of ice thickness $z$) for mirrors M1 and M2 for nominal operating temperatures (Table \ref{euclid_temperatures}). M1 always self-decontaminates, that is ${\rm d}z/{\rm d}t<0$, whereas M2 slowly contaminates. The radial dependence is fairly uniform across mirror surfaces.} 
\label{fig:mirror_surfaces}
\end{figure}

\section{\label{apdx_contamination_NISPVIS}Contamination estimation for the instrument cavity}
The operating temperatures inside NISP are around 132\,K for the optics, filters, grisms and the calibration lamp, 120\,K for the baseplate, and 95\,K for the detectors. We assume that the interior surface of the MLI that covers NISP has a temperature of 126\,K (mean of the baseplate and internal temperatures). The instrument cavity is geometrically very complex, with different optical surfaces seeing different parts of their surrounding. 

For a simple model, we assume that detectors and lenses have flat surfaces, and sit at the centre of a hemisphere (implying $\theta_{\rm S}=0$) from which an omni-directional flux of water is incoming. The hemisphere shall have two constituents with temperatures $T_{\rm MLI}=126$\,K (25\% of the hemisphere) and $T_{\rm baseplate}=120$\,K (75\%), with the constituents evenly distributed in the hemisphere. The flux (or net sublimation-condensation rate) is computed in analogy to Eq.~(\ref{eq:phitots1}), and is found independent of the radius of the hemisphere,
\begin{equation}
    \Phi_{\rm NISP} = \pi\,\left[\frac{1}{4}\,\Phi_0(T_{\rm baseplate}) + \frac{3}{4}\,\Phi_0(T_{\rm MLI}) - \Phi_0(T_{\rm NISP})\right]\;.
\end{equation}
For a NISP lens at $T=132$\,K, this model implies a net contamination rate (change of ice thickness $z$) of 
\begin{equation}
\left(\frac{{\rm d}z}{{\rm d}t}\right)_{\rm LENS} = -238\,{\rm nm\,month}^{-1}\;.
\end{equation}
For the detector at $T=88$\,K we have
\begin{equation}
\left(\frac{{\rm d}z}{{\rm d}t}\right)_{\rm NISP\_DET} = +26.4\,{\rm nm\,month}^{-1}\;,
\end{equation}
independent of detector temperature (sublimation is negligible). The corresponding contamination rates are summarised in Table \ref{contamination_rate_table}.

The same hemispheric model as for NISP is assumed for the other optical surfaces in the instrument cavity, that is the dichroic, FoM1 to FoM3, and the VIS detectors. We assume that 75\% of the subliming surfaces are at 
$T_{\rm baseplate}=120$\,K and 25\% at $T_{\rm MLI}=126$\,K,
\begin{equation}
    \Phi_{\rm VIS} = \pi\,\left[\frac{3}{4}\,\Phi_0(T_{\rm baseplate}) + \frac{1}{4}\,\Phi_0(T_{\rm MLI}) - \Phi_0(T_{\rm VIS})\right]\;.
\end{equation}
Here, the term `VIS' refers to everything that is outside NISP. The corresponding contamination rates are summarised in Table \ref{contamination_rate_table}.

\end{appendix}

\end{document}